\documentclass[letterpaper, 11 pt, onecolumn]{IEEEtran}
\usepackage{epsfig,subfigure,latexsym,amssymb,amsmath}
\usepackage{bbm}

\newcommand{\eqdef} {\mathrel{\mathop:}=}
\newtheorem{definition}{Definition}[section]
\newtheorem{remark}{Remark}[section]
\newtheorem{theorem}{Theorem}[section]

\newtheorem{lemma}{Lemma}[section]
\newtheorem{proposition}{Proposition}[section]
\newtheorem{corollary}{Corollary}[section]

\newcommand{\be}{\begin{equation}}
\newcommand{\ee}{\end{equation}}
\newcommand{\ben}{\begin{enumerate}}
\newcommand{\een}{\end{enumerate}}
\newcommand{\bea}{\begin{eqnarray}}
\newcommand{\eea}{\end{eqnarray}}
\newcommand{\bean}{\begin{eqnarray*}}
\newcommand{\eean}{\end{eqnarray*}}

\def\squarebox#1{\hbox to #1{\hfill\vbox to #1{\vfill}}}
\def\eor{\hskip 3pt $\Diamond$}

\newcommand{\cX}{{\cal X}}
\newcommand{\cY}{{\cal Y}}
\newcommand{\cZ}{{\cal Z}}

\newcommand{\cC}{{\cal C}}
\newcommand{\cP}{{\cal P}}

\newcommand{\cS}{{\cal S}}

\newcommand{\cU}{{\cal U}}
\newcommand{\cT}{{\cal T}}
\newcommand{\cM}{{\cal M}}

\def\b0{\mathbf{0}}

\def\bbR{\mathbb{R}}
\def\bbZ{\mathbb{Z}}
\def\bx{\mathbf{x}}
\def\by{\mathbf{y}}

\def\mE{\textrm{E}}
\def\me{\textrm{e}}
\def\mI{\textrm{I}}
\def\mSP{\textrm{SP}}
\def\mD{\textrm{D}}
\def\mV{\mbox{Var}}
\def\b1{\mathbbm{1}}
\def\W1{W_{1^{-},R,P}}
\def\qs{Q^{\ast}_{R,P}}

\begin{document}
\baselineskip=1.5\normalbaselineskip

\title{Refinement of the Sphere--Packing Bound: Asymmetric Channels}
\author{Y\"{u}cel Altu\u{g}~\IEEEmembership{Student Member }and Aaron B.~Wagner~\IEEEmembership{Member}
\thanks{The material in this paper was presented in part at the 2012 IEEE International Symposium on Information Theory (ISIT), Boston, MA.

The authors are with the School of Electrical and Computer Engineering, Cornell University, 
Ithaca, NY, 14853, USA.
E-mail: {\em ya68@cornell.edu, wagner@ece.cornell.edu}.}}

\maketitle
\begin{abstract}
We provide a refinement of the sphere-packing bound for constant composition codes over asymmetric discrete memoryless channels that improves the pre-factor in front of the exponential term. The order of our pre-factor is $\Omega(N^{-\frac{1}{2}\left( 1+\epsilon + \rho_{R}^{\ast}\right)})$ for any $\epsilon >0$, where $\rho_{R}^{\ast}$ is the maximum absolute-value subdifferential of the sphere-packing exponent at rate $R$ and $N$ is the blocklength. 
\end{abstract}

\section{Introduction}
\label{sec:intro}

Characterizing the interplay between the rate, blocklength and error probability of the best block code(s) on a discrete memoryless channel (DMC) is a central problem of information theory. Although it has been investigated since the early days of the field~\cite{shannon48}--\cite{csiszar-korner81}, it is still an active research topic~\cite{val-fas04}--\cite{altug2012-b}. In a broad sense, there are two approaches to this problem:
\begin{enumerate}
\item[(i)] \textit{Finite blocklength results:} Because of the significance of the short to moderate blocklengths in practice, one can seek finite blocklength bounds on the error probability for a given rate. This can be done for a general class of channels (e.g. \cite{val-fas04}, \cite{wei-sason08}) or particular channels (e.g. \cite[Theorem~35]{polyanskiy10}, \cite[Theorem~38]{polyanskiy10}). Although these bounds are useful to assess the performance of practical codes, they are typically not conceptually illuminating.     

\item[(ii)]\textit{Asymptotic results:} An alternative to finite blocklength results is resorting to an infinite blocklength limit to derive more insightful results. Although such results do not give ``hard'' bounds that are valid for small blocklengths, they do provide memorable rules of thumb. Furthermore, finite blocklength bounds can often be extracted from their proofs.  
\end{enumerate}

We shall adopt the asymptotic approach in this paper. There exist several asymptotic regimes in the literature, such as \emph{error exponents} (e.g. \cite{fano61}--\cite{haroutunian68}), the \emph{normal approximation} (e.g. \cite{strassen62,polyanskiy10}) and \emph{moderate deviations} (e.g. \cite{altug10,polyanskiy10-a}). We call error exponents, the normal approximation and moderate deviations the \emph{small error probability}, \emph{large error probability}, and \emph{medium error probability} regimes, respectively. In this paper, our focus will be on the small error probability regime. This regime not only has theoretical significance, but has practical value in those applications, such as data storage, that require extremely small error probabilities without the aid of feedback. 

Classical asymptotic results on the small error probability regime focus only on determining the exponents. In particular, until recently, the tightest pre-factor for the upper bound on the error probability was $\Theta(1)$, due to Fano~\cite{fano61} and Gallager~\cite{gallager65}. The best pre-factor in the lower bound for constant composition codes was $\Theta(N^{-|\cX||\cY|})$, 
due to Haroutunian~\cite{haroutunian68}, \cite[Theorem 2.5.3]{csiszar-korner81}, where
$|\cX|$ and $|\cY|$ are the cardinalities of the input and output alphabets, respectively.
(The original sphere-packing bound, derived by Shannon-Gallager-Berlekamp~\cite[Theorem 2]{SGB67}, had an $\Theta(e^{-\sqrt{N}})$ pre-factor.) Clearly, there is a considerable gap between the orders of the pre-factors in the upper and lower bounds. 

Recently, the authors have been working to reduce the gap between the pre-factors. The recent paper \cite{altug11} considers symmetric channels and refines the sphere-packing lower bound by proving a pre-factor of $\Theta(N^{-\frac{1}{2}(1+|\mE_{\mSP}^{\prime}(R)|)})$, where $\mE_{\mSP}^{\prime}(R)$ is the slope of the sphere-packing exponent at point $R$. The paper \cite{altug12} proves a refined random coding bound with a pre-factor of $O(N^{-\frac{1}{2}(1 - \epsilon + \tilde{\rho}^\ast_R)})$ for any $\epsilon >0$, for a broad class of channels, which includes all positive channels with positive dispersion. Here, $\tilde{\rho}^\ast_R$ is related to the  subgradient of the random coding exponent, which reduces to $|\mE_{\textrm{r}}^\prime(R)|$ for the case of completely symmetric or positive and symmetric channels; hence the optimal order of the pre-factor is determined, up to the sub-polynomial terms.

This work is a generalization of \cite{altug11} to asymmetric channels. We prove a lower bound for constant composition codes with a pre-factor of $\Omega(N^{-\frac{1}{2}\left( 1+ \epsilon + \rho_{R}^{\ast}\right)})$ for any $\epsilon >0$, where $\rho_{R}^{\ast}$ is the maximum absolute-value subgradient of the sphere-packing exponent. While the essential approach is similar to that of \cite{altug11}, the asymmetry of the channel results in a significantly more involved argument compared to its symmetric counterpart. Although some improved finite-$N$ bounds could be extracted from the proofs in this paper, the task of optimizing these bounds and numerically comparing them to the existing bounds is not pursued, since we focus on the asymptotic characterization. 

An analogy to sums of i.i.d. random variables is instructive. The small, medium, and large error probability regimes of channel coding correspond to large deviations, moderate deviations, and central limit theory of i.i.d. sums of random variables, respectively. Along the same analogy, the setup of this work resembles the \emph{exact asymptotics} problem in large deviations \cite{bahadur-rao60}, \cite[Theorem~3.7.4]{dembo-zeitouni98}. This problem aims to determine the pre-factor of the exponentially vanishing term in the large deviations theorem. Bahadur and Ranga Rao \cite{bahadur-rao60} characterized this pre-factor, $\Theta(1/\sqrt{N})$, including the constant, under some regularity conditions. Their result, in the form stated by Dembo and Zeitouni~\cite[Theorem 3.7.4]{dembo-zeitouni98}, is the following: 
\begin{theorem}(Bahadur-Ranga Rao)
Let $\mu_N$ denote the law of $\hat{S}_N = \frac{1}{N}\sum_{i=1}^N Z_i$, where $Z_i$ are i.i.d. real valued random variables with logarithmic moment generating function $\Lambda(\lambda) = \log \mE[e^{\lambda Z_1}]$. Consider the set $A = [a, \infty)$, where $a = \Lambda^\prime(\eta)$ for some positive $\eta \in \{ \lambda : \Lambda(\lambda) < \infty\}^\circ$. If the law of $X_1$ is non-lattice, then $ \lim_{N \rightarrow \infty} J_N \mu_N(A)=1$, where 
\[
J_N \eqdef e^{N \Lambda^\ast(a)} \eta \sqrt{\Lambda^{\prime \prime}(\eta) 2 \pi N}
\] 
and $\Lambda^\ast(\cdot)$ is the Fenchel-Legendre transform of $\Lambda(\cdot)$.
\label{thrm:2}
\end{theorem}
If $X_{1}$ is a lattice random variable, then the order of the pre-factor is the same, but the constant is different. Hence, $\Theta(N^{-\frac{1}{2}})$ is the correct order of the pre-factor for i.i.d.\ sums of random variables, and this factor will appear in our channel coding result. When one reduces the error event of a code to a sum of independent random variables, however, the threshold $a$ must vary slightly with $N$, as will be evident in the sequel. This complicates the proof by preventing one from directly applying
the Bahadur-Ranga Rao result (the random variables are also not i.i.d.\, but 
merely independent). More importantly, the slow variation of the 
threshold changes the order of the pre-factor slightly, to include
the slope term mentioned above.

The remainder of the paper is devoted to the statement and then the 
proof of our result. 


\section{Notation, Definitions and Main Result}
\label{sec:notation-definitions}

\subsection{Notation}
\label{ssec:notation}
Boldface letters denote vectors, regular letters with subscripts denote individual elements of vectors. Furthermore, capital letters represent random variables and lowercase letters denote individual realizations of the corresponding random variable. Throughout the paper, all logarithms are base-$e$ unless otherwise is stated. For a finite set $\cX$, $\cP(\cX)$ denotes the set of all probability measures on $\cX$. Similarly, for two finite sets $\cX$ and $\cY$, $\cP(\cY|\cX)$ denotes the set of all stochastic matrices from $\cX$ to $\cY$. Given any $P \in \cP(\cX)$, $\cS(P) \eqdef \{ x \in \cX \, : \, P(x) >0\}$. $\b1_{\{ \cdot \} }$ denotes the standard indicator function. Given two probability measures $\lambda_{1}, \lambda_{2}$, $\lambda_{1} \ll \lambda_{2}$ means `$\lambda_{1}$ is absolutely continuous with respect to $\lambda_{2}$' and $\lambda_{1} \equiv \lambda_{2}$ is equivalent to saying $\lambda_{1} \ll \lambda_{2} $ and $\lambda_{2} \ll \lambda_{1}$. $\Phi$ (resp. $\phi$) denotes the distribution (resp. density) of the standard Gaussian random variable. For a set $\cS$; $\cS^{c}$, $\textrm{cl}(S)$, $\cS^{\circ}$ and $\mbox{ri}(\cS)$ denotes complementary set, closure, interior and relative interior, respectively. $\bbR_{+}, \bbR^{+}$ and $\bbZ^{+}$ denotes the set of non-negative real numbers, positive real numbers and positive integers, respectively.  

\subsection{Definitions}
\label{ssec:defn}
Throughout the paper, let $W$ be a DMC satisfying\footnote{For the definition of $R_{\infty}$, see \cite[pg. 170]{csiszar-korner81}.} $R_{\infty} < C$. For any $P \in \cP(\cX)$, define
\begin{equation*}
\mE_{\mSP}(R,P) \eqdef \min_{V \, \in \, \cP(\cY|\cX) \, : \, \mI(P;V) \leq R}\mD(V||W|P), \label{eq:opt-SP-primal}
\end{equation*}
and $\mE_{\mSP}(R) \eqdef \max_{P \in \cP(\cX)} \mE_{\mSP}(R,P)$. 

The following can be shown\footnote{Since $\mE_{\mSP}(\cdot, P)$ is convex for all $P \in \cP(\cX)$, $\mE_\mSP(\cdot, \cdot)$ is continuous on $(R_\infty,\infty) \times \cP(\cX)$ (cf. Lemma~\ref{lem:opt-cla12} in the Appendix~\ref{app:trivial-types}) and $\cP(\cX)$ is compact, one can invoke the characterization of the subdifferential of the maximum function (e.g. \cite[Theorem 2.87]{Ruszczynski2006}) to deduce that $\partial \mE_{\mSP}(R) = \textrm{conv}\left( \cup_{P : \mE_{\mSP}(R,P) = \mE_\mSP(R)} \{ \partial \mE_{\mSP}(\cdot, P)(R)\}\right)$, where $\textrm{conv}(\cdot)$, $\partial \mE_{\mSP}(R)$ and $\partial \mE_{\mSP}(\cdot, P)(R)$ denotes the \emph{convex hull}, \emph{subdifferential of $\mE_{\mSP}(\cdot)$ at point $R$} and \emph{subdifferential of $\mE_{\mSP}(\cdot,P)$ at point $R$}, respectively. This observation, coupled with the differentiability of $\mE_{\mSP}(\cdot, P)$, i.e. Proposition~\ref{prop:diff-ESP}, and the continuity of $\mE_{\mSP}^\prime(R,\cdot)$, i.e. Proposition~\ref{prop:cont-saddle-point}, suffices to conclude the claim.} to be the \emph{maximum absolute value subgradient} of the sphere packing exponent at point $R$
\begin{equation}
\rho_{R}^{\ast} \eqdef \max_{P \, \in \, \cP(\cX)  :  \mE_{\mSP}(R,P) = \mE_{\mSP}(R)}|\textrm{E}^{\prime}_{\textrm{SP}}(R,P)|,
\label{eq:worst-slope}
\end{equation}
where $\textrm{E}^{\prime}_{\textrm{SP}}(R,P)$ denotes the slope\footnote{One can show that $\mE_{\mSP}(R,P)$ is differentiable with respect to $R$, for given $P$, provided that $R_\infty < R < C$ and $\mE_{\mSP}(R,P) >0$, e.g. Proposition~\ref{prop:diff-ESP}.} of $\textrm{E}_{\textrm{SP}}(\cdot,P)$ at point $R$.

Given any $(N,R)$ code $(f, \varphi)$, let $e(f, \varphi)$ (resp. $e_{m}(f, \varphi)$) denotes its maximal error probability (resp. error probability of the $m$-th message).  

Let $\cZ$ be a finite set and $Q, \hat{Q} \in \cP(\cZ)$. A \emph{deterministic hypothesis test}, $T \, : \, \cZ \rightarrow \{ 0,1\}$, over the set $\cZ$ in which $Q$ is the null hypothesis ($H_0$) and $\hat{Q}$ is the alternate hypothesis ($H_1$) is defined as 
\begin{equation*}
 T(z) = 
 \begin{cases}
 0, & \mbox{ if } z \in \cU_T, \\
 1, & \mbox{ if } z \in \cU_T^c,
 \end{cases}
\end{equation*}
where $\{ \cU_T, \cU_T^c\}$ are called the \emph{decision regions} of the test. Let $\cT(Q,\hat{Q})$ denote the set of all deterministic tests between $Q$ and $\hat{Q}$. The error probabilities associated with $T$ are defined as $\alpha_T \eqdef Q\{ \cU_T^c\}$ and $\beta_T \eqdef \hat{Q}\{ \cU_T\}$. For any $r >0$, define 
\begin{equation}
\alpha^\ast_{Q,\hat{Q}}(r) \eqdef \min_{T \in \cT(Q, \hat{Q}) : \beta_T \leq e^{-r}}\alpha_T.
\label{eq:opt-err-prob}
\end{equation}

\subsection{Main Result}

\begin{theorem}
Consider any $R \in (R_\infty , C)$ and $\zeta \in \bbR^+$. Then, for any sufficiently large $N$, depending on $R$, $W$ and $\zeta$ and any $(N,R)$ constant composition code $(f,\varphi)$,
\begin{equation} 
e(f,\varphi) \geq K \frac{e^{-N\textrm{E}_{\textrm{SP}}(R)}}{N^{\frac{1}{2}\left( 1+ (1+\zeta)\rho_{R}^{\ast}\right)}}, 
\label{eq:thrm-main}
\end{equation}
where $K \in \bbR^+$ is a constant that depends on $R$, $W$ and $\zeta$. 
\label{thrm:main}
\end{theorem}

\section{Proof of Theorem~\ref{thrm:main}}
\label{sec:pf}

\subsection{Overview}
\label{ssec:hyp-test}

There are at least three proofs of the sphere-packing bound
in the literature: that of Shannon-Gallager-Berlekamp~\cite{SGB67}, Haroutunian~\cite{haroutunian68} and Blahut~\cite{blahut74}. Of these, Blahut's
argument seems to be the most natural starting point for obtaining
improved pre-factors, as it allows one to convert the error event
of a code into an event involving a sum of i.i.d.\ random variables,
to which one can apply the Bahadur-Ranga Rao result. The Shannon-Gallager-Berlekamp
 argument is similar to Blahut's in some ways, but it is less amenable
to exact asymptotics. The Haroutunian argument is combinatorial and
even farther removed from i.i.d.\ sums.

Blahut's argument proceeds as follows.
Assume $R_{\infty} < R < C$ and let $(f, \varphi)$ be an $(N,R)$ code. Let $\{ \cU_{m}\}_{m \in \cM}$ denote the decision regions of $\varphi$ corresponding to each message $m \in \cM$. Let $Q \in \cP(\cY)$ be an auxiliary output distribution. Let $W(\by^{N}|\bx^{N}) \eqdef \prod_{n=1}^{N}W(y_{n}|x_{n})$ and $Q(\by^{N}) \eqdef \prod_{n=1}^{N}Q(y_{n})$.
Since $\sum_{\by^N \in \cY^N}Q(\by^N)=1$ and $|\cM| \geq e^{NR}$, there must be a message $m \in \cM$ such that  $Q\{ \cU_{m} \} \leq e^{-NR}$. Let $\bx^{N} \eqdef f(m)$ be the codeword for this message. It is clear that $e(f, \varphi) \geq e_{m}(f, \varphi) = W\left\{ \cU_{m}^{c} | \bx^{N}\right\}$.

Now consider the hypothesis test over the set
$\cY^{N}$ in which $W(\cdot|\bx^{N})$ is the
null hypothesis ($H_0$) and the i.i.d.\ output distribution $Q$ is the alternate hypothesis ($H_1$).
One feasible test is to accept $H_0$ on $\cU_{m}$ and $H_1$ on
$\cU_{m}^{c}$, resulting in type-I and type-II error probabilities
of $W(\cU_m^{c}|\bx^N) = e_{m}(f, \varphi)$ and
$Q\{ \cU_{m} \}$, respectively. Since $\alpha^\ast_{W(\cdot|\bx^N), Q}(NR)$ denotes the minimum type-I error probability,
optimized over all tests, subject to the constraint that
the type-II error probability does not exceed $e^{-NR}$ (cf. \eqref{eq:opt-err-prob}), we evidently must have
\begin{equation}
e(f, \varphi) \geq \alpha^\ast_{W(\cdot|\bx^N), Q}(NR).
\label{eq:hyp-test-4-q}
\end{equation}

The error exponent of this test can be expressed via the following
definition.
For any $V \in \cP(\cY|\cX)$, $P \in \cP(\cX)$ and $Q \in \cP(\cY)$, define $\mD(V||Q|P) \eqdef \sum_{x \in \cX}P(x) \mD(V(\cdot|x)||Q)$.

\begin{definition} For any $r \in \bbR_{+}$, $P \in \cP(\cX)$ and $Q \in \cP(\cY)$ 
\begin{equation}
\me_{\mSP}(Q,P,r) \eqdef \inf_{V \in \cP(\cY|\cX) \, : \, \mD(V||Q|P) \leq r}\mD(V||W|P).
\label{eq:opt-hyp-exp}
\end{equation}
\label{def:opt-hyp-exp}
\end{definition}
\vspace{-.25cm}
Then the optimal type-I error exponent can be shown to be (e.g. \cite[Section V]{blahut74}) $\me_{\mSP}(Q,P,R)$, where $P$ is the empirical distribution of $\bx^N$. 

Note that this exponent depends on the output distribution $Q$, 
which is to be selected. This distribution can be chosen to 
depend on $P$, since it can depend on the code, although
allowing such dependence necessitates a restriction to
constant composition codes. In the original
argument~\cite[Section V]{blahut74}, this freedom is not used, and $Q$ depends
on $R$ (and the channel) but not $P$. Pre-factors aside,
it is not clear that this choice yields the standard
sphere-packing exponent when \eqref{eq:opt-hyp-exp} is maximized over $P$.
This is asserted to be the case in \cite[Theorem 19]{blahut74} and \cite[Theorem 10.1.4]{blahut87}, but each of
these proofs has a nontrivial gap\footnote{Specifically, the argument for \cite[Theorem 19]{blahut74} seems to proceed as if Lagrange multipliers of $\max_P \mE_{\mSP}(R,P)$ and $\max_{P}\me_{\mSP}(Q,P,R)$ are the same, which is not evident. For \cite[Theorem 10.1.4]{blahut87}, only $\me_{\mSP}(Q,P_{R}^{\ast},R) = \max_{P}\mE_{\mSP}(R,P)$ is shown, where $P^\ast_R$ attains $\max_{P}\mE_{\mSP}(R,P)$, which does not imply the claim.}. Moreover, a
numerical study indicates that for the Z-channel and for
this choice of $Q$,
$\mE_{\mSP}(R) < \max_{P}\me_{\mSP}(Q,P,R)$, 
for a broad range of rates. For symmetric channels,
$Q$ can indeed be chosen independently of $P$~\cite{altug11},
and so the code need not be constant composition.
But in the general case, it appears that some dependence
is necessary if one hopes to obtain the sphere-packing
exponent. 

Our choice of $Q$ will depend on $P$ and give the
sphere-packing exponent. Thus, one of the ancillary
contributions of this paper is to give a complete
proof that the hypothesis testing reduction described
can be used to obtain the sphere-packing exponent.
In fact, using the hypothesis testing reduction, we shall prove
the stronger result that the exponent on the error probability of any
constant-composition code with composition $P$ 
is upper bounded by $\mE_{\mSP}(R,P)$; previously,
the only proof of this fact used combinatorial techniques.

It is worth noting that the Shannon-Gallager-Berlekamp
proof also involves the choice of an output distribution.
Their choice of output distribution also depends on $P$, but 
it is defined differently from ours. Our choice yields the
$\mE_{\mSP}(R,P)$ exponent, whereas Shannon-Gallager-Berlekamp
only establish an exponent of $\mE_{\mSP}(R)$. 

Before concluding this section, it is instructive to consider a binary symmetric channel (BSC) with crossover probability $p \in (0,1/2)$ in order to see why the slope related term arises in Theorem~\ref{thrm:main}. One can check that the output distribution mentioned in \cite[Eq. 9]{altug11} reduces to the uniform distribution and for these particular choices, 
\begin{equation}
\alpha^\ast_{W(\cdot|\bx^N), Q}(NR) \geq \sum_{n = n^\ast_R+1}^N \binom{N}{n}p^n(1-p)^{N-n} = \Pr\left\{ \frac{1}{N}\sum_{n=1}^{N}Z_{n} \geq \frac{n^{\ast}_{R}+1}{N}\right\},
\label{eq:overview-rem2-1}
\end{equation}
where $\{ Z_{n}\}_{n=1}^{N}$ are i.i.d. Bernoulli random variables with parameter $p$ and $n^\ast_R$ is the largest $k \in \bbZ^+$ satisfying
\begin{equation}
e^{- NR} \geq \sum_{n=0}^{k} \binom{N}{n}2^{-N} = \Pr \left\{ \frac{1}{N}\sum_{n=1}^{N} \tilde{Z}_{n} \leq \frac{k}{N}\right\}, \label{eq:overview-rem2-2}
\end{equation}
where $\{ \tilde{Z}_{n}\}_{n=1}^{N}$ are i.i.d. Bernoulli random variables with parameter $1/2$. 
Provided that $k/N < 1/2$, one can apply Theorem~\ref{thrm:2} to the right side of \eqref{eq:overview-rem2-2} to have 
\begin{equation}
\Pr \left\{ \frac{1}{N}\sum_{n=1}^{N} \tilde{Z}_{n} \leq \frac{k}{N}\right\} \geq \frac{K_{1}}{\sqrt{N}}e^{-N \mD\left(\frac{k}{N}||\frac{1}{2}\right)},
\label{eq:overview-rem2-3}
\end{equation} 
where $\mD\left( k/n||1/2\right) := k/n \log \frac{k/n}{1/2} +(1- k/n)\log \frac{1-k/n}{1/2}$ and $K_{1}$ is a positive constant. Plugging \eqref{eq:overview-rem2-3} into \eqref{eq:overview-rem2-2} and recalling the definition of $n^{\ast}_{R}$, one can verify that 
\begin{equation}
\frac{n_{R}^{\ast}}{N} \leq \textrm{h}^{-1}\left(\log 2 - R + \frac{\log \sqrt{N}}{N} - \frac{\log K_{1}}{N} \right) \label{eq:overview-rem2-4}
\end{equation}
By plugging \eqref{eq:overview-rem2-4} into \eqref{eq:overview-rem2-1}, applying Theorem~\ref{thrm:2} on the right side of \eqref{eq:overview-rem2-1} and carrying out the algebra, one can verify that 
\begin{equation}
\alpha^\ast_{W(\cdot|\bx^N), Q}(NR) \geq \frac{K_{2}}{\sqrt{N}}e^{-N \mE_{\mSP}\left( R - \frac{\log \sqrt{N}}{N} \right)} \geq \frac{K_{3}}{N^{0.5(1+|\mE_{\mSP}^{\prime}(R)|)}}e^{-N\mE_{\mSP}(R)}, \label{eq:overview-rem2-5}
\end{equation}
where $K_{2}, K_{3}$ are positive constants and the last inequality
follows by expanding $\mE_{\mSP}(\cdot)$ as a power series about $R$.
Note that if $\frac{n_{R}^{\ast}}{N}$ were constant in $N$, then
applying Theorem~\ref{thrm:2} to (\ref{eq:overview-rem2-1}) would 
give a pre-factor with
an order of $1/\sqrt{N}$. But Eq.~(\ref{eq:overview-rem2-4}) shows that 
$\frac{n_{R}^{\ast}}{N}$ increases with $N$ at a rate of $(\log N)/N$.
While this increase is too slow to affect the exponent, it does affect
the order of the pre-factor.

Finally, note that the arguments leading to \eqref{eq:overview-rem2-5} are nothing but the ``packing of Hamming spheres''. To be specific, one can check that (e.g. \cite{elias55}) for this channel, the error probability of any $(N,R)$ code is lower bounded by that of a hypothetical ``sphere-packed code'' with the same parameters. A sphere-packed code is a code such that the decoding region of each codeword is an Hamming sphere of a certain radius, say $\lceil N \delta(R) \rceil$ with $\delta(R) >0$, possibly excluding some strings in the outermost layer and the union of these spheres equals $\{ 0,1\}^N$. For the sphere-packed code, an error occurs when the noise pushes the received signal outside of the Hamming ball of radius $n^\ast_R$ centered at the codeword, whose probability is precisely the right side of \eqref{eq:overview-rem2-1}. By employing the upper bound given in \eqref{eq:overview-rem2-4}, one can deduce \eqref{eq:overview-rem2-5}. 

By continuing this sphere-packing analogy, one can intuitively view the lower bound obtained via the hypothesis testing reduction as the error probability of a hypothetical sphere-packed $(N,R)$ code on $\cY^N$ with $\log \frac{Q(\cdot)}{W(\cdot|\bx^N)}$ used instead of Hamming distance. Note that the extra term in the pre-factor essentially stems from the approximation of the ``maximal packing radius'' of the spheres under this metric.


\subsection{Selecting the output distribution}
\label{ssec:output-dist}

In order to describe our output distribution, we require the following technical results.

For any $Q \in \cP(\cY)$ and $\lambda \in [0,1)$, define
\begin{equation*}
\Lambda_{Q, P}(\lambda) \eqdef 
\begin{cases}
\mE_{P}\left[\log \mE_{W(\cdot|X)}\left[ \left(\frac{Q(Y)}{W(Y|X)}\right)^{\lambda}\right]\right], &  \lambda \in (0,1), \\
0, &  \lambda=0.
\end{cases}
\label{eq:opt-moment-gen}
\end{equation*} 
For any $R \in \bbR^+$, define 
\begin{align}
\cP_{R}(\cX) & \eqdef \{ P \in \cP(\cX) : \mE_{\mSP}(R,P) >0 \}, \label{eq:non-trivial-types} \\
\cP_{P,W}(\cY) & \eqdef \{ Q \in \cP(\cY) \, : \, \forall x \in \cS(P), \, \cS(Q) \cap \cS(W(\cdot|x)) \neq \emptyset\}, \label{eq:P-PW} \\
\tilde{\cP}_{P,W}(\cY) & \eqdef \{ Q \in \cP(\cY) \, : \, \forall x \in \cS(P), \, Q \gg W(\cdot|x)\}. \label{eq:tilde-P-PW}
\end{align}
Further, given any $R > R_\infty$ and $P \in \cP(\cX)$, 
\begin{equation}
K_{R,P}\, : \, \bbR_{+} \times \cP_{P,W}(\cY) \rightarrow \bbR, \mbox{ s.t. } K_{R,P}(\rho, Q) = -\rho R - (1+\rho)\Lambda_{Q,P}\left( \rho/(1+\rho)\right), \label{eq:K-RP}
\end{equation}
for all $(\rho, Q) \in \bbR_{+} \times \cP_{P,W}(\cY)$. 

\begin{proposition}(Saddle-point)
Consider any $R_\infty < R < C$ and $P \in \cP_R(\cX)$.
\begin{enumerate}
\item[(i)] $K_{R,P}(\cdot, \cdot)$ has a saddle-point with the saddle-value $\mE_{\mSP}(R,P)$.
\item[(ii)] Any saddle-point of $K_{R,P}(\cdot, \cdot)$, say $(\rho^\ast, Q^\ast)$, satisfies $(\rho^\ast, Q^\ast) \in \bbR^+ \times \tilde{\cP}_{P,W}(\cY)$. 
\end{enumerate}
\label{prop:saddle-point}
\end{proposition}

\begin{IEEEproof}
The proof is provided in the Appendix~\ref{app:prop-saddle-point}.
\end{IEEEproof}

Let $S(R,P)$ denote the set of saddle-points of $K_{R,P}(\cdot, \cdot)$. Moreover, 
\begin{align}
S(R,P)|_{\bbR_+} & \eqdef  \{\rho \in \bbR_+ : \exists \, Q \in \cP_{P,W}(\cY), \mbox{ s.t. } (\rho, Q) \in S(R,P) \} \label{eq:S_positive}, \\
S(R,P)|_{\cP_{P,W}(\cY)} & \eqdef  \{ Q \in \cP_{P,W}(\cY)  : \exists \, \rho \in \bbR_+, \mbox{ s.t. } (\rho, Q) \in S(R,P) \}, \label{eq:S_Q}
\end{align}

\begin{proposition}(Uniqueness of the saddle-point) For any $R_{\infty} < R < C$ and $P \in \cP_{R}(\cX)$, $S(R,P)$ is a singleton.
\label{prop:unique-saddle-point}
\end{proposition}

\begin{IEEEproof}
The proof is given in the Appendix~\ref{app:prop-unique-saddle-point}.
\end{IEEEproof}

\begin{definition}
Fix any $ R_{\infty} < R < C$. 
\begin{align}
& \rho^{\ast}_{R,\cdot} \, : \, \cP_{R}(\cX) \rightarrow \bbR_{+}, \mbox{ s.t. } \rho^{\ast}_{R,P} = \left.S(R,P)\right|_{\bbR_{+}},  \label{eq:opt-rho-star} \\
& Q^{\ast}_{R,\cdot} \, : \, \cP_{R}(\cX) \rightarrow \cP_{P,W}(\cY), \mbox{ s.t. } Q^{\ast}_{R,P} = \left.S(R,P)\right|_{\cP_{P,W}(\cY)}. \label{eq:opt-q-star}
\end{align}
\label{def:opt-rho-q-star}
\end{definition}
\vspace{-0.5cm}

Observe that owing to Proposition~\ref{prop:unique-saddle-point}, both \eqref{eq:opt-rho-star} and \eqref{eq:opt-q-star} are well-defined. 
The distribution $Q^{\ast}_{R,\cdot}$ in \eqref{eq:opt-q-star} will be
our output distribution.

\begin{proposition}(Differentiability of $\mE_{\mSP}(\cdot,P)$) Consider any $R_\infty < R < C$ and $P \in \cP_{R}(\cX)$. $\mE_{\mSP}(\cdot,P)$ is differentiable with $\rho^\ast_{R,P} = - \left.\frac{\partial \mE_{\mSP}(r,P)}{\partial r}\right|_{r = R}$.
\label{prop:diff-ESP}
\end{proposition}

\begin{IEEEproof}
The proof is given in the Appendix~\ref{app:prop-diff-ESP}. 
\end{IEEEproof}

\begin{proposition}(Continuity of the saddle-point) Consider any $R_\infty < R < C$. Both $\rho^\ast_{R, \cdot}$ and $Q^\ast_{R, \cdot}$ are continuous on $\cP_R(\cX)$. 
\label{prop:cont-saddle-point}
\end{proposition}

\begin{IEEEproof}
The proof is provided in the Appendix~\ref{app:prop-cont-saddle-point}.
\end{IEEEproof}

For any $R_\infty < R < C$ and $P \in \cP_{R}(\cX)$, let $\me_{\mSP}(R,P,r) \eqdef \me_{\mSP}(Q^\ast_{P,R}, P,r)$ and $\me_{\mSP}(R,P) \eqdef \me_{\mSP}(R,P,R)$.

\begin{theorem}(Equality of the exponents)
For any $R_{\infty} < R < C$ and $P \in \cP_{R}(\cX)$, 
\begin{equation}
\me_{\mSP}(R,P) = \mE_{\mSP}(R,P) . 
\label{eq:opt-thrm-hyp-exp}
\end{equation}
\label{lem:lem-opt3}
\end{theorem}
\vspace{-0.5cm}
\begin{IEEEproof}
The proof is given in the Appendix~\ref{app:lem:lem-opt3}.
\end{IEEEproof}

\begin{remark}
Recalling the discussion in the previous section, the equality of the exponents theorem, i.e. Theorem~\ref{lem:lem-opt3}, ensures that the exponent of the lower bound on the error probability emerging as a result of binary hypothesis testing reduction in which $Q^\ast_{R,\cdot}$ is the alternate distribution matches the sphere-packing exponent. \eor
\end{remark}


\subsection{Hypothesis testing reduction}
\label{ssec:hypothesis-testing-reduction}

For any $\nu,R \in \bbR^+$, define $\cP_{R,\nu}(\cX) \eqdef \{ P \in \cP(\cX) \, : \, \mE_{\mSP}(R,P) \geq \nu\}$. Fix some $R \in (R_\infty, C)$ and some sufficiently small $\nu >0$ that only depends on $W$ and $R$. Application of the hypothesis testing reduction of Section~\ref{ssec:hyp-test} to an $(N,R)$ constant composition code $(f, \varphi)$ with common composition\footnote{If $P \in \cP_{R, \nu}(\cX)^c$, then it is possible to prove that \eqref{eq:thrm-main} is true. See Lemma~\ref{lem:SCA-lem} in the Appendix~\ref{app:trivial-types}.} $P \in \cP_{R, \nu}(\cX)$ by using $Q_{R,P}^\ast$ as the auxiliary output distribution yields (recall \eqref{eq:hyp-test-4-q})
\begin{equation} e(f, \varphi) \geq \alpha_{N}(R),
\label{eq:hyp-test-4}
\end{equation}
where $\alpha_N(R) \eqdef \alpha_{W(\cdot|\bx^N), Q^\ast_{R,P}}(NR)$. On account of \eqref{eq:hyp-test-4}, in order to lower bound the maximal error probability of our code, it suffices to evaluate $\alpha_N(R)$. 

However, since $Q^\ast_{R,P} \gg W(\cdot | \bx^N)$ (cf. item (ii) of the saddle-point proposition, i.e. Proposition~\ref{prop:saddle-point}), but not necessarily\footnote{We have this equivalence if we consider a positive channel, for example.} $Q^\ast_{R,P} \equiv W(\cdot | \bx^N)$, we need to do little more work. To this end, we define
\begin{equation}
\tilde{\cT}(Q, \hat{Q}) \eqdef \left\{ T \in \cT(Q, \tilde{Q}) \, : \, \cU_{T} \cap [\cS(\hat{Q}) \backslash \cS(Q, \hat{Q})]=\emptyset  \, \mbox{ and } \, \cU_{T}^{c} \cap [\cS(Q) \backslash \cS(Q, \hat{Q})]=\emptyset \right\},
\label{eq:hyp-test-eff-set}
\end{equation}
where $\cS(Q,\hat{Q}) \eqdef \cS(Q) \cap \cS(\hat{Q})$. 

The proof of the following result is straightforward. 
\begin{lemma}
For any $r \in \bbR^{+}$, 
\begin{equation}
\alpha^{\ast}_{Q,\hat{Q}}(r)  = \min_{T \in \tilde{\cT}(Q, \hat{Q}) \, : \, \beta_{T}\leq e^{-r}}\alpha_{T}.
\label{eq:hyp-test-lemma-1}
\end{equation}
\label{lem:hyp-test-lemma-1}
\end{lemma}
\vspace{-0.5cm}
\begin{lemma}
For any $T \in \tilde{\cT}(Q, \hat{Q})$, we have
\begin{equation}
\alpha_{T} = Q\left\{ \cS(Q, \tilde{Q})\right\}Q\left\{ \cU_{T}^{c}\left. \right|\cS(Q, \hat{Q})\right\}, \quad \beta_{T} = \hat{Q}\left\{ \cS(Q, \hat{Q}) \right\}\hat{Q}\left\{ \cU_{T}\left. \right| \cS(Q, \hat{Q})\right\}, \label{eq:hyp-test-lemma-2}
\end{equation}
where the conditional probabilities are induced by $Q$ and $\hat{Q}$, respectively. 
\label{lem:hyp-test-lemma-2}
\end{lemma}
\begin{IEEEproof}
The result is obvious from the law of total probability and recalling \eqref{eq:hyp-test-eff-set}.
\end{IEEEproof}
Observe that owing to \eqref{eq:hyp-test-lemma-1} we have\footnote{$\tilde{\cT}(W(\cdot|\bx^{N}), \qs)$ is defined as in \eqref{eq:hyp-test-eff-set}.}
\begin{equation}
\alpha_{N}(R) = \min_{T \in \tilde{\cT}(W(\cdot|\bx^{N}), \qs) \, : \, \beta_{T} \leq e^{-NR}} \alpha_{T}.
\label{eq:hyp-test-5}
\end{equation}

In order to apply Lemmas~\ref{lem:hyp-test-lemma-1} and \ref{lem:hyp-test-lemma-2} to our particular case, we need the following definition. 	

\begin{definition}
Given any $C >R > R_{\infty}$ and $P \in \cP_{R}(\cX)$, 
\begin{equation}
W^{-}_{R,P}(\cdot|x) \eqdef 
\begin{cases}
\tilde{W}_{1^{-} , Q^{\ast}_{R,P}}(\cdot|x), \, & \mbox{ if } x \in \cS(P), \\
W(\cdot|x), & \mbox{ else},
\end{cases}
\label{eq:opt-w-minus}
\end{equation}
where 
\begin{equation}
\tilde{W}_{1^{-} , Q^{\ast}_{R,P}}(\cdot|x) \eqdef \lim_{\lambda \uparrow 1}\tilde{W}_{\lambda , Q^{\ast}_{R,P}}(\cdot|x), \, \forall \, x \in \cS(P). 
\label{eq:opt-w-minus-1}
\end{equation}
and $\tilde{W}_{\lambda , Q^{\ast}_{R,P}}(\cdot|x)$ is the \emph{tilted distribution} as defined in \eqref{eq:opt-tilted-channel} in the Appendix~\ref{app:prop-saddle-point}.
\label{def:opt-w-minus}
\end{definition}

\begin{remark}
One can check that for any $x \in \cS(P)$, 
\begin{equation}
\tilde{W}_{1^{-} , Q^{\ast}_{R,P}}(\cdot|x) = 
\begin{cases}
\frac{Q^{\ast}_{R,P}(y)}{Q^{\ast}_{R,P}\left\{ \cS(W(\cdot|x))\right\}}, & \mbox{ if } y \in \cS(W(\cdot|x)), \\ 
0, & \mbox{ else}. 
\end{cases}
\label{eq:opt-rem2}
\end{equation}
Equation \eqref{eq:opt-rem2} and the fact that $Q^{\ast}_{R,P} \gg W(\cdot|x)$, for all $x \in \cS(P)$, ensure that \eqref{eq:opt-w-minus} is a well-defined stochastic matrix from $\cX$ to $\cY$. Moreover, it is clear that $W^{-}_{R,P}(\cdot|x) \equiv W(\cdot|x)$, for all $x \in \cX$. \eor
\label{rem:opt-rem2}
\end{remark}

Returning to our application, since $Q^{\ast }_{R,P} \gg W(\cdot | \bx^N)$, \eqref{eq:hyp-test-lemma-2} implies that for any $\tilde{T} \in \tilde{\cT}(W(\cdot|\bx^{N}), \qs)$, we have
\begin{equation}
\alpha_{\tilde{T}} = W\left\{ \cU^c_{\tilde{T}} | \bx^N\right\}, \quad \beta_{\tilde{T}} = Q_{R,P}^{\ast }\left\{ \cS(W(\cdot|\bx^N))\right\} W_{R,P}^{- }\left\{ \cU_{\tilde{T}} | \bx^N\right\}, 
\label{eq:hyp-test-6}
\end{equation}
where $W_{R,P}^{- }(\by^N|\bx^N) \eqdef \prod_{n=1}^N W_{R,P}^{-}(y_n|x_n)$ and $W_{R,P}^{-}$ is defined in \eqref{eq:opt-w-minus}.

Also, 
\begin{align}
\log Q_{R,P}^{\ast }\left\{ \cS(W^N(\cdot|\bx^N))\right\} & = \sum_{n=1}^N \log Q_{R,P}^{\ast }\left\{ \cS(W(\cdot|x_n))\right\}  \label{eq:hyp-test-7}\\
 & = N \sum_{x \in \cS(P)}P(x)\log Q_{R,P}^{\ast }\left\{ \cS(W(\cdot|x))\right\}  \nonumber \\
 & = - N \mD(W_{R,P}^-||Q_{R,P}^\ast|P), \label{eq:hyp-test-8}
\end{align}
where \eqref{eq:hyp-test-7} follows since $\cS(W(\cdot|\bx^N)) =\cS(W(\cdot|x_1)) \times \ldots \times \cS(W(\cdot|x_N))$ and \eqref{eq:hyp-test-8} follows by noting 
\[
\log  Q_{R,P}^{\ast }\left\{ \cS(W(\cdot|x))\right\} = - \mD(W_{R,P}^-(\cdot|x)||Q_{R,P}^\ast), 
\]
which is a direct consequence of \eqref{eq:opt-w-minus}. 

Combining \eqref{eq:hyp-test-6} and \eqref{eq:hyp-test-8}, we conclude that for any $\tilde{T} \in \tilde{\cT}(W(\cdot|\bx^{N}), \qs)$
\begin{equation}
\left[ \beta_{\tilde{T}} \leq e^{-NR}\right] \Longleftrightarrow \left[ W_{R,P}^{-}\left\{ \cU_{\tilde{T}} | \bx^N \right\} \leq e^{-N r(R,P)}\right],
\label{eq:hyp-test-9}
\end{equation}
where 
\begin{equation}
r(R,P) \eqdef R - \mD(W^{-}_{R,P}||Q^{\ast}_{R,P}|P).
\label{eq:r}
\end{equation} 
Observe that the right side of \eqref{eq:hyp-test-9} defines a non-trivial constraint only if $r(R,P) >0$, which we establish next. To this end, we first define the following set:
\begin{equation}
\tilde{\cP}_{P,W}(\cY|\cX) \eqdef \{ V \in \cP(\cY|\cX) : \forall \, x \in \cS(P), \, V(\cdot|x) \ll W(\cdot|x)\}. 
\label{eq:opt-e-tilde-e-equ-pf0.5}
\end{equation}

\begin{lemma}(Positivity of $r(R,P)$) Given any $R_{\infty} < R < C$ and $P \in \cP_R(\cX)$, 
\begin{itemize}
\item[(i)] $\forall \, V \in \tilde{\cP}_{P,W}(\cY|\cX), \, \mD(V||Q^{\ast}_{R,P}|P) = \mD(V||W_{R,P}^{-}|P) + \mD(W^{-}_{R,P}||Q^{\ast}_{R,P}|P)$.
\item[(ii)] $r(R,P) >0$.
\end{itemize}
\label{lem:lem-R-new}
\end{lemma}

\begin{IEEEproof}
The proof is given in the Appendix~\ref{app:lem-R-new}.
\end{IEEEproof}

Now, consider a binary hypothesis testing setup with the null hypothesis (resp. alternate hypothesis) $W(\cdot|\bx^N)$ (resp. $W_{R,P}^{- }(\cdot|\bx^N)$). Owing to \eqref{eq:hyp-test-4}, \eqref{eq:hyp-test-5}, \eqref{eq:hyp-test-6} and \eqref{eq:hyp-test-9}, we deduce that 
\begin{equation}
e(f, \varphi) \geq \tilde{\alpha}_N(r(R,P)) \eqdef \min_{T^\prime \in \tilde{T}(W(\cdot|\bx^N), W^{-}_{R,P}(\cdot|\bx^N)) \, : \, \beta_{T^\prime} \leq e^{-N r(R,P)}}\alpha_{T^\prime}.
\label{eq:hyp-test-10}
\end{equation}

On account of \eqref{eq:hyp-test-10}, in order to lower bound the maximal error probability of our constant composition code, it suffices to evaluate $\tilde{\alpha}_N(r(R,P))$. Instead of directly characterizing $\tilde{\alpha}_N(r(R,P))$, we give a lower bound on it by means of a test that is easier to analyze. In order to define this test, we need the following ``shifted exponent". 

\begin{definition}
Given any $C > R > R_{\infty}$, $r \in \bbR_{+}$ and $P \in \cP_{R}(\cX)$,
\begin{equation}
\tilde{\me}_{\mSP}(R,P,r) \eqdef \inf_{V \in \cP(\cY|\cX) \, : \, \mD(V||W^{-}_{R,P}|P) \leq r}\mD(V||W|P).
\label{eq:opt-tilde-e-def}
\end{equation}
\label{def:opt-tilde-e-def}
\end{definition}
\vspace{-.65cm}
\begin{lemma}(Shifted exponent)
For any $R > R_{\infty}$, $P \in \cP_{R}(\cX)$ and $r > \mD(W_{R,P}^{-}||Q_{R,P}^{\ast}|P)$, we have 
\[
\tilde{\me}_{\mSP}(R,P, r - \mD(W_{R,P}^{-}||Q_{R,P}^{\ast}|P)) = \me_{\mSP}(R,P,r).\]
\label{lem:opt-e-tilde-e-equ}
\end{lemma}
\vspace{-.5cm}
\begin{IEEEproof}
Fix an arbitrary $ R > R_{\infty}$, $P \in \cP_{R}(\cX)$ and $r > \mD(W_{R,P}^{-}||Q_{R,P}^{\ast}|P)$. Define $\tilde{r} \eqdef r - D(W^{-}_{R,P}||Q^{\ast}_{R,P}|P)$. Clearly, $\tilde{r} \in \bbR^{+}$. 
On account of the fact that $\tilde{\me}_{\mSP}(R,P,\tilde{r}) \leq \tilde{e}_{\mSP}(R,P,0) = \mD(W^{-}_{R,P}||W|P) < \infty$, it is easy to see that
\begin{equation}
\tilde{\me}_{\mSP}(R ,P, \tilde{r}) = \min_{V \in \tilde{\cP}_{P,W}(\cY|\cX) \, : \, \mD(V||W^{-}_{R,P}|P) \leq \tilde{r}}\mD(V||W|P).
\label{eq:opt-e-tilde-e-equ-pf1}
\end{equation}

Similarly, 
\begin{equation}
\me_{\mSP}(R,P,r) = \min_{V \in \tilde{\cP}_{P,W}(\cY|\cX) \, : \, \mD(V||Q^{\ast}_{R,P}|P) \leq r}\mD(V||W|P).
\label{eq:opt-e-tilde-e-equ-pf2}
\end{equation}
The item (i) of Lemma~\ref{lem:lem-R-new} ensures that the feasible regions of the right sides of \eqref{eq:opt-e-tilde-e-equ-pf1} and \eqref{eq:opt-e-tilde-e-equ-pf2} are the same. Since the cost functions of the two problems are the same, the lemma follows. 
\end{IEEEproof}

Fix an arbitrary $\zeta \in \bbR^{+}$ and let $\epsilon_{N} \eqdef \left( \frac{1}{2} + \zeta \right)\frac{\log N}{N}$ (resp. $\tilde{\epsilon}_{N} \eqdef \epsilon_{N} - \frac{1}{N}$) and define $R_{N} \eqdef R - \epsilon_{N}$ (resp. $\tilde{R}_{N} \eqdef R - \tilde{\epsilon}_{N}$). Note that for all sufficiently large $N \in \bbZ^{+}$, $C > \tilde{R}_{N} > R_{N} > R_{\infty}$. Throughout, we consider such an $N \in \bbZ^{+}$. Further, similar to \eqref{eq:r}, define $r_{N}(P,R) \eqdef R_N - \mD(W^{-}_{R,P}||Q^{\ast}_{R,P}|P)$ (resp. $\tilde{r}_{N}(P,R) \eqdef \tilde{R}_N - \mD(W^{-}_{R,P}||Q^{\ast}_{R,P}|P)$). Also,
\begin{align}
A_{N} & \eqdef \left\{ \by^{N} \, : \, \frac{1}{N} \sum_{n=1}^{N}\log\frac{W(y_{n}|x_{n})}{W_{R,P}^{-}(y_{n}|x_{n})} >  r_{N}(R,P) - \tilde{\me}_{\mSP}(R, P,r_{N}(R,P)) \right\}, \label{eq:easier-test-AN} \\
A_{N}^{c} & = \left\{ \by^{N} \, : \, \frac{1}{N} \sum_{n=1}^{N}\log\frac{W_{R,P}^{-}(y_{n}|x_{n})}{W(y_{n}|x_{n})} \geq \tilde{\me}_{\mSP}(R, P,r_{N}(R,P)) -r_{N}(R,P) \right\}. \label{eq:easier-test-ANc} 
\end{align}
Equations \eqref{eq:easier-test-AN} and \eqref{eq:easier-test-ANc} are the decision regions of the test, i.e. the test decides $W(\cdot|\bx^{N})$ if $\by^{N} \in A_{N}$ and $W^{-}_{R,P}(\cdot|\bx^{N})$ if $\by^{N} \in A_{N}^{c}$. Let
\begin{equation}
\alpha_{N} \eqdef W\left\{ A_{N}^{c} | \bx^{N}\right\}, \quad \beta_{N} \eqdef W^{- }_{R,P}\left\{ A_{N}| \bx^{N}\right\},
\label{eq:easier-test-alpbeta}
\end{equation}
denote the error probabilities of the aforementioned test. 

\begin{remark}
The analysis of the events $A_N$ and $A_N^c$ would be direct applications of
Bahadur-Ranga Rao but for two complications: First, the random 
variables in the sum are not i.i.d., only independent. This 
does not present a major difficulty, as one can prove a
version of the Bahadur-Ranga Rao result for independent random
variables that is weaker but sufficient for our purposes, which is given in the next section.
The second complication is that the threshold in both
events depends on $N$. One could define constant-threshold
versions of these events by replacing $r_N(R,P)$ with $r(R,P)$.
Applying exact asymptotics to the resulting events would
yield a lower bound on $\alpha_N$ of the order $
\frac{1}{\sqrt{N}} \exp(-N \mE_{\mSP}(R,P))$
and show that $\beta_N$ is of the order $\frac{1}{\sqrt{N}} \exp(-N r(R,P))$.
The problem with this approach is that $e(f,\varphi)$ is
lower bounded by the type-I error probability of the
optimal test whose type-II probability does not exceed 
$e^{-Nr(R,P)}$. From the above expression of $\beta_N$, we see that the aforementioned test is not optimal because, although it is a likelihood ratio test,
it is ``undershooting'' the type-II constraint due to
the $1/\sqrt{N}$ pre-factor. By replacing $r(R,P)$ with $r_N(R,P)$,
we ensure that $\beta_N$ does not undershoot the 
constraint (in fact, it will violate it by a small
amount). The $r_N(R,P)$ fluctuations will give rise to the
slope term in the pre-factor of the probability of $A_N$. \eor
\label{rem:varying-threshold}
\end{remark}

\subsection{Analysis of the hypothesis test}
\label{ssec:analysis-of-the-hypothesis-test}
In this section, we analyze the hypothesis test stated in the previous section. In order to accomplish this, we begin with the following generalization of the Bahadur-Ranga Rao theorem.  

\subsubsection{Sharp Lower Bound} 
\label{ssec:SLB} 
The content of this section resembles Dembo-Zeitouni's proof of Theorem~\ref{thrm:2} (cf. \cite[Theorem 3.7.4]{dembo-zeitouni98}). Here, we essentially use the same ideas but generalize them to cover non-identical case.

Let $\{ Z_i\}_{i=1}^n$ be a sequence of independent, real-valued random variables and $\lambda_i$ be the law of $Z_i$. Assume $\sum_{i=1}^n \mV[Z_i] \in \bbR^+$. Define $\Lambda_i(\delta) \eqdef \log \mE\left[ e^{\delta Z_i}\right]$, $M_i(\delta) \eqdef e^{\Lambda_i(\delta)}$ and the Fenchel-Legendre transform of $\frac{1}{n}\sum_{i=1}^n \Lambda_i(\cdot)$ as:
\begin{equation}
\forall x \in \bbR, \, \Lambda_n^\ast(x) \eqdef \sup_{\delta \in \bbR} \left\{ x \delta - \frac{1}{n} \sum_{i=1}^n \Lambda_i(\delta) \right\}. 
\label{eq:SLB-fenchel-def}
\end{equation}

Let $q \in \bbR$ be such that $\exists \, \eta \in (0,1]$ with the following properties:
\begin{enumerate}
\item[(i)] There exists a neighborhood of $\eta$ such that $\frac{1}{n}\sum_{i=1}^n \Lambda_i(\delta) < \infty$, for all $\delta$ in this neighborhood.
\item[(ii)] $\frac{1}{n}\sum_{i=1}^n \Lambda_i^\prime(\eta) = q$. 
\end{enumerate}

\begin{remark}
The reason to choose $1$ as an upper bound on $\eta$ above is just for our application of the main result of this section in the sequel. One may use an arbitrary constant and modify the result accordingly. \eor
\label{rem:SLB-remark-0}
\end{remark}

Define $\hat{S}_n \eqdef \frac{1}{n}\sum_{i=1}^n Z_i$ and let $\mu_n$ denote the law of $\hat{S}_n$. Also, define the probability measure $\tilde{\lambda}_i$ via
\begin{equation}
\frac{d\tilde{\lambda}_i}{d\lambda_i}(z_i) \eqdef e^{\eta z_i - \Lambda_i(\eta)}. \label{eq:SLB-lambda-tilde}
\end{equation}

Let $\tilde{\mu}_n$ denote the law of $\hat{S}_n$ when $Z_i$ are independent with the marginal law $\tilde{\lambda}_i$. Further, define\footnote{We shall show that all of the following quantities are well-defined in the proof of the following proposition, given in the Appendix~\ref{app:SLB-pf}.} $T_{i}  \eqdef Z_{i} - \mE_{\tilde{\lambda}_i}[Z_i]$, $m_{2,n} \eqdef \sum_{i=1}^{n} \mV_{\tilde{\lambda}_{i}}[T_{i}]$, $m_{3,n} \eqdef \sum_{i=1}^{n}\mE_{\tilde{\lambda}_{i}}\left[|T_{i}|^{3}\right]$ and 
$W_{n}  \eqdef \frac{1}{\sqrt{m_{2,n}}}\sum_{i=1}^{n}T_{i}.$ Also, $K_{n}(\eta) \eqdef \frac{15 \sqrt{2 \pi} m_{3,n} }{m_{2,n}}.$ 

\begin{proposition}(Sharp lower bound) Provided that 
\begin{equation}
\sqrt{m_{2,n}} \geq 1+ (1 + K_n(\eta))^2
\label{eq:SLB-final-0}
\end{equation}
holds, 
\begin{equation}
\mu_{n}([q, \infty)) \geq e^{-n\Lambda^{\ast}_{n}(q)} \frac{e^{-K_n(\eta)}}{2\sqrt{2 \pi m_{2,n}}}. 
\label{eq:SLB-final}
\end{equation}
\label{prop:SLB}
\end{proposition} 
 
\begin{IEEEproof}
The proof is given in the Appendix~\ref{app:SLB-pf}.
\end{IEEEproof}

\subsubsection{Analysis of $\alpha_N$ and $\beta_N$}
\label{ssec:SLB-application}
In this section, we apply Proposition~\ref{prop:SLB} to lower bound $\alpha_{N}$ and $\beta_{N}$ given in \eqref{eq:easier-test-alpbeta}. To this end, we begin with the following technical results. 

\begin{definition}
Let $C > R > R_{\infty}$ and $P \in \cP_R(\cX)$ be arbitrary but fixed. Let $\lambda \in \bbR$ be arbitrary. 
\begin{align}
\Lambda_{0, P, x}(\lambda) & \eqdef \log \mE_{W(\cdot|x)}\left[ e^{\lambda \log\frac{W^-_{R,P}(Y|x)}{W(Y|x)}}\right], \label{eq:zerovar-lambda0-1} \\
\Lambda_{0,P}(\lambda) & \eqdef \sum_{x \in \cX}P(x) \Lambda_{0, P, x}(\lambda), \label{eq:zerovar-lambda0-2} \\
\Lambda_{1, P, x}(\lambda) & \eqdef \log \mE_{W^-_{R,P}(\cdot|x)}\left[ e^{\lambda \log\frac{W(Y|x)}{W^-_{R,P}(Y|x)}}\right], \label{eq:zerovar-lambda01-1} \\
\Lambda_{1,P}(\lambda) & \eqdef \sum_{x \in \cX}P(x) \Lambda_{1, P, x}(\lambda), \label{eq:zerovar-lambda1-2} 
\end{align}
\label{def:zerovar-lambda}
\end{definition}
\vspace{-0.5cm}
\begin{remark}$ $ \\
\vspace{-0.87cm}
\begin{enumerate}
\item[(i)] Since $W^-_{R,P}(\cdot|x) \equiv W(\cdot|x)$ for all $x \in \cX$, each quantity given in Definition~\ref{def:zerovar-lambda} is well-defined. Also, one can check that $\Lambda_{1, P, x}(\lambda) = \Lambda_{0,P,x}(1-\lambda)$, which, in turn, implies that $\Lambda_{1,P}(\lambda) = \Lambda_{0,P}(1-\lambda)$.

\item[(ii)] The fact that $W^-_{R,P}(\cdot|x) \equiv W(\cdot|x)$ for all $x \in \cX$ also ensures that $\Lambda_{0,P}(\lambda), \Lambda_{1,P}(\lambda) \in \bbR$ and hence $\Lambda_{0,P}(\cdot), \Lambda_{1,P}(\cdot) \in \mathcal{C}^\infty(\bbR)$.

\item[(iii)] Consider any $\lambda \in \bbR$. It is easy to verify the following (for the sake of notational convenience, we denote partial derivatives with respect to $\lambda$ as the ordinary ones):
\begin{align}
\Lambda_{0,P,x}^\prime(\lambda) & = \mE_{\tilde{W}_{\lambda, P}(\cdot|x)}\left[ \log \frac{W^-_{R,P}(Y|x)}{W(Y|x)}\right], \label{eq:zerovar-0-der1-1}\\
\Lambda_{0,P}^\prime(\lambda) & = \sum_{x \in \cX}P(x)\Lambda_{0,P,x}^\prime(\lambda), \label{eq:zerovar-0-der1-2} \\
\Lambda_{0,P,x}^{\prime \prime}(\lambda) & = \mV_{\tilde{W}_{\lambda, P}(\cdot|x)}\left[ \log \frac{W^-_{R,P}(Y|x)}{W(Y|x)}\right], \label{eq:zerovar-0-der2-1}\\
\Lambda_{0,P}^{\prime \prime}(\lambda) & = \sum_{x \in \cX}P(x)\Lambda_{0,P,x}^{\prime \prime}(\lambda), \label{eq:zerovar-0-der2-2}
\end{align}
where $\tilde{W}_{\lambda, P} (\cdot|x) \eqdef \tilde{W}_{\lambda, W^-_{R,P}(\cdot|x)}$ (cf. \eqref{eq:opt-tilted-channel}) for the sake of notational convenience. 

Further, item (ii) above ensures that
\begin{align}
& \Lambda_{1,P,x}^\prime(\lambda) = -\Lambda_{0,P,x}^\prime(1 - \lambda), \quad  \Lambda_{1,P}^\prime(\lambda) = - \Lambda_{0,P}^\prime(1-\lambda), \label{eq:zerovar-1-der1} \\
& \Lambda_{1,P,x}^{\prime \prime}(\lambda) = \Lambda_{0,P,x}^{\prime \prime}(1 - \lambda), \quad  \Lambda_{1,P}^{\prime \prime}(\lambda) =  \Lambda_{0,P}^{\prime \prime}(1-\lambda), \label{eq:zerovar-1-der2} 
\end{align}
for any $\lambda \in \bbR$.

\item[(iv)] We have 
\begin{align}
\Lambda_{0,P}^{\prime}(0) & = - \Lambda^{\prime}_{1,P}(1) = -\mD(W||W_{R,P}^-|P), \label{eq:zerovar-der1-0}\\
\Lambda_{0,P}^{\prime}(1) & = - \Lambda^{\prime}_{1,P}(0) = \mD(W_{R,P}^-||W|P), \label{eq:zerovar-der1-1}
\end{align}
as a direct consequence of \eqref{eq:zerovar-0-der1-2} and \eqref{eq:zerovar-1-der1}. \eor
\end{enumerate}
\label{rem:zerovar-rem1}
\end{remark}

\begin{lemma}(Positive variance)
Let $C > R > R_{\infty}$ and $P \in \cP_R(\cX)$ be arbitrary. For all $\lambda \in [0,1]$, $\Lambda_{0,P}^{\prime \prime}(\lambda) > 0$. 
\label{thrm:zerovar-prop4}
\end{lemma}
\begin{IEEEproof}
Consider any $C> R > R_{\infty}$, $P \in \cP_R(\cX)$ and recall that $r(R,P) = R - \mD(W_{R,P}^-||Q_{R,P}^\ast|P)$ (cf. \eqref{eq:r}).

For contradiction, suppose there exists $\lambda \in [0,1]$ such that $\Lambda_{0,P}^{\prime \prime}(\lambda) =0$. We have
\begin{align}
\left[ \Lambda_{0,P}^{\prime \prime}(\lambda) =0 \right] & \Longleftrightarrow \left[ \forall x \in \cS(P), \, \log\frac{W^-_{R,P}(Y|x)}{W(Y|x)} = \Lambda_{0,P,x}^\prime(\lambda), \, W(\cdot|x)-\mbox{(a.s.)}\right]  \label{eq:zerovar-prop4-pf1} \\
 & \Longleftrightarrow \left[ \forall x \in \cS(P), \, W(Y|x) = W_{R,P}^-(Y|x)e^{-\Lambda_{0,P,x}^\prime(\lambda)}, \, W(\cdot|x)-\mbox{(a.s.)}  \right], \label{eq:zerovar-3.b.5} 
\end{align}
where \eqref{eq:zerovar-prop4-pf1} follows from \eqref{eq:zerovar-0-der1-1}, \eqref{eq:zerovar-0-der2-1} and \eqref{eq:zerovar-0-der2-2}.
Summing the right side of \eqref{eq:zerovar-3.b.5} over $y \in \cS(W(\cdot|x))$ yields 
\begin{equation}
\forall x \in \cS(P), \, \Lambda_{0,P,x}^\prime(\lambda) =0.
\label{eq:zerovar-3.b.6}
\end{equation}
Combining \eqref{eq:zerovar-3.b.5} and \eqref{eq:zerovar-3.b.6} and recalling the definition of $W_{R,P}^-$ (cf. \eqref{eq:opt-w-minus}), we deduce that 
\begin{equation}
\left[ \Lambda_{0,P}^{\prime \prime}(\lambda) =0\right] \Longleftrightarrow \left[ \forall (x,y) \in \cX \times \cY, \, W(y|x) = W_{R,P}^-(y|x)\right].
\label{eq:zerovar-prop4-pf2}
\end{equation}
The right side of \eqref{eq:zerovar-prop4-pf2} implies that $\tilde{e}_{\mSP}(R,P,r) = 0$ for all $r \in \bbR_+$ and in particular $\tilde{e}_{\mSP}(R, P,r(R,P)) =0$. This observation, coupled with the equality of the exponents theorem, i.e. Theorem~\ref{lem:lem-opt3}, and the shifted exponent exponent lemma, i.e. Lemma~\ref{lem:opt-e-tilde-e-equ}, implies that $\mE_{\mSP}(R,P) =0$ that contradicts the fact that $P \in \cP_R(\cX)$. 
\end{IEEEproof}

\begin{definition}Let $C >R > R_{\infty}$ be arbitrary. For any $(\lambda, P) \in [0,1] \times \cP_{R}(\cX)$,
\begin{align}
m_{0,3}(\lambda, P) & \eqdef \sum_{x \in \cS(P)}P(x) \mE_{\tilde{W}_{\lambda, P}(\cdot|x)}\left[ \left| \log\frac{W_{R,P}^{-}(Y|x)}{W(Y|x)} - \Lambda_{0,P,x}^{\prime}(\lambda) \right|^{3}\right], \label{eq:m03}\\
m_{1,3}(\lambda, P) &\eqdef \sum_{x \in \cS(P)}P(x) \mE_{\tilde{W}_{1-\lambda, P}(\cdot|x)}\left[ \left| \log\frac{W(Y|x)}{W_{R,P}^{-}(Y|x)} - \Lambda_{1,P,x}^{\prime}(\lambda) \right|^{3}\right]. \label{eq:m13}
\end{align} 
\label{def:m3-def}
\end{definition}

Note that owing to \eqref{eq:zerovar-1-der1}, \eqref{eq:m03} and \eqref{eq:m13}, one can verify that 
\begin{equation}
\forall \, (\lambda,P) \in [0,1] \times \cP_{R}(\cX), \, m_{0,3}(\lambda, P) = m_{1,3}(1-\lambda, P).
\label{eq:m3-equal}
\end{equation}

\begin{lemma}(Continuity)
\begin{enumerate}
\item[(i)] $\Lambda_{0,\cdot}^{\prime}(\cdot)$ is continuous on $(0,1] \times \cP_{R}(\cX)$. 
\item[(ii)] $\Lambda_{0,\cdot}^{\prime \prime}(\cdot)$ is continuous on $(0,1] \times \cP_{R}(\cX)$. 
\item[(iii)] $m_{0,3}(\cdot, \cdot)$ is continuous on $(0,1] \times \cP_{R}(\cX)$. 
\item[(iv)] $\mD(W^{-}_{R,\cdot}||Q^{\ast}_{R,\cdot}|\cdot)$ is continuous on $\cP_{R}(\cX)$.
\end{enumerate}
\label{thrm:cont-1}
\end{lemma}

\begin{IEEEproof}
The proof is given in the Appendix~\ref{app:continuity-pf}.
\end{IEEEproof}

\begin{lemma}
Fix arbitrary $C> R > R_{\infty}$ and $P \in \cP_{R}(\cX)$. For any $r \in \bbR^{+}$, we have
\begin{equation}
\tilde{\me}_{\mSP}(R,P,r) = \max_{ s \in \bbR_{+}} \left\{ -s r + e_{0}(s,P)\right\}, 
\label{eq:opt-lem5}
\end{equation}
where 
\begin{equation}
e_{0}(s,P) \eqdef -(1+s)\sum_{x \in \cS(P)}P(x) \log\sum_{y \in \cS(W(\cdot|x))}W(y|x)^{1/(1+s)}W^{-}_{R,P}(y|x)^{s/(1+s)}
\label{eq:opt-e0-def}
\end{equation}
for any $s \in \bbR_{+}$. 
\label{lem:opt-lem5}
\end{lemma}

\begin{IEEEproof}
We have, 
\begin{align}
\tilde{\me}_{\mSP}(R,P,r) & = \inf_{V \in \cP(\cY|\cX) \, : \, \mD(V||W^{-}_{R,P}|P) \leq r} \mD(V||W|P)  \nonumber \\
 & = \max_{s \in \bbR_{+}} \min_{V \in \cP(\cY|\cX)} \left\{ \mD(V||W|P) + s(\mD(V||W^{-}_{R,P}|P) - r)\right\}  \label{eq:opt-lem5-pf2} \\
 & = \max_{s \in \bbR_{+}}\left\{ -s r + \sum_{x \in \cS(P)} P(x) \min_{V(\cdot|x) \in \cP(\cY)}\left[ \mD(V(\cdot|x)||W(\cdot|x)) +s \mD(V(\cdot|x)||W^{-}_{R,P}(\cdot|x))\right] \right\}  \label{eq:opt-lem5-pf3}\\
 & = \max_{s \in \bbR_{+}}\left\{ -sr + e_{0}(s,P)\right\}, \label{eq:opt-lem5-pf4} 
\end{align}
where \eqref{eq:opt-lem5-pf2} follows since Slater's condition holds (cf. \cite[Corollary 28.2.1]{rockafellar70}), \eqref{eq:opt-lem5-pf4} follows by noting that 
\begin{equation*}
V^\ast_{P,s}(y|x) \eqdef \frac{W(y|x)^{1/(1+s)}W_{R,P}^{-}(y|x)^{s/(1+s)}}{\sum_{\tilde{y} \in \cY} W(\tilde{y}|x)^{1/(1+s)}W_{R,P}^{-}(\tilde{y}|x)^{s/(1+s)}},
\label{eq:opt-lem5-pf5}
\end{equation*}
attains the minimum in \eqref{eq:opt-lem5-pf3} for any $x \in \cS(P)$ and recalling \eqref{eq:opt-e0-def}. 
\end{IEEEproof}

\begin{corollary}
Consider any $C > R > R_{\infty}$, $P \in \cP_{R}(\cX)$. For all $r \in \bbR^{+}$, the set of maximizers of \eqref{eq:opt-lem5} is exactly $\partial \tilde{\me}_{\mSP}(R, P,\cdot)(r)$. 
\label{cor:subdiff}
\end{corollary}

\begin{IEEEproof}
Proof follows exactly the same lines as that of Lemma~\ref{lem:opt-cla14}. 
\end{IEEEproof}

\begin{lemma}(Differentiability of the shifted exponent)
Let $C >R > R_{\infty}$ and $r \in \bbR^{+}$ be given. 
\begin{equation}
s^{\ast}(R, \cdot,r) \, : \, \cP_{R}(\cX) \rightarrow \bbR_{+}, \mbox{ s.t. } s^{\ast}(R, P,r) \eqdef -\frac{\partial \tilde{\me}_{\mSP}(R,P,r)}{\partial r}, \, \forall P \in \cP_{R}(\cX),
\label{eq:thrm-sstar-def}
\end{equation}
is a well-defined function. 
\label{thrm:thrm-sstar-def}
\end{lemma}

\begin{IEEEproof}
Consider any $P \in \cP_{R}(\cX)$. For any $s \in \bbR_+$, \eqref{eq:opt-e0-def}, \eqref{eq:zerovar-lambda0-1}, \eqref{eq:zerovar-lambda0-2}, \eqref{eq:zerovar-0-der1-2} and \eqref{eq:zerovar-0-der2-2} imply that
\begin{equation}
\frac{\partial^2 e_o(s, P)}{\partial s^2}  = -\frac{1}{(1+s)^3}\Lambda_{0,P}^{\prime \prime}\left(\frac{s}{1+s}\right) < 0. 
\label{eq:zerovar-3.b.4}
\end{equation}
where the inequality follows from the positive variance lemma, i.e. Lemma~\ref{thrm:zerovar-prop4}. Equation \eqref{eq:zerovar-3.b.4} ensures the strict concavity of the cost function of \eqref{eq:opt-lem5} and hence the uniqueness of the maximizer. Recalling Corollary~\ref{cor:subdiff}, this implies that \eqref{eq:thrm-sstar-def} is well-defined. 
\end{IEEEproof}

The shifted exponent lemma, i.e. Lemma~\ref{lem:opt-e-tilde-e-equ}, and the differentiability of the shifted exponent, i.e. Lemma~\ref{thrm:thrm-sstar-def}, immediately implies the following result. 

\begin{corollary}
Given any $C > R > R_{\infty}$ and $P \in \cP_{R}(\cX)$ and , 
\[
\left.\frac{\partial \me_{\mSP}(R,P,\tilde{r})}{\partial \tilde{r}}\right|_{\tilde{r}=r} = -s^{\ast}(R,P,r-\mD(W_{R,P}^{-}||Q_{R,P}^{\ast}|P)),
\]
for any $r > \mD(W_{R,P}^{-}||Q_{R,P}^{\ast}|P)$.
\label{cor:cor-eSP-der}
\end{corollary}

Throughout this section, unless stated otherwise, suppose $C(W) >R > R_{\infty}$ and $P \in \cP_{R}(\cX)$ be arbitrary and fixed.

\begin{definition} Consider any $C >R > R_{\infty}$ and $P \in \cP_{R}(\cX)$. Given any $z \in \bbR$, 
\begin{align}
\Lambda^\ast_{0, P}(z) & \eqdef \sup_{\lambda \in \bbR}\left\{ \lambda z -  \Lambda_{0,P}(\lambda) \right\}, \label{eq:fenchel-0}\\
\Lambda^\ast_{1, P}(z) & \eqdef \sup_{\lambda \in \bbR}\left\{ \lambda z -  \Lambda_{1,P}(\lambda) \right\}. \label{eq:fenchel-1}
\end{align}
\label{def:fenchel}
\end{definition}
\vspace{-0.5cm}
\begin{lemma} (Regularity) Fix any $C >R > R_{\infty}$ and $P \in \cP_{R}(\cX)$. For any $0 < r < \mD(W||W_{R,P}^{-}|P)$,  
\begin{itemize}
\item[(i)] $\Lambda^\ast_{0,P}(\tilde{\me}_{\mSP}(R,P,r) - r) = \tilde{\me}_{\mSP}(R,P,r)$.
\item[(ii)] $\Lambda_{1,P}^\ast(r - \tilde{\me}_{\mSP}(R,P,r))= r$.
\item[(iii)] There exists a unique $\eta(R,P,r) \in (0,1)$, such that $\Lambda_{0,P}^{\prime}(\eta(R,P,r)) = \tilde{\me}_{\mSP}(R,P,r) - r$. In particular, $\eta(R,P,r) = \frac{s^\ast(R,P,r)}{1+s^\ast(R,P,r)}$. 
\end{itemize}
\label{lem:lem-easier-test-2}
\end{lemma}

\begin{IEEEproof}
The proof is given in the Appendix~\ref{app:regularity-lemma-pf}.
\end{IEEEproof}

Next, we claim that 
\begin{equation}
0< r(R,P) < \mI(P;W) - \mD(W_{R,P}^{-}||Q_{R,P}^{\ast}|P) \leq \mD(W||W_{R,P}^-|P).
\label{eq:r(P)-rel}
\end{equation}

The first inequality follows from the positivity of $r(R,P)$ lemma, i.e. Lemma~\ref{lem:lem-R-new}. The second inequality is clear from the definition of $r(R,P)$ and the fact that $P \in \cP_{R}(\cX)$. The last inequality follows by noting
\[
\mD(W_{R,P}^{-}||Q_{R,P}^{\ast}|P) + \mD(W||W_{R,P}^-|P) = \mD(W||Q_{R,P}^{\ast}|P) \geq \min_{Q \in \cP(\cY)}\mD(W||Q|P) = \mI(P;W),
\]
where the first equality follows from the item (i) of Lemma~\ref{lem:lem-R-new} and the last one follows from \eqref{eq:mutual-inf}. Hence, \eqref{eq:r(P)-rel} follows. 

Further, define
\[
\Upsilon(W, R, \nu) \eqdef \max_{P \in \cP_{R,\nu}(\cX)}\mD(W||Q_{R,P}^{\ast}|P), \quad H \eqdef \left[ \frac{\frac{\nu}{2\Upsilon(W,R,\nu)}}{1+\frac{\nu}{2\Upsilon(W,R,\nu)}}, 1\right].
\]
Since $\mE_{\mSP}(\cdot, \cdot)$ is continuous (cf. Lemma~\ref{lem:opt-cla12}), $\cP_{R, \nu}$ is closed and therefore, by noting the boundedness of $\cP(\cX)$, is compact. Further, owing to the continuity of $\mD(W||Q_{R,\cdot}^{\ast}|\cdot)$ (cf. the item (iv) of the continuity lemma, i.e. Lemma~\ref{thrm:cont-1}) and the compactness of $\cP_{R, \nu}(\cX)$, $\Upsilon(W, R, \nu)$ is well-defined and finite. 

\begin{lemma}
For any $P \in \cP_{R, \nu}(\cX)$ 
\begin{equation}
\eta(R,P, r) \in H, \, \forall \, r \in (0,  r(P,R)]. 
\label{eq:lem-eta-lb}
\end{equation}
\label{lem:eta-lb}
\end{lemma} 
\vspace{-0.75cm}
\begin{IEEEproof} Let $P \in \cP_{R, \nu}(\cX)$ be arbitrary. Owing to the item (iii) of the regularity lemma, i.e. Lemma~\ref{lem:lem-easier-test-2}, it suffices to prove that for all $r \in (0,  r(P,R)]$
\begin{equation}
\eta(R,P, r) \geq \frac{\frac{\nu}{2\Upsilon(W,R,\nu)}}{1+\frac{\nu}{2\Upsilon(W,R,\nu)}}.
\label{eq:lem-eta-lb-pf}
\end{equation} 
Moreover, the fact that $\eta(R,P,r) = s^{\ast}(R,P,r)/(1+s^{\ast}(R,P,r))$ (cf. item (iii) of Lemma~\ref{lem:lem-easier-test-2}), \eqref{eq:r(P)-rel}, the convexity and the non-increasing property of $\tilde{\me}_{\mSP}(R,P,\cdot)$, it suffices to show \eqref{eq:lem-eta-lb-pf} for $r = r(R,P)$. The differentiability of the shifted exponent lemma, i.e. Lemma~\ref{thrm:thrm-sstar-def}, and Corollary~\ref{cor:cor-eSP-der} imply that
\begin{equation}
s^{\ast}(R,P,r(R,P)) = -\left.\frac{\partial \tilde{\me}_{\mSP}(R,P,r)}{\partial r}\right|_{r = r(R,P)} = -\left.\frac{\partial \me_{\mSP}(R,P,r)}{\partial r}\right|_{r =R}.
\label{eq:lem-eta-lb1.0}
\end{equation}
Moreover, using the convexity and the non-increasing property of $\me_{\mSP}(R,P,\cdot)$, one can see that 
\begin{equation}
-\left.\frac{\partial \me_{\mSP}(R,P,r)}{\partial r}\right|_{r =R} \geq \frac{\nu}{2}\frac{1}{(\me^{-1}_{\mSP}(R,P,\cdot)(\nu/2) - R)} \geq \frac{\nu}{2\Upsilon(W,R,\nu)},
\label{eq:lem-eta-lb1.1}
\end{equation}
where the last inequality follows by noting that $\me_{\mSP}(R,P,r) = 0$ for all $r \geq \mD(W||Q_{R,P}^{\ast}|P)$. By combining \eqref{eq:lem-eta-lb1.0} and \eqref{eq:lem-eta-lb1.1}, we deduce that 
\begin{equation}
s^{\ast}(R,P,r(R,P))  \geq \frac{\nu}{2\Upsilon(W,R,\nu)}.
\label{eq:lem-eta-lb1}
\end{equation}
Since $\eta(R,P,r) = s^{\ast}(R,P,r)/(1+s^{\ast}(R,P,r))$, \eqref{eq:lem-eta-lb1} implies \eqref{eq:lem-eta-lb}. 
\end{IEEEproof}

Finally, we define the following:
\begin{align}
\overline{M}(\nu, W, R) & \eqdef \max_{(\lambda, P) \in H \times \cP_{R,\nu}}\frac{m_{0,3}(\lambda,P)}{\Lambda_{0,P}^{\prime \prime}(\lambda)}, \label{eq:M-max-def}\\
\overline{V}(\nu, W, R) & \eqdef \max_{(\lambda, P) \in H \times \cP_{R,\nu}}\Lambda_{0,P}^{\prime \prime}(\lambda), \label{eq:V-max-def}\\
\underline{V}(\nu, W, R) & \eqdef \min_{(\lambda, P) \in H \times \cP_{R,\nu}}\Lambda_{0,P}^{\prime \prime}(\lambda)\label{eq:V-min-def}, 
\end{align}
where $H$ is as defined prior to Lemma~\ref{lem:eta-lb}. Recalling the compactness of $H$ and $\cP_{R, \nu}(\cX)$, the positive variance lemma, i.e. Lemma~\ref{thrm:zerovar-prop4} and the continuity lemma, i.e. Lemma~\ref{thrm:cont-1} ensures that \eqref{eq:M-max-def}, \eqref{eq:V-max-def} and \eqref{eq:V-min-def} are well-defined, positive and finite. 

Define $K_{\textrm{max}} \eqdef 2\sqrt{2 \pi}c \overline{M}(\nu, W,R)$ with $c = 30/4$. Note that $K_{\textrm{max}} \in \bbR^{+}$. Also, let $N \in \bbZ^+$ be sufficiently large, such that 
\begin{equation}
\sqrt{N} \geq \frac{1+(1+K_{\textrm{max}})^2}{\sqrt{\underline{V}(\nu, W, R)}}, \label{eq:easier-test-cond} 
\end{equation}
and consider such an $N$ from now on. 

Next, we apply the sharp lower bound proposition, i.e. Proposition~\ref{prop:SLB}, to $\alpha_N$ to deduce a lower bound. Observe that \eqref{eq:zerovar-0-der2-1}, \eqref{eq:zerovar-0-der2-2} and the positive variance proposition, i.e. Proposition~\ref{thrm:zerovar-prop4} and the item (iii) of the regularity lemma, i.e. Lemma~\ref{lem:lem-easier-test-2} ensures the fulfillment of the assumptions under which Proposition~\ref{prop:SLB} is stated. Moreover, \eqref{eq:easier-test-cond} guarantees that \eqref{eq:SLB-final-0} holds, and hence we can apply Proposition~\ref{prop:SLB} to $W\left\{ A_N^c | \bx^N \right\}$ (cf. \eqref{eq:easier-test-ANc} and \eqref{eq:easier-test-alpbeta}) to deduce 
\begin{equation}
\alpha_{N} \geq \frac{K}{\sqrt{N}}\exp\{-N\Lambda^\ast_{0,P}(\tilde{\me}_{\mSP}(R,P,r_{N}(R,P)) - r_{N}(R,P))\},
\label{eq:easier-test1}
\end{equation}
where we define 
\begin{equation}
K \eqdef \frac{e^{-K_{\textrm{max}}}}{2\sqrt{2 \pi \overline{V}(\nu, W, R)}}.
\label{eq:K}
\end{equation}
Note that $K$ only depends on $W$, $R$ and $\nu$.

Further, recalling the definition of $\beta_N$ (cf. \eqref{eq:easier-test-AN} and \eqref{eq:easier-test-alpbeta}) one can check that 
\begin{equation}
\beta_{N} \geq W_{R,P}^{-}\left\{\frac{1}{N}\sum_{n=1}^{N}\log\frac{W(Y_{n}|x_{n})}{W^{-}_{R,P}(Y_{n}|x_{n})} \geq \tilde{r}_{N}(R,P) - \tilde{\me}_{\mSP}(R,P,\tilde{r}_{N}(R,P)) \left. \right| \bx^{N}\right\}.
\label{eq:easier-test1.5}
\end{equation}

Next, we apply the sharp lower bound proposition, i.e. Proposition~\ref{prop:SLB}, to the right side of \eqref{eq:easier-test1.5} by noting the fact that the explanations provided prior to \eqref{eq:easier-test1} are still valid (recall \eqref{eq:zerovar-1-der1} and \eqref{eq:zerovar-1-der2}) and infer the following 
\begin{equation}
\beta_{N} \geq \frac{K}{\sqrt{N}}e^{-N\Lambda^\ast_{1,P}( \tilde{r}_{N}(R,P) - \tilde{\me}_{\mSP}(R,P,\tilde{r}_{N}(R,P)))} = \frac{K}{\sqrt{N}} e^{-N\tilde{r}_{N}(R,P)} = \frac{K N^{\zeta}}{e}e^{-Nr(R,P)},
\label{eq:easier-test2}
\end{equation}
where the first equality follows from the item (ii) of the regularity lemma, i.e. Lemma~\ref{lem:lem-easier-test-2}. 

If we let $N \in \bbZ^{+}$ to be sufficiently large, so that 
\[
\frac{K N^{\zeta}}{e} > 1,
\]
then \eqref{eq:easier-test2} implies that $\beta_{N} > e^{-Nr(P,R)}$. Since our test is a likelihood ratio test, by violating the constraint we can only improve the optimal error performance, and hence (cf. \eqref{eq:easier-test1})
\begin{equation*}
\tilde{\alpha}_{N}(r(P,R)) \geq \alpha_{N} \geq \frac{K}{\sqrt{N}}e^{-N\Lambda^\ast_{0,P}(\tilde{\me}_{\mSP}(R,P,r_{N}(R,P)) - r_{N}(R,P))},
\end{equation*} 
which, in turn, implies that (cf. \eqref{eq:hyp-test-10})
\begin{equation}
e(f,\varphi) \geq \frac{K}{\sqrt{N}}e^{-N\Lambda^\ast_{0,P}(\tilde{\me}_{\mSP}(R,P,r_{N}(R,P)) - r_{N}(R,P))}. 
\label{eq:easier-test3}
\end{equation}

\subsection{Approximation of the exponent}
\label{ssec:approximation-of-exponent}
In this final section, we approximate the exponent in \eqref{eq:easier-test3} to conclude the proof. 

To begin with, we note that (e.g. \cite[Exercise 2.2.24]{dembo-zeitouni98}) $\Lambda_{0,P}^{\ast}(\cdot) \in \mathcal{C}^{\infty}(-\mD(W||W_{R,P}^{-}|P), \mD(W_{R,P}^{-}||W|P))$. Moreover, with the aid of the inverse function theorem and the item (iii) of the regularity lemma, i.e. Lemma~\ref{lem:lem-easier-test-2}, one can check that for any $r \in (0, \mD(W||W_{R,P}^{-}|P))$,
\begin{equation}
\Lambda_{0,P}^{\ast \, \prime}(\tilde{\me}_{\mSP}(R,P,r) - r) = \eta(R, P,r), \quad \Lambda_{0,P}^{\ast \, \prime \prime}(\tilde{\me}_{\mSP}(R,P,r) - r) = \frac{1}{\Lambda_{0,P}^{\prime \prime}(\eta(R,P,r))}. \label{eq:fenchel-der}
\end{equation}

Define\footnote{Owing to the item (iv) of the continuity lemma, i.e. Lemma~\ref{thrm:cont-1}, and the compactness of $\cP_{R, \nu}(\cX)$, the maximum is well-defined.} 
\[
\delta(R,\nu,W) \eqdef R - \max_{P \, \in \, \cP_{R, \nu}(\cX)}\mD(W_{R,P}^{-}||Q_{R,P}^{\ast}|P).
\]
Observe that owing to Lemma~\ref{lem:lem-R-new}, $\delta(R,\nu,W) >0$. Hence, one can choose $N \in \bbZ^{+}$ to be sufficiently large, such that $\epsilon_{N} \leq \delta(R,\nu,W)/2$. Consider such an $N$ from now on. 

Using Taylor's theorem, for some $\bar{x} \in (\tilde{\me}_{\mSP}(R,P,r_{N}(R,P)) - r_{N}(R,P) , \tilde{\me}_{\mSP}(R,P,r(R,P)) - r(R,P))$, we get
\begin{align}
\Lambda_{0, P}^{\ast}(\tilde{\me}_{\mSP}(R,P,r_{N}(R,P)) - r_{N}(R,P)) & = \Lambda_{0, P}^{\ast}(\tilde{\me}_{\mSP}(R,P,r(R,P)) - r(R,P)) + \left\{ (\tilde{\me}_{\mSP}(R,P,r_{N}(R,P))- r_{N}(R,P)) \right. \nonumber \\
& \left.  - (\tilde{\me}_{\mSP}(R,P,r(R,P)) - r(R,P))\right\} \Lambda_{0, P}^{\ast \, \prime}(\tilde{\me}_{\mSP}(R,P,r(R,P)) - r(R,P))  \nonumber \\
&  + \frac{ \Lambda_{0,P}^{\ast \, \prime \prime}(\bar{x})\left\{ (\tilde{\me}_{\mSP}(R,P,r_{N}(R,P)) - r_{N}(R,P)) - (\tilde{\me}_{\mSP}(R,P,r(R,P)) - r(R,P))\right\}^{2}}{2} \nonumber \\
& = \Lambda_{0, P}^{\ast}(\tilde{\me}_{\mSP}(R,P,r(R,P)) - r(R,P)) + \epsilon_{N}\Lambda_{0, P}^{\ast \, \prime}(\tilde{\me}_{\mSP}(R,P,r(R,P)) - r(R,P))   \nonumber \\
& + [\tilde{\me}_{\mSP}(R,P,r_{N}(R,P)) - \tilde{\me}_{\mSP}(R,P,r(R,P)) ]\Lambda_{0, P}^{\ast \, \prime}(\tilde{\me}_{\mSP}(R,P,r(R,P)) - r(R,P)) \nonumber \\
& + \frac{ \Lambda_{0,P}^{\ast \, \prime \prime}(\bar{x})\left\{ (\tilde{\me}_{\mSP}(R,P,r_{N}(R,P)) - r_{N}(R,P)) - (\tilde{\me}_{\mSP}(R,P,r(R,P)) - r(R,P))\right\}^{2}}{2}  \label{eq:ach-taylor-0} \\
& = \Lambda_{0, P}^{\ast}(\tilde{\me}_{\mSP}(R,P,r(R,P)) - r(R,P)) + \epsilon_{N} \eta(R,P,r(R,P)) \nonumber \\
& + (\tilde{\me}_{\mSP}(R,P,r_{N}(R,P)) - \tilde{\me}_{\mSP}(R,P,r(R,P)) ) \eta(R,P,r(R,P))\nonumber \\
& + \frac{\Lambda_{0,P}^{\ast \, \prime \prime}(\bar{x}) \left\{ (\tilde{\me}_{\mSP}(R,P,r_{N}(R,P)) - r_{N}(R,P)) - (\tilde{\me}_{\mSP}(R,P,r(R,P)) - r(R,P))\right\}^{2}}{2} , \label{eq:ach-taylor-1}
\end{align}
where \eqref{eq:ach-taylor-0} follows by recalling the fact that $r_N(R,P) = r(R,P) - \epsilon_N$ and \eqref{eq:ach-taylor-1} follows from \eqref{eq:fenchel-der} by recalling \eqref{eq:r(P)-rel}.

Recalling the item (i) of the regularity lemma, i.e. Lemma~\ref{lem:lem-easier-test-2}, \eqref{eq:ach-taylor-1} implies that
\begin{align}
&\Lambda_{0, P}^{\ast}(\tilde{\me}_{\mSP}(R,P,r_{N}(R,P)) - r_{N}(R,P)) \tilde{\me}_{\mSP}(R,P, r_{N}(R,P))   = \tilde{\me}_{\mSP}(R,P,r_{N}(R,P)) + \frac{\eta(R,P,r(R,P))}{1-\eta(R,P,r(R,P))}\epsilon_{N}  \nonumber \\
&  + \frac{\Lambda_{0,P}^{\ast \, \prime \prime}(\bar{x})\epsilon_{N}^{2}}{2 (1-\eta(R,P,r(R,P)))}\left( 1 + \frac{\tilde{\me}_{\mSP}(R,P, r_{N}(R,P)) - \tilde{\me}_{\mSP}(R,P, r(R,P)) }{\epsilon_{N}}\right)^{2}, \label{eq:taylor1}
\end{align}
for some $\bar{x} \in (\tilde{\me}_{\mSP}(R,P,r(R,P)) - r(R,P), \, \tilde{\me}_{\mSP}(R,P,r_{N}(R,P)) - r_{N}(R,P))$. 

Note that, since $\tilde{\me}_{\mSP}(R,P, \cdot) - (\cdot)$ is strictly decreasing and continuous, there exists a unique $\bar{r} \in (r(R,P)-\delta(R,\nu,W)/2, r(R,P))$ such that\footnote{Actually, $ \bar{r} \in (r_{N}(R,P),r(R,P))$.} $\bar{x} = \tilde{\me}_{\mSP}(R,P, \bar{r}) - \bar{r}$ and hence (recall \eqref{eq:fenchel-der} and \eqref{eq:r(P)-rel})
\begin{equation}
\Lambda_{0,P}^{\ast \, \prime \prime}(\bar{x}) = 1/\Lambda_{0,P}^{\prime \prime}(\eta(R,P,\bar{r})). 
\label{eq:var-1}
\end{equation}
Moreover, the item (iii) of the regularity lemma, i.e. Lemma~\ref{lem:lem-easier-test-2}, implies that 
\begin{equation}
\frac{\eta(R,P,r(R,P))}{1-\eta(R,P,r(R,P))} = s^{\ast}(R,P,r(R,P)). 
\label{eq:var-1.1}
\end{equation}
Plugging \eqref{eq:var-1} and \eqref{eq:var-1.1} into \eqref{eq:taylor1}, we deduce that 
\begin{align}
\Lambda_{0, P}^{\ast}(\tilde{\me}_{\mSP}(R,P,r_{N}(R,P)) - r_{N}(R,P)) & = \tilde{\me}_{\mSP}(R,P, r_{N}(R,P)) \nonumber \\
& = \tilde{\me}_{\mSP}(R,P,r(R,P)) + s^{\ast}(R,P,r(R,P)) \epsilon_{N} \nonumber \\ 
& + \frac{ 1+s^{\ast}(R,P,r(R,P))}{2\Lambda_{0,P}^{\prime \prime}(\eta(R,P,\bar{r}))}\epsilon_{N}^{2}\left( 1 + \frac{\tilde{\me}_{\mSP}(R,P, r_{N}(R,P)) - \tilde{\me}_{\mSP}(R,P, r(R,P)) }{\epsilon_{N}}\right)^{2}, \label{eq:taylor1.1}
\end{align}

Moreover, using exactly the same arguments as above, but this time with a first-order Taylor series, we infer that 
\begin{equation}
\tilde{\me}_{\mSP}(R,P,r_{N}(R,P)) = \tilde{\me}_{\mSP}(R,P,r(R,P)) + \epsilon_{N} \frac{\eta(R,P,\tilde{r})}{1- \eta(R,P,\tilde{r})}, \label{eq:taylor3}
\end{equation}
for some $\tilde{r} \in (r_{N}(R,P), r(R,P))$. 

On account of the convexity and the non-increasing property of $\tilde{\me}_{\mSP}(R,P,\cdot)$, we have
\begin{equation}
\left|\frac{\partial \tilde{\me}_{\mSP}(R,P,r^{\prime})}{\partial r^{\prime}}\right| \leq \frac{\tilde{\me}_{\mSP}(R,P,0)}{\delta(R,\nu,W)/2},
\label{eq:easier-test4}
\end{equation}
for any $r_{N}(R,P) \leq r^{\prime} \leq r(R,P)$.

By noting that $\tilde{\me}_{\mSP}(R,P,0) = \mD(W_{R,P}^{-}||W|P) = \Lambda_{0,P}^{\prime}(1)$ and letting\footnote{Owing to the continuity lemma, i.e. Lemma~\ref{thrm:cont-1}, the maximum is well-defined and finite.} $F \eqdef \max_{P \in \cP_{R, \nu}(\cX)}\Lambda_{0,P}^{\prime}(1) < \infty$, \eqref{eq:easier-test4} further implies that 
\begin{equation}
\frac{\eta(R,P,r^{\prime})}{1-\eta(R,P,r^{\prime})} = s^{\ast}(R,P,r^{\prime}) =\left|\frac{\partial \tilde{\me}_{\mSP}(R,P,r^{\prime})}{\partial r^{\prime}}\right| \leq \frac{F}{\delta(R,\nu,W)/2} =: \tilde{s} < \infty,
\label{eq:easier-test5}
\end{equation}
for any $r_{N}(R,P) \leq r^{\prime} \leq r(R,P)$.

Plugging \eqref{eq:V-min-def}, \eqref{eq:taylor3} and \eqref{eq:easier-test5} into \eqref{eq:taylor1.1} yields 
\begin{align}
\Lambda_{0, P}^{\ast}(\tilde{\me}_{\mSP}(R,P,r_{N}(R,P)) - r_{N}(R,P)) & = \tilde{\me}_{\mSP}(R,P,r_{N}(R,P)) \nonumber \\
& \leq \tilde{\me}_{\mSP}(R,P,r(R,P)) \nonumber \\
& + s^{\ast}(R,P,r(R,P)) \epsilon_{N}\left[ 1 + \epsilon_{N}\frac{(1+\tilde{s})^{2}[1+s^\ast(R,P,r(R,P))]}{2\underline{V}(\nu, W, R) s^\ast(R,P,r(R,P))}\right] \nonumber \\
 & = \mE_{\mSP}(R,P)+ s^{\ast}(R,P,r(R,P)) \epsilon_{N}\left[ 1 + \epsilon_{N}\frac{(1+\tilde{s})^{2}[1+s^\ast(R,P,r(R,P))]}{2\underline{V}(\nu, W, R)s^\ast(R,P,r(R,P))}\right] \label{eq:taylor4.0}\\
 & \leq \mE_{\mSP}(R,P)+ s^{\ast}(R,P,r(R,P)) \epsilon_{N}\left[ 1 + \epsilon_{N}\frac{(1+\tilde{s})^{2}}{2\underline{V}(\nu, W, R)} \left( 1 + \frac{2\Upsilon(W,R,\nu)}{\nu}\right) \right], \label{eq:taylor4}
\end{align}
where \eqref{eq:taylor4.0} follows from the equality of the exponents theorem, i.e. Theorem~\ref{lem:lem-opt3} and the shifted exponent lemma, i.e. Lemma~\ref{lem:opt-e-tilde-e-equ}, and \eqref{eq:taylor4} follows from \eqref{eq:var-1.1} and Lemma~\ref{lem:eta-lb}.

Consider $\zeta \in \bbR^{+}$ that is fixed in the definition of $\epsilon_{N}$. Since $\tilde{s}$ is bounded, $\underline{V}(\nu, W, R)$ and $\Upsilon(W,R,\nu)$ and the fact that $\nu >0$, one can deduce that for all sufficiently large $N$, 
\[
\epsilon_{N}\frac{(1+\tilde{s})^{2}}{2\underline{V}(\nu, W, R)} \left( 1 + \frac{2\Upsilon(W,R,\nu)}{\nu}\right) \leq \zeta,
\]
and hence \eqref{eq:taylor4} reduces to the following, for all sufficiently large $N$,
\begin{equation}
\Lambda_{0, P}^{\ast}(\tilde{\me}_{\mSP}(R,P,r_{N}(R,P)) - r_{N}(R,P)) \leq \mE_{\mSP}(R,P)+ s^{\ast}(R,P,r(R,P)) \epsilon_{N}(1+\zeta).
\label{eq:taylor4.1}
\end{equation}

Next, we claim that 
\begin{equation}
s^{\ast}(R,P,r(R,P)) = \rho^{\ast}_{R,P}. 
\label{eq:der-eq}
\end{equation}

To prove this, we first claim that $\rho^{\ast}_{R,P}$ is a Lagrange multiplier of $\me_\mSP(R,P)$. To see this, first note that
\begin{align}
\me_{\mSP}(R,P) & = \mE_{\mSP}(R,P) \label{eq:last1} \\
 & = K_{R,P}(\rho^{\ast}_{R,P}, Q^{\ast}_{R,P}) \label{eq:last2} \\
 & = \max_{\rho \in \bbR_{+}} K_{R,P}(\rho, Q^{\ast}_{R,P}) \label{eq:last3} \\
 & = \max_{\rho \in \bbR_{+}} \min_{V \in \cP(\cY|\cX)} \left[ \mD(V||W|P) + \rho(\mD(V||Q_{R,P}^{\ast}|P) - R)\right], \label{eq:last4}
\end{align}
where \eqref{eq:last1} follows from the equality of the exponents theorem, i.e. Theorem~\ref{lem:lem-opt3}, \eqref{eq:last2} follows from the saddle-point proposition, i.e. Proposition~\ref{prop:saddle-point}, and the uniqueness of the saddle-point proposition, i.e. Proposition~\ref{prop:unique-saddle-point}, \eqref{eq:last3} follows by noting that $(\rho_{R,P}^{\ast}, Q_{R,P}^{\ast})$ is the unique saddle-point of $K_{R,P}(\cdot, \cdot)$ and \eqref{eq:last4} follows by solving the convex minimization problem. Hence, \eqref{eq:last4} gives the Lagrangian dual of $\me_{\mSP}(R,P)$. 

Further, one can also check that 
\begin{equation}
\max_{\rho \in \bbR_{+}} \min_{V \in \cP(\cY|\cX)} \left[ \mD(V||W|P) + \rho(\mD(V||Q_{R,P}^{\ast}|P) - R)\right] = \min_{V \in \cP(\cY|\cX)} \left[ \mD(V||W|P) + \rho^\ast_{R,P}(\mD(V||Q_{R,P}^{\ast}|P) - R)\right]. 
\label{eq:last4.0}
\end{equation}

\eqref{eq:last4} and \eqref{eq:last4.0} implies that $\rho^{\ast}_{R,P}$ is a Lagrange multiplier of $\me_\mSP(R,P)$. Moreover, the sub-differential characterization of the Lagrange multipliers (e.g. \cite[Theorem~29.1]{rockafellar70}) along with the differentiability of the shifted exponent lemma, i.e. Lemma~\ref{thrm:thrm-sstar-def}, and Corollary~\ref{cor:cor-eSP-der}, implies \eqref{eq:der-eq}. 

Plugging \eqref{eq:der-eq} into \eqref{eq:taylor4.1}, we deduce that  
\begin{equation}
\Lambda_{0, P}^{\ast}(\tilde{\me}_{\mSP}(R,P,r_{N}(R,P)) - r_{N}(R,P)) \leq \mE_{\mSP}(R,P)+ \rho^{\ast}_{R,P} \epsilon_{N}(1+\zeta).
\label{eq:taylor4.2}
\end{equation}

Define $\cP_{R}^{\ast}(\cX) \eqdef \left\{ P \in \cP(\cX) : \mE_{\mSP}(R,P) = \mE_{\mSP}(R)\right\} \neq \emptyset$. Observe that $\cP_{R}^{\ast}$ is a compact set. Also, for any $P \in \cP(\cX)$, $|P - \cP_{R}^{\ast}| \eqdef \inf_{Q \in \cP_{R}^{\ast}}|| Q - P||_{1}$. For any $\theta \in \bbR^{+}$, $\cP_{\theta}(\cX) \eqdef \left\{ P \in \cP_{R,\nu}(\cX) : |P - \cP_{R}^{\ast}(\cX)| \geq \theta \right\}$. 

Observe that (recall \eqref{eq:worst-slope} and the differentiability of $\mE_{\mSP}(\cdot,P)$ proposition, i.e. Proposition~\ref{prop:diff-ESP})
\begin{equation}
\rho_{R}^{\ast} = \max_{P \in \cP_{R}^{\ast}(\cX)}\rho^{\ast}_{R,P},
\label{eq:rho-ast}
\end{equation}
where owing to the compactness of $\cP_{R}^{\ast}(\cX)$ and the continuity of $\rho^{\ast}_{R,\cdot}$, the maximum is well-defined and finite. 

Since $\cP_{R,\nu}(\cX)$ is compact, $\rho_{R,\cdot}^{\ast}$ is uniformly continuous on this set, equivalently
\begin{equation}
\forall \upsilon \in \bbR^{+}, \, \exists \, a(\upsilon) \in \bbR^{+}, \mbox{ s.t. } \forall P,Q \in \cP_{R,\nu}(\cX), \, ||P-Q||_{1}< a(\upsilon) \Rightarrow |\rho_{R,P}^{\ast}-\rho_{R,Q}^{\ast}| < \zeta. 
\label{eq:cont}
\end{equation}

Consider $\zeta \in \bbR^{+}$ that is fixed in the definition of $\epsilon_{N}$ and let $a(\zeta) \in \bbR^{+}$ be chosen such that \eqref{eq:cont} holds. 

If $P \in \cP_{R, \nu}(\cX)-\cP_{a(\zeta)}(\cX)$, then \eqref{eq:cont} ensures that $\rho^{\ast}_{R,P} \leq \rho^{\ast}_{R} + \zeta$, which, in turn, implies that 
\begin{equation}
\exp(-N\epsilon_{N}(1+\zeta)\rho_{R,P}^{\ast}) \geq N^{-(1+\zeta)\left(\frac{1}{2}+\zeta\right)(\rho_{R}^{\ast}+\zeta)}.
\label{eq:taylor5} 
\end{equation}

Suppose $P \in \cP_{a(\zeta)}(\cX)$. Since $\mE_{\mSP}(R) - \max_{P \, \in \, \textrm{cl}(\cP_{a(\zeta)})}\mE_{\mSP}(R,P) \in \bbR^{+}$, one can check that for all sufficiently large $N$, uniformly over $\cP_{a(\zeta)}(\cX)$, we have 
\begin{equation}
\exp\left( -N \left[ \mE_{\mSP}(R,P) + \epsilon_{N}(1+\zeta)\rho^{\ast}_{R,P}\right]\right) \geq \frac{e^{-N\mE_{\mSP}(R)}}{N^{(1+\zeta)\left( \frac{1}{2}+\zeta\right)\rho_{R}^{\ast}}}. 
\label{eq:taylor6}
\end{equation}

Equations \eqref{eq:easier-test3}, \eqref{eq:taylor4.2}, \eqref{eq:taylor5} and \eqref{eq:taylor6} imply \eqref{eq:thrm-main}, hence we conclude the proof of Theorem~\ref{thrm:main}. 


\appendices


\section{Proof of Proposition~\ref{prop:saddle-point}}
\label{app:prop-saddle-point}

\begin{lemma}
For any $R > R_{\infty}$ and $P \in \cP(\cX)$, 
\begin{equation}
\mE_{\mSP}(R,P) = \max_{\rho \, \in \, \bbR_{+}} \min_{q \, \in \, \cP(\cY)} \left\{ -\rho R - (1+\rho) \Lambda_{Q,P}\left( \frac{\rho}{1+\rho}\right)\right\} .
\label{eq:opt-lem1.1}
\end{equation}
\label{lem:opt-lem1.1}
\end{lemma}
\vspace{-0.5cm}
\begin{IEEEproof} The proof is clear from basic optimization theoretic arguments, (e.g. \cite[Exercise 2.5.23]{csiszar-korner81}), we just reproduce the steps for the sake of completeness. 
\begin{align}
\mE_{\mSP}(R,P) & = \max_{\rho \, \in \, \bbR_{+}} \min_{V \, \in \, \cP(\cY|\cX)} \left\{ \mD(V||W|P) + \rho[\mI(P;V) - R]\right\} \label{eq:opt-lem1.1-pf1}\\
& = \max_{\rho \, \in \, \bbR_{+}} \min_{V \, \in \, \cP(\cY|\cX)} \left\{ \mD(V||W|P) +\rho\left[ \min_{Q \, \in \, \cP(\cY)}\mD(V||Q|P) - R \right] \right\} \nonumber\\
& = \max_{\rho \, \in \, \bbR_{+}} \left\{ - \rho R + \min_{Q \, \in \, \cP(\cY)} \min_{V \, \in \, \cP(\cY|\cX) }\left[ \mD(V||W|P) + \rho \mD(V||Q|P) \right] \right\} \nonumber \\
& = \max_{\rho \, \in \, \bbR_{+}}\min_{Q \, \in \, \cP(\cY)} \left\{ -\rho R -(1+\rho) \Lambda_{Q,P}\left(\frac{\rho}{1+\rho}\right)\right\}. \nonumber
\end{align}
\end{IEEEproof}

\begin{remark} Recalling the definitions of $\cP_{P,W}(\cY)$ and $\tilde{\cP}_{P,W}(\cY)$ (cf. \eqref{eq:P-PW} and \eqref{eq:tilde-P-PW}), we note the following facts:
\begin{enumerate}
\item[(i)] $\cP_{P,W}(\cY)$ and $\tilde{\cP}_{P,W}(\cY)$ are convex sets and $\tilde{\cP}_{P,W}(\cY) \subset \cP_{P,W}(\cY) $. 

\item[(ii)] From the basic facts about convex sets (e.g. \cite[Proposition 1.4.1 (c), Proposition 1.4.3 (b)]{bertsekas2003}), $\mbox{ri}(\bbR_{+}) = \bbR^{+}$ and $\mbox{ri}(\cP_{P,W}(\cY)) = \mbox{ri}(\cP(\cY)) = \{ Q \in \cP(\cY) : Q(y) >0, \, \forall y \in \cY\}.$

\item[(iii)] For any $Q \in \cP_{P,W}(\cY)$, $\Lambda_{Q,P}(\lambda) \in \bbR$, for all $\lambda \in [0,1)$. 

\item[(iv)] For any $Q \in \cP(\cY) \backslash \cP_{P,W}(\cY)$, $\Lambda_{Q,P}(\lambda) = -\infty $, for all $\lambda \in (0,1)$ and hence given any $R > R_{\infty}$, $P \in \cP(\cX)$ and $Q \in \cP(\cY) \backslash \cP_{P,W}(\cY)$, $K_{R,P}(\rho, Q) = \infty$ for all $\rho \in \bbR^{+}$. 
\end{enumerate}
\label{rem:opt-rem1}
\end{remark}

\begin{lemma}Consider any $R > R_{\infty}$ and $P \in \cP(\cX)$. 
\begin{enumerate}
\item[(i)] Given any $\rho \in \bbR_{+}$ (resp. $\rho \in \bbR^{+}$), $K_{R,P}(\rho, \cdot)$ is (resp. strictly) convex on $\cP_{P,W}(\cY)$ (resp. $\tilde{\cP}_{P,W}(\cY)$).
\item[(ii)] Given any $Q \in \cP_{P,W}(\cY)$, $K_{R,P}(\cdot, Q)$ is concave on $\bbR_{+}$.    
\end{enumerate}
\label{lem:opt-cla1}
\end{lemma}

\begin{IEEEproof} Let $R > R_{\infty}$ and $P \in \cP(\cX)$ be arbitrary. 
\begin{enumerate}
\item[(i)] Given any $x \in \cS(P)$ and $\lambda \in [0,1)$ define $f_{x, \lambda} \, : \, \cP_{P,W}(\cY) \rightarrow \bbR^{+}$ such that 
\[
f_{x, \lambda}(Q) \eqdef 
\begin{cases}
\sum_{y \in \cY} W(y|x)^{1-\lambda} Q(y)^{\lambda}, & \mbox{ if } \lambda \in (0,1), \\
1, & \mbox{ if } \lambda =0,
\end{cases}
\]
for any $Q \in \cP_{P,W}(\cY)$. Let $Q_{1}, Q_{2} \in \cP_{P,W}(\cY)$ and $\theta \in (0,1)$ be arbitrary. For any $\lambda \in (0,1)$, we have 
\begin{align}
f_{x, \lambda}(\theta Q_{1} + (1-\theta)Q_{2}) & = \sum_{y \in \cY} W(y|x)^{1-\lambda}[\theta Q_{1}(y) + (1-\theta)Q_{2}(y)]^{\lambda} \nonumber \\
& \geq \sum_{y \in \cY} W(y|x)^{1-\lambda} [\theta Q_{1}(y)^{\lambda} + (1-\theta)Q_{2}(y)^{\lambda}] \label{eq:opt-cla1-pf1}\\
& = \theta f_{x,\lambda}(Q_{1}) + (1-\theta)f_{x,\lambda}(Q_{2}), \label{eq:opt-cla1-pf2}
\end{align}
where \eqref{eq:opt-cla1-pf1} follows from the concavity of $(\cdot)^{\lambda}$ on $\bbR_{+}$ for any $\lambda \in (0,1)$. Clearly, \eqref{eq:opt-cla1-pf2} is true for $\lambda =0$.

Since $\log(\cdot)$ is strictly increasing and strictly concave on $\bbR^{+}$, \eqref{eq:opt-cla1-pf2} implies that
\begin{align}
\log(f_{x, \lambda}(\theta Q_{1} + (1-\theta)Q_{2})) & \geq \log(\theta f_{x,\lambda}(Q_{1}) + (1-\theta)f_{x,\lambda}(Q_{2})) \nonumber \\
 & \geq \theta \log(f_{x,\lambda}(Q_{1})) + (1 - \theta)\log(f_{x,\lambda}(Q_{2})). \label{eq:opt-cla1-pf3}
\end{align}
\eqref{eq:opt-cla1-pf3} implies that given any $\rho \in \bbR_{+}$, $\Lambda_{\cdot, P}\left(\frac{\rho}{1+\rho}\right)$ is concave on $\cP_{P,W}(\cY)$. By recalling the definition of $K_{R,P}$ (cf. \eqref{eq:K-RP}), this implies that $K_{R,P}(\rho, \cdot)$ is convex on $\cP_{P,W}(\cY)$. 

Strict concavity follows by noting that for any $Q_{1}, Q_{2} \in \tilde{\cP}_{P,W}(\cY)$ such that $Q_{1} \neq Q_{2}$ and $\lambda \in (0,1)$, the inequality in \eqref{eq:opt-cla1-pf1} is strict owing to the strict concavity of $(\cdot)^{\lambda}$ on $\bbR^{+}$ for any $\lambda \in (0,1)$. 

\item[(ii)] For any $\lambda \in (0,1)$, $Q \in \cP_{P,W}(\cY)$ and $x \in \cS(P)$ define
\begin{equation}
\forall y \in \cY, \, \tilde{W}_{\lambda, Q}(y|x) \eqdef \frac{W(y|x)^{1-\lambda}Q(y)^{\lambda}}{\sum_{\tilde{y}\in \cY} W(\tilde{y}|x)^{1-\lambda}Q(\tilde{y})^{\lambda}}.
\label{eq:opt-tilted-channel}
\end{equation}
Recalling the definition of $\cP_{P,W}(\cY)$, $\tilde{W}_{\lambda, Q}(\cdot|x)$ is a well-defined probability measure on $\cY$. It is easy to check that\footnote{For the sake of notational convenience $\Lambda_{Q,P}^{\prime}(\lambda) $ (resp. $\Lambda_{Q,P}^{\prime \prime}(\lambda) $) denotes $\frac{\partial \Lambda_{Q,P}(\lambda)}{\partial \lambda}$ (resp. $\frac{\partial^{2} \Lambda_{Q,P}(\lambda)}{\partial \lambda^{2}}$) in the sequel.}
\begin{align}
\Lambda_{Q,P}^{\prime}(\lambda) & = \sum_{x \in \cS(P)} P(x)\mbox{E}_{\tilde{W}_{\lambda,Q}(\cdot|x)}\left[ \log \frac{Q(Y)}{W(Y|x)}\right], \label{eq:opt-der1}\\
\Lambda_{Q,P}^{\prime \prime}(\lambda) & = \sum_{x \in \cS(P)}P(x) \mbox{Var}_{\tilde{W}_{\lambda,Q}(\cdot|x)}\left[ \log \frac{Q(Y)}{W(Y|x)}\right], \label{eq:opt-der2}
\end{align}
for any $Q \in \cP_{P,W}(\cY)$ and $\lambda \in (0,1)$. Recalling the definition of $K_{R,P}$ (cf. \eqref{eq:K-RP}), \eqref{eq:opt-der2} implies that 
\begin{equation}
\frac{\partial^2 K_{R,P}(\rho, q)}{\partial \rho^2}  = -\frac{1}{(1+\rho)^{3}} \Lambda_{Q,P}^{\prime \prime}\left(\frac{\rho}{1+\rho}\right) \leq 0, 
\label{eq:opt-cla1-pf4}
\end{equation}
for any $Q \in \cP_{P,W}(\cY)$ and $ \rho \in \bbR^{+}$. 

Now, fix any $Q \in \cP_{P,W}(\cY)$. \eqref{eq:opt-cla1-pf4} implies that $-K_{R,P}(\cdot, Q)$ is convex on $\bbR^{+}$, equivalently, the epigraph of $-K_{R,P}(\cdot, Q)$ with its domain restricted to $\bbR^{+}$ is a convex set. Furthermore, 
\[
\lim_{\rho \downarrow 0} -K_{R,P}(\rho, Q) \leq 0 = -K_{R,P}(0,Q).
\]
Hence, after adding $0$ into the domain of $K_{R,P}(\cdot, Q)$, its epigraph remains to be convex.  
\end{enumerate}
\end{IEEEproof}

\begin{definition}
Let $G \subset \bbR^{n}$ and $f \, : \, G \rightarrow \bbR$. $(G,f)$ is ``convex and closed in Fenchel's sense'' (cf. \cite[pg. 151]{rockafellar64}, \cite [end of Section 2]{fenchel49}) (resp. ``concave and closed in Fenchel's sense'') provided that:
\begin{enumerate}
\item[(i)] $G$ is convex.

\item[(ii)] $f$ is convex (resp. concave) and lower (resp. upper) semi-continuous. 

\item[(iii)] Any accumulation point of $G$ that does not belong to $G$ satisfies $\lim f(\cdot) = \infty$ (resp. $\lim f(\cdot) = - \infty$).
\end{enumerate}
\label{def:opt-closed-convex}
\end{definition}

\begin{lemma} Let $R > R_{\infty}$ and $P \in \cP(\cX)$ be arbitrary. For any $Q \in \mbox{ri}(\cP_{P,W}(\cY))$ (resp. $\rho \in \mbox{ri}(\bbR_{+})$), $(\bbR_{+}, K_{P,R}(\cdot, Q))$ (resp. $(\cP_{P,W}(\cY), K_{P,R}(\rho, \cdot))$) is concave (resp. convex) and closed in Fenchel's sense.
\label{lem:opt-cla3}
\end{lemma}

\begin{IEEEproof} Fix any $R > R_{\infty}$ and $P \in \cP(\cX)$.

First, fix an arbitrary $Q \in \tilde{P}_{P,W}(\cY)$. Observe that $\Lambda_{Q,P}(\lambda) \in \bbR$ for all $\lambda \in (0, 1)$, which in turn implies that $\Lambda_{q,P}(\lambda)$ is infinitely differentiable with respect to $\lambda$ for all $\lambda \in (0,1)$. Moreover, recalling the definition of $\tilde{P}_{P,W}(\cY)$, it is easy to check that for any $Q \in \tilde{P}_{P,W}(\cY)$, $\lim_{\lambda \downarrow 0}\Lambda_{Q,P}(\lambda) = 0 = \Lambda_{Q, P}(0)$. These two observations ensure the continuity (and \emph{a fortiori} upper semi-continuity) of $K_{R,P}(\cdot, Q)$ on $\bbR_{+}$. By noting (recall item (ii) of Remark~\ref{rem:opt-rem1}) $\mbox{ri}(\cP_{P,W}(\cY)) \subset \tilde{P}_{P,W}(\cY)$, the fact that $\bbR_{+}$ is closed and convex and the concavity of $K_{R,P}(\cdot, Q)$ (cf. item (ii) of Lemma~\ref{lem:opt-cla1}) this suffices to conclude that $(\bbR_{+}, K_{R,P}(\cdot, Q))$ is concave and closed in Fenchel's sense. 

Next, fix an arbitrary $\rho \in \mbox{ri}(\bbR_{+}) = \bbR^{+}$ (cf. the item (ii) of Remark~\ref{rem:opt-rem1}). Observe that any accumulation point of $\cP_{P,W}(\cY)$ which does not belong to $\cP_{P,W}(\cY)$, say $Q_{0}$, satisfies $Q_{0} \in \cP(\cY) \backslash \cP_{P,W}(\cY) $, owing to the compactness of $\cP(\cY)$, and hence $K_{R,P}(\rho, Q_{0}) = \infty$. Further, item (i) of Remark~\ref{rem:opt-rem1} and item (i) of Lemma~\ref{lem:opt-cla1} ensures that in order to conclude that $K_{R,P}(\rho, \cdot)$ is convex and closed in Fenchel's sense, we only need to verify the lower semi-continuity. Implied by its convexity, $K_{R,P}(\rho, \cdot)$ is continuous on $\mbox{ri}(\cP(\cY))$. Let $Q_{0} \in \cP_{P,W}(\cY) \backslash \mbox{ri}(\cP(\cY))$ be arbitrary. Consider an arbitrary sequence $\{ Q_{k}\}_{k \geq 1}$  such that $Q_{k} \in \cP_{P,W}(\cY)$ and $\lim_{k \rightarrow \infty}Q_{k} = Q_{0}$. Lastly, define $\lambda \eqdef \frac{\rho}{1+\rho} \in (0,1)$. We have
\begin{align}
\lim_{k \rightarrow \infty} \Lambda_{Q_{k},P}(\lambda) & = \lim_{k \rightarrow \infty} \sum_{x \in \cS(P)} P(x) \log \sum_{y \in \cY} W(y|x)^{1-\lambda}Q_{k}(y)^{\lambda} \nonumber \\
 & = \sum_{x \in \cS(P)} P(x) \log \sum_{y \in \cY} W(y|x)^{1-\lambda}Q_{0}(y)^{\lambda} \label{eq:opt-cla3-pf1} \\
 & = \Lambda_{Q_{0},P}(\lambda), \nonumber 
\end{align}
where \eqref{eq:opt-cla3-pf1} follows from the continuity of $\log(\cdot)$ and $(\cdot)^{\lambda}$. 
\end{IEEEproof}

Now, we are ready to prove the existence of a saddle-point. To this end, fix arbitrary $R > R_{\infty}$ and $P \in \cP(\cX)$ from now on. 

We first establish 
\begin{equation}
-\infty < \max_{\rho \, \in \, \bbR_{+}} \inf_{Q \, \in \, \cP_{P,W}(\cY)} K_{R,P}(\rho,Q) = \min_{Q \, \in \, \cP_{P,W}(\cY)}\sup_{\rho \, \in \, \bbR_{+}} K_{R,P}(\rho,Q) < \infty.
\label{eq:opt-cla4-pf1}
\end{equation}
In order to prove \eqref{eq:opt-cla4-pf1}, we use a minimax theorem of Rockafellar, \cite[Theorem 8]{rockafellar64}. Lemma~\ref{lem:opt-cla3} ensures that $(\bbR_{+}, \cP_{P,W}(\cY), K_{R,P})$ is a ``closed saddle-element'' (cf. \cite[pg. 151]{rockafellar64}) and the boundedness of $\cP_{P,W}(\cY)$ guarantees the fulfillment of condition (II) for the validity of the aforementioned theorem (cf. \cite[pg. 172]{rockafellar64}). Therefore \cite[eq. (7.2)]{rockafellar64} implies that
\begin{equation}
- \infty < \sup_{\rho \, \in \, \bbR_{+}}\inf_{Q \, \in \, \cP_{P,W}(\cY)}K_{R,P}(\rho, Q) = \min_{Q\, \in \, \cP_{P,W}(\cY)} \sup_{\rho \, \in \, \bbR_{+}}K_{R,P}(\rho, Q).
\label{eq:opt-cla4-pf2}
\end{equation} 
Next, we claim that 
\begin{equation}
\forall \, \rho \in \bbR_{+}, \, \inf_{Q \, \in \, \cP_{P,W}(\cY)}K_{R,P}(\rho, Q) = \inf_{Q \, \in \, \cP(\cY)}K_{R,P}(\rho, Q). 
\label{eq:opt-cla4-pf3}
\end{equation}
Since $\Lambda_{Q,P}(0) = 0$, for all $q \in \cP(\cY)$, \eqref{eq:opt-cla4-pf3} is trivially true for $\rho=0$. On the other hand, for any $\rho \in \bbR^{+}$, item (iv) of Remark~\ref{rem:opt-rem1} implies that  
\[
\forall \, Q \in \cP(\cY) \backslash \cP_{P,W}(\cY), \, K_{R,P}(\rho,Q) = \infty,  
\]
which, in turn, implies \eqref{eq:opt-cla4-pf3}. Equation \eqref{eq:opt-lem1.1} and \eqref{eq:opt-cla4-pf3} imply that 
\begin{equation}
\mE_{\mSP}(R,P) = \max_{\rho \, \in \, \bbR_{+}} \min_{Q \in \cP(\cY)} K_{R,P}(\rho, Q) = \max_{\rho \, \in \, \bbR_{+}} \inf_{Q \in \cP_{P,W}(\cY)} K_{R,P}(\rho, Q) < \infty. \label{eq:opt-cla4-pf4}
\end{equation}
Equation \eqref{eq:opt-cla4-pf2} and \eqref{eq:opt-cla4-pf4} imply that 
\begin{equation*}
-\infty < \max_{\rho \, \in \, \bbR_{+}} \inf_{Q \in \cP_{P,W}(\cY)} K_{R,P}(\rho, Q) = \min_{Q \, \in \, \cP_{P,W}(\cY)} \sup_{\rho \, \in \, \bbR_{+}}K_{R,P}(\rho, Q) < \infty, 
\end{equation*}
which is \eqref{eq:opt-cla4-pf1}.

From \cite[Lemma 36.2]{rockafellar70}, \eqref{eq:opt-cla4-pf1} ensures the existence of a saddle-point on $\bbR_{+} \times \cP_{P,W}(\cY)$ and \eqref{eq:opt-cla4-pf4} implies the saddle-value is $\mE_{\mSP}(R,P)$. Hence we conclude the proof of the first assertion of the proposition. 

Next, we prove the second assertion. 

\begin{lemma} Consider any $R > R_{\infty}$ and $P \in \cP(\cX)$. If $0 \in \left.S(R,P)\right|_{\bbR_{+}}$, then $\mE_{\mSP}(R,P) =0$, equivalently, if $\mE_{\mSP}(R,P) > 0$, then $0 \notin \left.S(R,P)\right|_{\bbR_{+}}$.
\label{lem:opt-cla5}
\end{lemma}

\begin{IEEEproof}
Consider any $R > R_{\infty}$ and $P \in \cP(\cX)$. Assume $0 \in \left.S(R,P)\right|_{\bbR_{+}}$. We clearly have $K_{R,P}(0, Q) = 0$, for all $Q \in \cP_{P,W}(\cY)$, which in turn implies that (recall the definition of the saddle-point) $K_{R,P}(0,\hat{Q}) =0$ for any $\hat{Q} \in \cP_{P,W}(\cY)$ satisfying $(0, \hat{Q}) \in S(R,P)$. From the first assertion of Proposition~\ref{prop:saddle-point}, this implies the claim. 
\end{IEEEproof}
Recalling the definition of $\cP_{R}(\cX)$ (cf. \eqref{eq:non-trivial-types}), Lemma~\ref{lem:opt-cla5} immediately implies the following result.
\begin{corollary}
For any $C > R > R_{\infty}$ and $P \in \cP_{R}(\cX)$, $\left.S(R,P)\right|_{\bbR_{+}} \subset \bbR^{+}$.
\label{cor:opt-cor1}
\end{corollary}

\begin{lemma}
For any $C > R > R_{\infty}$ and $P \in \cP_{R}(\cX)$, $\left. S(R,P) \right|_{\cP_{P,W}(\cY)} \subset \tilde{\cP}_{P,W}(\cY)$. 
\label{lem:opt-cla6}
\end{lemma}

\begin{IEEEproof} Fix any $C > R > R_{\infty}$ and $P \in \cP_{R}(\cX)$. Let $\hat{\rho} \in \left. S(R,P) \right|_{\bbR_{+}}$ be arbitrary. Note that owing to Corollary~\ref{cor:opt-cor1}, $\hat{\rho} \in \bbR^{+}$. Define $\lambda \eqdef \frac{\hat{\rho}}{1+\hat{\rho}} \in (0,1)$ and recall that (cf. proof of Lemma~\ref{lem:opt-cla1}) $\Lambda_{\cdot, P}(\lambda)$ is concave on $\cP_{P,W}(\cY)$. 

For any $\hat{Q} \in \cP_{P,W}(\cY)$ such that $(\hat{\rho},\hat{Q}) \in S(R,P)$ we have
\begin{equation}
K_{R,P}(\hat{\rho}, \hat{Q}) = \min_{Q \, \in \, \cP_{P,W}(\cY)}K_{R,P}(\hat{\rho}, Q) = -\hat{\rho}R - (1+\hat{\rho}) \max_{Q \in \cP_{P,W}(\cY)}\Lambda_{Q, P}\left( \frac{\hat{\rho}}{1+\hat{\rho}}\right), \label{eq:opt-cla6-pf1}
\end{equation}
from the definition of the saddle-point. 

Now, consider any $Q \in \cP_{P,W}(\cY)$ and for any $x \in \cS(P)$, define $\Lambda_{Q,x}(\lambda) \eqdef \log \sum_{y \in \cY}W(y|x)^{1-\lambda}Q(y)^{\lambda}$. Note that we have 3 possibilities for the partial derivatives of $\Lambda_{Q,x}(\lambda)$ with respect to $Q(y)$:
\begin{enumerate}
\item If $y \in \cS(W(\cdot|x)) \cap \cS(Q)$, then 
\begin{equation}
\frac{\partial \Lambda_{Q, x}(\lambda)}{\partial Q(y)} = \frac{\lambda W(y|x)^{1-\lambda}Q(y)^{\lambda -1}}{\sum_{\tilde{y}\in \cY} W(\tilde{y}|x)^{1-\lambda}Q(\tilde{y})^{\lambda}}, \label{eq:opt-cla6-pf2}
\end{equation}
which is continuous in $Q(y)$. 

\item If $y \notin \cS(W(\cdot|x))$, then (since any variation along this direction does not change the value of the function)
\begin{equation}
\frac{\partial \Lambda_{Q, x}(\lambda)}{\partial Q(y)} = 0, 
\label{eq:opt-cla6-pf3}
\end{equation}
which is continuous in $Q(y)$. 

\item  If $y \notin \cS(Q)$ and $y \in \cS(W(\cdot|x))$, then 
\begin{equation}
\frac{\partial \Lambda_{Q, x}(\lambda)}{\partial Q(y)} = \infty. 
\label{eq:opt-cla6-pf4}
\end{equation}
\end{enumerate}

Then, \cite[Theorem 4.4.1]{gallager68} implies that\footnote{Strictly speaking the statement of the aforementioned theorem requires the cost function of the maximization problem to be continuously differentiable (with possible infinite value on the boundary) on the whole probability simplex. However, it is easy to verify that the proof given by Gallager is also applicable to our case. Indeed, for sufficiency, the item (iv) of Remark~\ref{rem:opt-rem1} ensures that the value of the cost function evaluated at any $Q$ satisfying \eqref{eq:opt-cla6-pf5} and \eqref{eq:opt-cla6-pf6} is not smaller than its counterpart for any $Q \in \cP(\cY) \backslash \cP_{P,W}(\cY)$. For necessity, again the item (iv) of Remark~\ref{rem:opt-rem1} ensures that any optimizer cannot be in $\cP(\cY) \backslash \cP_{P,W}(\cY)$.} a necessary and sufficient condition for any $Q \in \cP_{P,W}(\cY)$ to achieve the maximum in \eqref{eq:opt-cla6-pf1} is:
\begin{align}
\frac{\partial \Lambda_{Q,P}(\lambda)}{\partial Q(y)}  = \delta, \, \forall \, y \in \cS(Q), \label{eq:opt-cla6-pf5} \\
\frac{\partial \Lambda_{Q,P}(\lambda)}{\partial Q(y)}  \leq \delta,  \forall \, y \notin \cS(Q), \label{eq:opt-cla6-pf6}
\end{align}
for some $\delta \in \bbR$. Clearly, if $Q \notin \tilde{\cP}_{P,W}(\cY)$ then it cannot satisfy \eqref{eq:opt-cla6-pf5} and \eqref{eq:opt-cla6-pf6} (cf. \eqref{eq:opt-cla6-pf4}). Hence, any minimizer of \eqref{eq:opt-cla6-pf1} belongs to $\tilde{\cP}_{P,W}(\cY)$. 
\end{IEEEproof}

Corollary~\ref{cor:opt-cor1} and Lemma~\ref{lem:opt-cla6} imply the second assertion of the proposition.


\section{Proof of Proposition~\ref{prop:unique-saddle-point}}
\label{app:prop-unique-saddle-point}

\begin{lemma}
Consider any $C > R > R_{\infty}$ and $P \in \cP_{R}(\cX)$. For any $ \hat{\rho} \in \left.S(R,P)\right|_{\bbR_{+}}$, there exists a unique $\hat{Q} \in \cP_{P,W}(\cY)$, such that $(\hat{\rho}, \hat{Q}) \in S(R,P)$.
\label{lem:opt-cla7}
\end{lemma}

\begin{IEEEproof} Consider any $C > R > R_{\infty}$ and $P \in \cP_{R}(\cX)$. Let $ \hat{\rho} \in \left.S(R,P)\right|_{\bbR_{+}}$ be arbitrary. Existence of a $\hat{Q} \in \cP_{P,W}(\cY)$, such that $(\hat{\rho}, \hat{Q}) \in S(R,P)$ is guaranteed by the item (i) of saddle-point proposition, i.e. Proposition~\ref{prop:saddle-point}, hence we prove the uniqueness. 

To this end, note that owing to the item (ii) of saddle-point proposition, (Corollary~\ref{cor:opt-cor1} to be precise), $\hat{\rho} \in \bbR^{+}$. Moreover, the same result (Lemma~\ref{lem:opt-cla6} to be precise) also implies that any $\hat{Q} \in \cP_{P,W}(\cY)$, such that $(\hat{\rho}, \hat{Q}) \in S(R,P)$ satisfies $Q \in \tilde{\cP}_{P,W}(\cY)$ and attains the minimum in the following expression
\begin{equation}
\min_{Q \, \in \, \tilde{\cP}_{P,W}(\cY)}K_{R,P}(\hat{\rho},Q),
\label{eq:opt-cla7-pf1} 
\end{equation}
as a direct consequence of the definition of the saddle-point. However, item (i) of Lemma~\ref{lem:opt-cla1} implies that $K_{R,P}(\hat{\rho}, \cdot)$ is strictly convex on $\tilde{\cP}_{P,W}(\cY)$ and hence the minimizer of \eqref{eq:opt-cla7-pf1} is unique. 
\end{IEEEproof}

\begin{lemma}
Consider any $C > R > R_{\infty}$ and $P \in \cP_{R}(\cX)$. For any $\hat{Q} \in \left.S(R,P)\right|_{\cP_{P,W}(\cY)}$, 
\begin{equation}
\forall \, \rho \in \bbR^{+}, \,  \frac{\partial^{2} K_{R,P}(\rho, \hat{Q})}{\partial \rho^{2}} = -\frac{1}{(1+\rho)^{3}}\Lambda^{\prime \prime}_{\hat{Q},P}\left( \frac{\rho}{1+\rho}\right) < 0, \label{eq:opt-cla8-pf8}
\end{equation}
and there exists a unique $\hat{\rho} \in \bbR_{+}$, such that $(\hat{\rho}, \hat{Q}) \in S(R,P)$. 
\label{lem:opt-cla8}
\end{lemma}

\begin{IEEEproof}
Consider any $C > R > R_{\infty}$ and $P \in \cP_{R}(\cX)$. Let $\hat{Q} \in \left.S(R,P)\right|_{\cP_{P,W}(\cY)}$ be arbitrary. The existence of a $\hat{\rho} \in \bbR_{+}$, such that $(\hat{\rho}, \hat{Q}) \in S(R,P)$ is guaranteed by the item (i) of saddle-point proposition, i.e. Proposition~\ref{prop:saddle-point}, hence we prove the uniqueness. 

To this end, note that on account of the item (ii) of saddle-point proposition, (Lemma~\ref{lem:opt-cla6}, in particular), $\hat{Q} \in \tilde{\cP}_{P,W}(\cY)$, and hence $\Lambda_{\hat{Q},P}(\lambda)$ is infinitely differentiable with respect to $\lambda$ on $(0,1)$.
 
We first claim that 
\begin{equation}
\Lambda_{\hat{Q},P}^{\prime \prime}(\lambda) >0, \, \forall \lambda \in (0,1). 
\label{eq:var}
\end{equation}

For contradiction, suppose there exists a $ \lambda \in (0,1)$ such that $\Lambda_{\hat{Q},P}^{\prime \prime}(\lambda) = 0$. Note that
\begin{align}
\left[ \, \exists \, \lambda \in (0,1), \mbox{ s.t. } \Lambda_{\hat{Q},P}^{\prime \prime}(\lambda) = 0 \, \right] & \Leftrightarrow \left[ \, \exists \, \lambda \in (0,1), \mbox{ s.t. } \sum_{x \in \cS(P)}P(x)\mbox{Var}_{\tilde{W}_{\lambda,\hat{Q}}(\cdot|x)}\left[ \log \frac{\hat{Q}(Y)}{W(Y|x)} \right] =0 \, \right] \label{eq:opt-cla8-pf1} \\
 & \Leftrightarrow \left[ \, \exists \, \lambda \in (0,1), \mbox{ s.t. } \forall \, x \in \cS(P), \,    \mbox{Var}_{\tilde{W}_{\lambda,\hat{Q}}(\cdot|x)}\left[ \log \frac{\hat{Q}(Y)}{W(Y|x)} \right] =0 \, \right] \nonumber \\
 & \Leftrightarrow \left[ \, \exists \, \lambda \in (0,1), \mbox{ s.t. } \forall \, x \in \cS(P), \, \hat{Q}(y) = W(y|x) e^{\Lambda^{\prime}_{\hat{Q},x}(\lambda)}, \, \forall \, y \in \cS(W(\cdot|x)) \, \right], \label{eq:opt-cla8-pf2} 
\end{align}
where $\Lambda^{\prime}_{\hat{Q},x}(\lambda) \eqdef \mE_{\tilde{W}_{\lambda,\hat{Q}}(\cdot|x)}\left[ \log\frac{\hat{Q}(Y)}{W(Y|x)}\right]$ (cf. \eqref{eq:opt-der1}) and \eqref{eq:opt-cla8-pf1} follows from \eqref{eq:opt-der2}.

By the contradiction assumption, the left side of \eqref{eq:opt-cla8-pf1} is true. Fix any such $\lambda \in (0,1)$. Then, for any $\rho \in \bbR^{+}$, we have
\begin{align}
\Lambda_{\hat{Q},P}\left( \frac{\rho}{1+\rho}\right) & = \sum_{x \in \cS(P)}P(x) \log \sum_{y \in \cS(W(\cdot|x))}W(y|x)^{1/(1+\rho)}\hat{Q}(y)^{\rho/(1+\rho)}, \nonumber \\
 & = \frac{\rho}{1+\rho}\sum_{x \in \cS(P)}P(x)\Lambda_{\hat{Q},x}^{\prime}(\lambda), \label{eq:opt-cla8-pf3}
\end{align}
where \eqref{eq:opt-cla8-pf3} follows from \eqref{eq:opt-cla8-pf2}. We further have,
\begin{align}
\mE_{\mSP}(R,P) & = \max_{\rho \in \bbR_{+}}K_{R,P}(\rho, \hat{Q}), \label{eq:opt-cla8-pf4} \\
& = \max\left\{0, \sup_{\rho \in \bbR^{+}}K_{R,P}(\rho, \hat{Q})\right\}, \label{eq:opt-cla8-pf5}
\end{align}
where \eqref{eq:opt-cla8-pf4} follows by recalling the definition of the saddle-point and the item (i) of saddle-point proposition, i.e. Proposition~\ref{prop:saddle-point}, and \eqref{eq:opt-cla8-pf5} follows by noting the fact that $K_{R,P}(0,Q)=0$ for all $Q \in \cP_{P,W}(\cY)$.

Also, \eqref{eq:opt-cla8-pf3} implies that 
\begin{align}
\sup_{\rho \in \bbR^{+}}K_{R,P}(\rho, \hat{Q}) & = \sup_{\rho \in \bbR^{+}}\left\{ -\rho R - \rho \sum_{x \in \cS(P)} P(x) \Lambda^{\prime}_{\hat{Q}, x}(\lambda)\right\} \nonumber \\
 & = \sup_{\rho \in \bbR^{+}}-\rho\left\{ R + \Lambda^{\prime}_{\hat{Q},P}(\lambda)\right\}, \label{eq:opt-cla8-pf6}
\end{align}
where \eqref{eq:opt-cla8-pf6} follows by recalling \eqref{eq:opt-der1}. Equations \eqref{eq:opt-cla8-pf5} and \eqref{eq:opt-cla8-pf6} clearly imply that either $\mE_{\mSP}(R,P) = \infty$, which is impossible since $R > R_{\infty}$, or $\mE_{\mSP}(R,P) =0$, which is impossible since $P \in \cP_{R}(\cX)$. Hence, \eqref{eq:var} follows. A direct calculation reveals that \eqref{eq:var} implies \eqref{eq:opt-cla8-pf8}. 

Next, recalling the definition of the saddle-point, we note that any $\hat{\rho} \in \bbR_{+}$ such that $(\hat{\rho}, \hat{Q}) \in S(R,P)$ satisfies
\begin{align}
K_{R,P}(\hat{\rho}, \hat{Q}) & = \max_{\rho \in \bbR_{+}}K_{R,P}(\rho, \hat{Q}) \nonumber \\
 & = \max_{\rho \in \bbR^+}K_{R,P}(\rho, \hat{Q}), \label{eq:opt-cla8-pf7}
\end{align}
where \eqref{eq:opt-cla8-pf7} follows by recalling the assumption that $P \in \cP_R(\cX)$. Equation \eqref{eq:opt-cla8-pf8} ensures that $K_{R,P}(\cdot, \hat{Q})$ is strictly concave on $\bbR^{+}$ and hence the maximizer of the right side of \eqref{eq:opt-cla8-pf7} is unique. 
\end{IEEEproof}

In order to conclude the proof, fix any $C > R > R_{\infty}$ and $P \in \cP_{R}(\cX)$ and observe that (e.g. \cite[Proposition VII.4.1.3]{urruty93}) $S(R,P) = \left.S(R,P)\right|_{\bbR_{+}} \times \left.S(R,P)\right|_{\cP_{P,W}(\cY)}$. Combining this fact with Lemmas~\ref{lem:opt-cla7} and \ref{lem:opt-cla8} implies that $S(R,P)$ is a singleton, which was to be shown.


\section{Proof of Proposition~\ref{prop:diff-ESP}}
\label{app:prop-diff-ESP}
First, we define the set of Lagrange multipliers of $\mE_{\mSP}(R,P)$ as follows:
For any $R>R_{\infty}$ and $P \in \cP(\cX)$, 
\begin{equation}
\mathcal{L}(R,P) \eqdef  \left\{ \hat{\rho} \in \bbR_{+} \, : \, \hat{\rho} \mbox{ attains } \max_{\rho \, \in \, \bbR_{+}}\min_{V \in \cP(\cY|\cX)} [\mD(V||W|P) + \rho (\mI(P;V) -R)]\right\}.
\label{eq:opt-lag-mult}
\end{equation}

\begin{lemma}
For any $R > R_{\infty}$ and $P \in \cP(\cX)$, we have $\mathcal{L}(R,P) = \left.S(R,P)\right|_{\bbR_{+}}$. 
\label{lem:opt-cla13}
\end{lemma}

\begin{IEEEproof}
First of all, owing to the positivity of the relative entropy, it is easy to verify that 
\begin{equation}
\mI(P;V) = \min_{Q \in \cP(\cY)}\mD(V||Q|P),
\label{eq:mutual-inf}
\end{equation}
which, in turn, implies that (by solving the convex optimization problem)
\begin{equation}
\forall \, \rho \in \bbR_{+}, \, \min_{V \in \cP(\cY|\cX)}\{ \mD(V||W|P) + \rho (\mI(P;V) - R)\} = \min_{Q \in \cP(\cY)}\left\{ -\rho R - (1+\rho)\Lambda_{Q,P}\left( \frac{\rho}{1+\rho}\right)\right\}.
\label{eq:opt-cla13-pf1}
\end{equation}
Further, since for any $Q \in \cP(\cY)$, $\Lambda_{Q,P}(0) = 0$ and for any $\rho \in \bbR^{+}$, $\Lambda_{Q, P}\left( \frac{\rho}{1+\rho}\right) = -\infty$, if $Q \notin \cP_{P,W}(\cY)$ (cf. item (iv) of Remark~\ref{rem:opt-rem1}), we have
\begin{equation}
\min_{Q \in \cP(\cY)}\left\{ -\rho R - (1+\rho)\Lambda_{Q,P}\left( \frac{\rho}{1+\rho}\right)\right\} = \inf_{Q \in \cP_{P,W}(\cY)} \left\{ -\rho R - (1+\rho)\Lambda_{Q,P}\left( \frac{\rho}{1+\rho}\right)\right\}.
\label{eq:opt-cla13-pf2}
\end{equation}
Lastly, \cite[Lemma 36.2]{rockafellar70} ensures that $\hat{\rho} \in \left.S(R,P)\right|_{\bbR_{+}}$ if and only if $\hat{\rho}$ attains $\max_{\rho \in \bbR_{+}}\left\{ \inf_{Q \in \cP_{P,W}(\cY)}K_{R,P}(\rho,Q)\right\}$, which (owing to \eqref{eq:opt-cla13-pf1} and \eqref{eq:opt-cla13-pf2}) implies that $\mathcal{L}(R,P) = \left.S(R,P)\right|_{\bbR_{+}}$.
\end{IEEEproof}

\begin{lemma}
For any $R > R_{\infty}$ and $P \in  \cP(\cX)$, we have $\left.S(R,P)\right|_{\bbR_{+}} = - \partial \mE_{\mSP}(\cdot,P)(R)$, where $\partial \mE_{\mSP}(\cdot,P)(R)$ is the \emph{subdifferential of $\mE_{\mSP}(\cdot,P)$ at $R$} (cf. \cite[page 215]{rockafellar70}).
\label{lem:opt-cla14}
\end{lemma}

\begin{IEEEproof}
We note that (cf. \cite[Theorem 29.1]{rockafellar70}\footnote{Strictly speaking, this result is stated for a finite dimensional Euclidean space. However, one can represent the stochastic matrices in $\bbR^{|\cX||\cY|}$ and update each function accordingly and easily check this representation obeys the conditions of the aforementioned theorem. This reasoning applies to the similar situations in the sequel.}) $\mathcal{L}(R,P) =  - \partial \mE_{\mSP}(\cdot,P)(R)$. The claim follows by recalling Lemma~\ref{lem:opt-cla13}.
\end{IEEEproof}

Uniqueness of the saddle-point proposition, i.e. Proposition~\ref{prop:unique-saddle-point}, and Lemma~\ref{lem:opt-cla14} immediately imply that for any $C > R > R_{\infty}$ and $P \in \cP_{R}(\cX)$, 
\begin{equation}
\left.S(R,P)\right|_{\bbR_{+}} = - \left.\frac{\partial \, \mE_{\mSP}(r,P)}{\partial r}\right|_{r = R}.
\label{eq:opt-cla14-pf1}
\end{equation}
By recalling the definition of $\rho^\ast_{R,P}$ (e.g. \eqref{eq:opt-rho-star}), \eqref{eq:opt-cla14-pf1} implies that
\[ 
\rho^\ast_{R,P} = - \left.\frac{\partial \, \mE_{\mSP}(r,P)}{\partial r}\right|_{r = R},
\]
which was to be shown. 


\section{Proof of Proposition~\ref{prop:cont-saddle-point}}
\label{app:prop-cont-saddle-point}
Let $C > R > R_{\infty}$ be arbitrary. Fix any $P_{0} \in \cP_{R}(\cX)$ and consider any $\{ P_{k}\}_{k \geq 1}$ such that $P_{k} \in \cP_{R}(\cX)$, $\forall \, k \in \bbZ^{+}$ and $\lim_{n \rightarrow \infty}P_{k} = P_{0}$.

We begin with showing the continuity of $\rho^{\ast}_{R,\cdot}$. Recalling \eqref{eq:opt-rho-star} and the differentiability of $\mE_{\mSP}(\cdot,P)$ proposition, i.e. Proposition~\ref{prop:diff-ESP}, we have
\begin{equation}
\forall k \in \bbZ_{+}, \, \rho_{R, P_{k}}^{\ast} = - \left.\frac{\partial \, \mE_{\mSP}(r,P_{k})}{\partial r}\right|_{r = R}. 
\label{eq:opt-cla15-pf1}
\end{equation}
Further, continuity of $\mE_{\mSP}(\cdot, \cdot)$ on $(R_\infty,\infty) \times \cP(\cX)$ (e.g. Lemma~\ref{lem:opt-cla12}) implies that
\begin{equation}
\lim_{k \rightarrow \infty}\mE_{\mSP}(R,P_{k}) = \mE_{\mSP}(R,P_{0}).
\label{eq:opt-cla15-pf2}
\end{equation}
On account of \eqref{eq:opt-cla15-pf1}, \eqref{eq:opt-cla15-pf2} and a continuity result of  Hiriart-Urruty and Lemar\'echal (\cite[Corollary VI.6.2.8]{urruty93}) we conclude that 
\begin{equation*}
\lim_{k \rightarrow \infty}\rho_{R,P_{k}}^{\ast}= \rho_{R,P_{0}}^{\ast},
\end{equation*}
which implies that $\rho^{\ast}_{R,\cdot}$ is continuous on $\cP_R(\cX)$.

Next, we claim the continuity of $Q_{R,\cdot}^{\ast}$. Owing to the compactness of $\cP(\cY)$, there exists a subsequence $\{k_{n}\}_{n \geq 1}$ such that $\lim_{n \rightarrow \infty} Q^{\ast}_{R,P_{k_{n}} } = Q_{0}$ for some $Q_{0} \in \cP(\cY)$. Consider such a subsequence. 

Recalling the saddle-point proposition, i.e. Proposition~\ref{prop:saddle-point}, and the definitions of $\rho^\ast_{R,\cdot}$ and $Q^\ast_{R,\cdot}$ (e.g. \eqref{eq:opt-rho-star} and \eqref{eq:opt-q-star}), we have
\begin{equation}
\forall \, n \in \bbZ^{+}, \, \mE_{\mSP}(R,P_{k_{n}}) = - R \rho_{R,P_{k_{n}}}^{\ast} - (1 + \rho_{R,P_{k_{n}}}^{\ast}) \Lambda_{Q^{\ast}_{R,P_{k_{n}}}, P_{k_{n}}}\left( \frac{\rho_{R,P_{k_{n}}}^{\ast}}{1+\rho_{R,P_{k_{n}}}^{\ast}}\right).
\label{eq:opt-cla16-pf1}
\end{equation}

Next, we define $f: \bbR_{+}\times \bbR^{+} \rightarrow \bbR$, such that $f(a,b) \eqdef a^{b}$ for any $(a,b) \in \bbR_{+} \times \bbR^{+}$ and note that $f$ is continuous on $\bbR_{+} \times \bbR^{+}$. Using this, the continuity of $\rho^\ast_{R,\cdot}$ and  $\log(\cdot)$, we deduce that 
\begin{equation}
\lim_{n \rightarrow \infty} \Lambda_{Q^{\ast}_{R,P_{k_{n}}},P_{k_{n}}}\left( \frac{\rho^{\ast}_{R,P_{k_{n}}}}{1+\rho^{\ast}_{R,P_{k_{n}}}}\right) = \Lambda_{Q_{0}, P_{0}}\left( \frac{\rho^{\ast}_{R,P_{k_{0}}}}{1+\rho^{\ast}_{R, P_{k_{0}}}}\right).
\label{eq:opt-cla16-pf3}
\end{equation}
\eqref{eq:opt-cla16-pf1}, \eqref{eq:opt-cla16-pf3} and the continuity of $\rho^\ast_{R,\cdot}$ implies that
\begin{align}
-R \rho_{R,P_{0}}^{\ast} - (1+\rho_{R,P_{0}}^{\ast} ) \Lambda_{Q_{0}, P_{0}}\left( \frac{\rho_{R,P_{0}}^{\ast}}{1+\rho_{R,P_{0}}^{\ast}}\right) & = \mE_{\mSP}(R,P_{0}) \nonumber \\
 & = \min_{Q \, \in \, \cP_{P,W}(\cY)}\left\{ -R \rho_{R,P_{0}}^{\ast}- (1+\rho_{R,P_{0}}^{\ast}) \Lambda_{Q, P_{0}}\left( \frac{\rho_{R,P_{0}}^{\ast}}{1+\rho_{R,P_{0}}^{\ast}}\right)\right\} \label{eq:opt-cla16-pf4} \\
 & = \min_{Q \, \in \, \cP(\cY)}\left\{ -R \rho_{R,P_{0}}^{\ast} - (1+\rho_{R,P_{0}}^{\ast}) \Lambda_{Q, P_{0}}\left( \frac{\rho_{R,P_{0}}^{\ast}}{1+\rho_{R,P_{0}}^{\ast}}\right)\right\}, \label{eq:opt-cla16-pf5} 
\end{align}
where \eqref{eq:opt-cla16-pf4} follows from recalling the definition of the saddle-point and \eqref{eq:opt-cla16-pf5} follows from item (iv) of Remark~\ref{rem:opt-rem1}. The uniqueness of the saddle-point proposition, i.e. Proposition~\ref{prop:unique-saddle-point}, the definition of $Q^\ast_{R,P}$ (e.g. \eqref{eq:opt-q-star}) and \eqref{eq:opt-cla16-pf5} imply that $Q_{0} = Q^{\ast}_{R,P_{0}}$. Since $\{ k_{n}\}_{n \geq 1}$ is arbitrary, we conclude that
\begin{equation*}
\lim_{k \rightarrow \infty}Q^{\ast}_{R,P_{k}} = Q^{\ast}_{R,P_{0}},
\end{equation*}
which implies that $Q^\ast_{R, \cdot}$ is continuous on $\cP_R(\cX)$. Hence, we conclude the proof.


\section{Proof of Theorem~\ref{lem:lem-opt3}}
\label{app:lem:lem-opt3}

Fix an arbitrary $C(W) > R > R_{\infty}$ and $P \in \cP_{R}(\cX)$. Define $L(V,\rho) \eqdef \mD(V||W|P) + \rho(\mI(P;V) - R)$, for any $ V \in \cP(\cY|\cX)$ and $\rho \in \bbR_{+}$. We have
\begin{equation}
\mE_{\mSP}(R,P)  = \min_{V \, \in \, \cP(\cY|\cX)}  \sup_{\rho \, \in \, \bbR_{+}} L(V, \rho)  = \max_{\rho \, \in \, \bbR_{+}} \min_{V \, \in \, \cP(\cY|\cX)} L(V, \rho), \label{eq:thrm-hyp-exp-pf1}
\end{equation}
where the second equality follows from \eqref{eq:opt-lem1.1-pf1}. \eqref{eq:thrm-hyp-exp-pf1} ensures that $L(\cdot, \cdot)$ has a saddle-point on $\cP(\cY|\cX) \times \bbR_{+}$. It is well-known that (e.g. \cite[Corollary 28.3.1]{rockafellar70}) $\hat{V} \in \cP(\cY|\cX)$ is a minimizer of $\mE_{\mSP}(R,P)$ if and only if there exists some $\hat{\rho} \in \bbR_{+}$, such that $(\hat{V}, \hat{\rho})$ is a saddle-point of $L(\cdot, \cdot)$. 

Recalling the definition of the saddle-point, the definition of $\rho^\ast_{R,P}$ (e.g. \eqref{eq:opt-rho-star}), \eqref{eq:opt-lag-mult} and Lemma~\ref{lem:opt-cla13}, we conclude that an equivalent condition for $V_{R,P}^{\ast}$ to be an optimizer of $\mE_{\mSP}(R,P)$ is 
\begin{equation}
V_{R,P}^{\ast} \, \in \, \arg \, \min _{V \in \cP(\cY|\cX)} L(V, \rho_{R,P}^{\ast}).
\label{eq:thrm-hyp-exp-pf2}
\end{equation}

Further, 
\begin{align}
\mE_{\mSP}(R,P) & = \min _{V \in \, \cP(\cY|\cX)} L(V, \rho_{R,P}^{\ast}) \label{eq:thrm-hyp-exp-pf3} \\
& = \min_{Q \, \in \, \cP(\cY)} \min_{V \in \, \cP(\cY|\cX)} \left\{ \mD(V||W|P) + \rho_{R,P}^{\ast}[\mD(V||Q|P) - R] \right\}  \label{eq:thrm-hyp-exp-pf4} \\
 & \leq \min_{V \in \, \cP(\cY|\cX)} \left\{ \mD(V||W|P) + \rho_{R,P}^{\ast}[\mD(V||Q^{\ast}_{R,P}|P) - R] \right\}  \nonumber \\
 & \leq K_{R,P}(\rho^{\ast}_{R,P}, Q^{\ast}_{R,P})  \label{eq:thrm-hyp-exp-pf5}\\
 & = \mE_{\mSP}(R,P), \label{eq:thrm-hyp-exp-pf6}
\end{align}
where \eqref{eq:thrm-hyp-exp-pf3} follows from \eqref{eq:thrm-hyp-exp-pf2}, \eqref{eq:thrm-hyp-exp-pf4} follows from \eqref{eq:mutual-inf}, \eqref{eq:thrm-hyp-exp-pf5} follows by plugging in $\tilde{W}_{\frac{\rho_{R,P}^{\ast}}{1+\rho_{R,P}^{\ast}}, Q^{\ast}_{R,P}}$ (cf. \eqref{eq:opt-tilted-channel}) and \eqref{eq:thrm-hyp-exp-pf6} follows from the saddle-point proposition, i.e. Proposition~\ref{prop:saddle-point} and the uniqueness of the saddle-point proposition, i.e. Proposition~\ref{prop:unique-saddle-point}. Hence, we deduce that
\begin{equation}
\min _{V \in \cP(\cY|\cX)} L(V, \rho_{R,P}^{\ast}) = \min_{ V \in \cP(\cY|\cX)} \left\{ \mD(V||W|P) + \rho^{\ast}_{R,P} [\mD(V||Q^{\ast}_{R,P}|P) - R]\right\} = \mE_{\mSP}(R,P),
\label{eq:thrm-hyp-exp-pf6.5}
\end{equation}
and $\tilde{W}_{\frac{\rho_{R,P}^{\ast}}{1+\rho_{R,P}^{\ast}}, Q^{\ast}_{R,P}}$ is an optimizer of $\min_{V \in \, \cP(\cY|\cX)} \left\{ \mD(V||W|P) + \rho_{R,P}^{\ast}[\mD(V||Q^{\ast}_{R,P}|P) - R] \right\}$. Moreover, since 
\[ 
L(V, \rho_{R,P}^{\ast}) \leq \mD(V||W|P) + \rho_{R,P}^{\ast}[\mD(V||Q^{\ast}_{R,P}|P) - R], \forall \, V \in \cP(\cY|\cX),
\]
\eqref{eq:thrm-hyp-exp-pf6.5} further implies that $\tilde{W}_{\frac{\rho_{R,P}^{\ast}}{1+\rho_{R,P}^{\ast}}, Q^{\ast}_{R,P}} \in \arg \, \min _{V \in \cP(\cY|\cX)} L(V, \rho_{R,P}^{\ast})$, and hence $\tilde{W}_{\frac{\rho_{R,P}^{\ast}}{1+\rho_{R,P}^{\ast}}, Q^{\ast}_{R,P}}$ is a minimizer of $\mE_{\mSP}(R,P)$, owing to \eqref{eq:thrm-hyp-exp-pf2}.

Next, we note that on account of \eqref{eq:opt-cla6-pf2}, for any $Q \in \tilde{\cP}_{P,W}(\cY)$, we have
\begin{equation}
\frac{\partial \Lambda_{Q, P}\left( \frac{\rho^{\ast}_{R,P}}{1+\rho^{\ast}_{R,P}}\right)}{\partial Q(y)} = \frac{\rho^{\ast}_{R,P}}{1+\rho^{\ast}_{R,P}} \sum_{x \in \cS(P)}P(x) \frac{W(y|x)^{1/(1+\rho_{R,P}^{\ast})} Q(y)^{-1/(1+\rho_{R,P}^{\ast})}}{\sum_{\tilde{y} \in \cY} W(\tilde{y}|x)^{1/(1+\rho_{R,P}^{\ast})}Q(\tilde{y})^{\rho_{R,P}^{\ast}/(1+\rho_{R,P}^{\ast})}}, 
\label{eq:thrm-hyp-exp-pf7}
\end{equation}
for all $y \in \cS(Q)$. Moreover, \eqref{eq:opt-cla6-pf3} implies that for any $Q \in \tilde{\cP}_{P,W}(\cY)$, 
\begin{equation}
\frac{\partial \Lambda_{Q, P}\left( \frac{\rho^{\ast}_{R,P}}{1+\rho^{\ast}_{R,P}}\right)}{\partial Q(y)} = 0, \, \forall \, y \notin \cS(Q).
\label{eq:thrm-hyp-exp-pf8}
\end{equation}
KKT conditions that $Q^\ast_{R,P}$ satisfies, i.e. \eqref{eq:opt-cla6-pf5} and \eqref{eq:opt-cla6-pf6}, coupled with \eqref{eq:thrm-hyp-exp-pf7} and \eqref{eq:thrm-hyp-exp-pf8} (by choosing $\delta = \frac{\rho_{R,P}^{\ast}}{1+\rho_{R,P}^{\ast}}$ to ensure that $Q_{R,P}^{\ast}$ sums to $1$) imply that
\begin{equation}
Q_{R,P}^{\ast}(y) = \sum_{x \in \cS(P)}P(x) \frac{W(y|x)^{1/(1+\rho_{R,P}^{\ast})} Q^{\ast}_{R,P}(y)^{\rho_{R,P}^{\ast}/(1+\rho_{R,P}^{\ast})}}{\sum_{\tilde{y} \in \cY} W(\tilde{y}|x)^{1/(1+\rho_{R,P}^{\ast})}Q^{\ast}_{R,P}(\tilde{y})^{\rho_{R,P}^{\ast}/(1+\rho_{R,P}^{\ast})}}, \, \forall \, y \in \cY.
\label{eq:thrm-hyp-exp-pf9}
\end{equation}
Clearly, \eqref{eq:thrm-hyp-exp-pf9} implies that
\[
\sum_{x \in \cS(P)}P(x) \tilde{W}_{\frac{\rho_{R,P}^{\ast}}{1+\rho_{R,P}^{\ast}}, Q^{\ast}_{R,P}}(y|x) = Q_{R,P}^{\ast}(y), \, \forall \, y \in \cY,
\]
which, in turn, implies that (since $\tilde{W}_{\frac{\rho_{R,P}^{\ast}}{1+\rho_{R,P}^{\ast}}, Q^{\ast}_{R,P}}$ is an optimizer of $\mE_{\mSP}(R,P)$)
\begin{equation}
\mI\left(P; \tilde{W}_{\frac{\rho_{R,P}^{\ast}}{1+\rho_{R,P}^{\ast}}, Q^{\ast}_{R,P}}\right) = \mD\left(\tilde{W}_{\frac{\rho_{R,P}^{\ast}}{1+\rho_{R,P}^{\ast}}, Q^{\ast}_{R,P}}||Q_{R,P}^{\ast}|P\right) \leq R.
\label{eq:thrm-hyp-exp-pf10}
\end{equation}

Next, we conclude the proof as follows. First, 
\begin{align}
\me_{\mSP}(R,P) & = \inf_{V \in \cP(\cY|\cX)} \sup_{\rho \in \bbR_{+}} \left\{ \mD(V||W|P) + \rho[\mD(V||Q^{\ast}_{R,P}|P) - R]\right\}  \nonumber \\
 & \geq \inf_{V \in \cP(\cY|\cX)} \left\{ \mD(V||W|P) + \rho_{R,P}^{\ast}[\mD(V||Q^{\ast}_{R,P}|P) - R]\right\}  \nonumber \\
 & = \mE_{\mSP}(R,P), \label{eq:thrm-hyp-exp-pf11}
\end{align}
where \eqref{eq:thrm-hyp-exp-pf11} follows from \eqref{eq:thrm-hyp-exp-pf6.5}. 

On the other hand, \eqref{eq:thrm-hyp-exp-pf10} and the fact that $\tilde{W}_{\frac{\rho_{R,P}^{\ast}}{1+\rho_{R,P}^{\ast}}, Q^{\ast}_{R,P}} $ is a minimizer of $\mE_{\mSP}(R,P)$ ensure that
\begin{equation}
\me_{\mSP}(R,P) \leq \mD\left(  \tilde{W}_{\frac{\rho_{R,P}^{\ast}}{1+\rho_{R,P}^{\ast}}, Q^{\ast}_{R,P}}  || W | P \right) = \mE_{\mSP}(R,P).
\label{eq:thrm-hyp-exp-pf12}
\end{equation}
Combining \eqref{eq:thrm-hyp-exp-pf11} and \eqref{eq:thrm-hyp-exp-pf12}, we infer that
\begin{equation*}
\me_{\mSP}(R,P) = \min_{V \in \cP(\cY|\cX) \, : \, \mD(V||Q^{\ast}_{R,P}|P) \leq R}\mD(V||W|P) = \mE_{\mSP}(R,P),
\end{equation*}
which was to be shown. 


\section{Analysis of the case $P \in \cP_{R,\nu}^c$}
\label{app:trivial-types}

First, we define the following set: $\cP_{W}(\cY|\cX) \eqdef \{ V \in \cP(\cY|\cX) \, : \, \forall \, x \in \cX, \, V(\cdot|x) \ll W(\cdot|x)\}$. One can check the following via elementary calculations. 

\begin{lemma}
$\cP_{W}(\cY|\cX)$ is convex and compact. 
\label{lem:opt-cla9}
\end{lemma}

Next result will also be used in different parts of the paper. 
\begin{lemma}
$\mE_{\mSP}(\cdot,\cdot)$ is continuous on $(R_{\infty}, \infty) \times \cP(\cX)$. 
\label{lem:opt-cla12}
\end{lemma}

\begin{IEEEproof} The proof follows similar lines to those of \cite[Lemma 2.2.2]{csiszar-korner81}, which proves continuity of the rate-distortion function.  

First, note that given any $P \in \cP(\cX)$, $\mE_{\mSP}(\cdot, P)$ is convex on $(R_{\infty}, \infty)$. Fix an arbitrary $(R_{0},P_{0}) \in (R_{\infty}, \infty) \times \cP(\cX)$ and a sequence $\left\{ (R_{n},P_{n}) \right\}_{n \geq 1}$ such that $(R_{n},P_{n}) \in (R_{\infty}, \infty) \times \cP(\cX)$ and $\lim_{n \rightarrow \infty}(R_{n},P_{n}) = (R_{0}, P_{0})$.

Because of the convexity, $\mE_{\mSP}(\cdot, P_{0})$ is continuous on $(R_{\infty}, \infty)$. Hence, for any $\epsilon \in \bbR^{+}$ one can choose $V \in \cP(\cY|\cX)$ such that $\mI(P_{0};V) < R_{0}$ and $\mD(V||W|P_{0}) < \mE_{\mSP}(R_{0},P_{0}) + \epsilon$. Moreover, on account of continuity of $\mD(V||W|\cdot)$ and $\mI(\cdot;V)$, we have
\[
\mD(V||W|P_{n}) < \mE_{\mSP}(R_{0},P_{0}) + 2 \epsilon, \quad \mI(P_{n};V) \leq R_{n}, 
\]
for sufficiently large $n$, which, in turn, implies that 
\begin{equation}
\limsup_{n \rightarrow \infty} \mE_{\mSP}(R_{n},P_{n}) \leq \mE_{\mSP}(R_{0},P_{0}). 
\label{eq:opt-cla12-pf1}
\end{equation}

Conversely, let $V_{n} \in \cP(\cY|\cX)$ be a minimizer of $\mE_{\mSP}(R_{n}, P_{n})$ and w.l.o.g. suppose\footnote{To see why this does not yield a loss of generality, first note that since $\mE_{\mSP}(R_{n},P_{n}) < \infty$, we necessarily have $V_{n}(\cdot|x) \ll W(\cdot|x)$, for all $x \in \cS(P_n)$. On the other hand, $x \notin \cS(P_n)$ does not affect  neither the cost nor the constraint and hence the corresponding rows of the alternate channel, i.e. optimization variable of $\mE_{\mSP}(R_n, P_n)$, can be chosen arbitrarily without affecting optimality.} $V_{n} \in \cP_{W}(\cY|\cX)$. Let $\{ n_{k}\}_{k \geq 1}$ be a subsequence  such that 
\begin{equation}
\lim_{k \rightarrow \infty} \mE_{\mSP}(R_{n_{k}},P_{n_{k}}) = \liminf_{n \rightarrow \infty} \mE_{\mSP}(R_{n}, P_{n}), 
\label{eq:opt-cla12-pf2}
\end{equation}
and 
\begin{equation}
\lim_{k \rightarrow \infty}V_{n_{k}} = V,
\label{eq:opt-cla12-pf3}
\end{equation}
for some $V \in \cP_{W}(\cY|\cX)$. Note that existence of such a subsequence is ensured by the compactness of $\cP_{W}(\cY|\cX)$ (cf. Lemma~\ref{lem:opt-cla9}). Equation \eqref{eq:opt-cla12-pf3} further implies that
\begin{align}
\lim_{k \rightarrow \infty}\mI(P_{n_{k}};V_{n_{k}}) & = \mI(P_0;V) \leq R_0, \label{eq:opt-cla12-pf4} \\
\lim_{k \rightarrow \infty} \mD(V_{n_{k}}||W|P_{n_{k}}) & = \mD(V||W|P_0), \label{eq:opt-cla12-pf5}
\end{align}
where \eqref{eq:opt-cla12-pf4} follows from the continuity of $\mI(\cdot;\cdot)$ and \eqref{eq:opt-cla12-pf5} follows from the continuity of $\mD(\cdot||W|\cdot)$ on $\cP_{W}(\cY|\cX) \times \cP(\cX)$. Equations \eqref{eq:opt-cla12-pf2}, \eqref{eq:opt-cla12-pf4} and \eqref{eq:opt-cla12-pf5} imply that
\begin{equation}
\mE_{\mSP}(R_{0},P_{0}) \leq \liminf_{n \rightarrow \infty}\mE_{\mSP}(R_{n},P_{n}). \label{eq:opt-cla12-pf6}
\end{equation}
Equations \eqref{eq:opt-cla12-pf1} and \eqref{eq:opt-cla12-pf6} imply that
\[
\lim_{n \rightarrow \infty}\mE_{\mSP}(R_{n},P_{n}) = \mE_{\mSP}(R_{0},P_{0}).
\]
\end{IEEEproof}

Consider any $R_{\infty} < R < C$. For any $\nu \in \bbR^{+}$,
\begin{equation}
\cP_{R,\nu}(\cX) \eqdef \{ P \in \cP(\cX) \, : \, \mE_{\mSP}(R,P) \geq \nu \}. 
\label{eq:SCA-Pdelta-def}
\end{equation}
Let 
\begin{equation}
\epsilon \eqdef (R - R_{\infty})/2,
\label{eq:epsilon}
\end{equation}
and fix an arbitrary $a \in (1,2)$. Note that since $\mE_{\mSP}(\cdot)$ is convex, it is easy to see that it is Lipschitz continuous on $[R- \epsilon, R]$ (e.g. \cite[Theorem 10.4]{rockafellar70}), i.e. there exists $L \in \bbR^{+}$, such that
\begin{equation}
 \forall \, r_{1}, r_{2} \in [R-\epsilon, R], \quad  |\mE_{\mSP}(r_{1}) - \mE_{\mSP}(r_{2})| \leq L |r_1 - r_2|
\label{eq:SCA-Lip}
\end{equation}

Next, we consider an arbitrary $\nu \in \bbR^{+}$ satisfying:
\begin{equation}
\nu \leq \min \left\{ (a-1) , \, \frac{\epsilon}{2}, \, \frac{\mE_{\mSP}(R)(2-a)}{a(2L+1)}\right\}.
\label{eq:SCA-nu-def}
\end{equation}

We claim that\footnote{Owing to  Lemma~\ref{lem:opt-cla12}, the max is well-defined.}
\begin{equation}
\max_{P \, \in \, \textrm{cl}(\cP_{R,\nu}(\cX)^c)}\mE_{\mSP}(R - \nu, P) \leq \frac{\mE_{\mSP}(R)}{a}.
\label{eq:SCA-upperbound}
\end{equation}

For contradiction, suppose 
\begin{equation}
\max_{P \, \in \, \textrm{cl}(\cP_{R,\nu}(\cX)^c)}\mE_{\mSP}(R - \nu, P) > \frac{\mE_{\mSP}(R)}{a},
\label{eq:SCA-upperbound-pf1}
\end{equation}
with a maximizer $\tilde{P}$. Since $\mE_{\mSP}(\cdot, \tilde{P})$ is convex and non-decreasing, \eqref{eq:SCA-upperbound-pf1} implies that 
\begin{equation}
\mE_{\mSP}(R-2 \nu, \tilde{P}) > \frac{\mE_{\mSP}(R)}{a} + \nu\left( \frac{\mE_{\mSP}(R)}{a \nu} - 1 \right) = \frac{2\mE_{\mSP}(R)}{a}-\nu. \label{eq:SCA-upperbound-pf2}
\end{equation}

Further, owing to \eqref{eq:SCA-nu-def}, we have
\begin{equation}
\frac{2\mE_{\mSP}(R)}{a}-\nu \geq \mE_{\mSP}(R) + 2L\nu. 
\label{eq:SCA-upperbound-pf3}
\end{equation}
Also, \eqref{eq:SCA-Lip} and \eqref{eq:SCA-nu-def} imply that
\begin{equation}
\mE_{\mSP}(R-2\nu) \leq \mE_{\mSP}(R) + 2L \nu.
\label{eq:SCA-upperbound-pf4}
\end{equation}

Plugging \eqref{eq:SCA-upperbound-pf3} and \eqref{eq:SCA-upperbound-pf4} into \eqref{eq:SCA-upperbound-pf2} yields
\begin{equation*}
\mE_{\mSP}(R-2 \nu, \tilde{P}) > \mE_{\mSP}(R-2\nu),
\end{equation*}
which is a contradiction, by recalling the definition of $\mE_{\mSP}(\cdot)$, and hence \eqref{eq:SCA-upperbound} follows.

Let $P \in \textrm{cl}(\cP_{R,\nu}(\cX)^c)$ be arbitrary. We have
\begin{align}
(1+\nu)\mE_{\mSP}(R-\nu, P) & \leq \frac{(1+\nu)\mE_{\mSP}(R)}{a} \label{eq:SCA-1} \\
 & \leq \mE_{\mSP}(R), \label{eq:SCA-2}
\end{align}
where \eqref{eq:SCA-1} follows from \eqref{eq:SCA-upperbound} and \eqref{eq:SCA-2} follows from \eqref{eq:SCA-nu-def}. 

Let $(f, \varphi)$ be an $(N,R)$ constant composition code with common composition $P \in \textrm{cl}(\cP_{R,\nu}(\cX)^c)$. For all sufficiently large $N$, which only depends on $\nu, |\cX|, |\cY|$, we have
\begin{align}
e(f, \varphi) & \geq \frac{1}{2}\exp(-N(1+\nu)\mE_{\mSP}(R-\nu, P)) \label{eq:SCA-3} \\
 & \geq \frac{1}{2}\exp(-N\mE_{\mSP}(R)), \label{eq:SCA-4}
\end{align} 
where \eqref{eq:SCA-3} follows from the sphere packing lower bound for constant composition codes (cf. \cite[Theorem 2.5.3]{csiszar-korner81}) and \eqref{eq:SCA-4} follows from \eqref{eq:SCA-2}. Hence, we have the following lemma. 

\begin{lemma} Fix $R_{\infty} < R < C$ and $\nu>0$ satisfying \eqref{eq:SCA-nu-def}. Then, for all sufficiently large $N$, which only depends on $\nu, |\cX|$, and $|\cY|$, any $(N,R)$ constant composition code with common composition $P \in \textrm{cl}(\cP_{R,\nu}(\cX)^c)$ satisfies
\begin{equation}
e(f, \varphi) \geq \frac{1}{2}\exp(-N \mE_{\mSP}(R)). 
\label{eq:SCA-lem}
\end{equation}
\label{lem:SCA-lem}
\end{lemma}

\vspace{-.5cm}
\section{Proof of Lemma~\ref{lem:lem-R-new}}
\label{app:lem-R-new}
We begin with the proof of item (i). First, note that 
\begin{align}
\mD(V||W_{R,P}^{-}|P) & = \sum_{x \in \cS(P)} P(x) \sum_{y \in \cS(V(\cdot|x))} V(y|x) \log \frac{V(y|x)}{W_{R,P}^{-}(y|x)} \nonumber \\
 & = \sum_{x \in \cS(P)} P(x) \left\{ \log Q^{\ast}_{R,P}\{ \cS(W(\cdot|x)) \} + \mD(V(\cdot|x)||Q^{\ast}_{R,P})\right\} \label{eq:opt-e-tilde-e-equ-pf4}\\
 & = \mD(V||Q^{\ast}_{R,P}|P) + \sum_{x \in \cS(P)} P(x) \log Q^{\ast}_{R,P}\{ \cS(W(\cdot|x)) \}, \label{eq:opt-e-tilde-e-equ-pf5}
\end{align}
where \eqref{eq:opt-e-tilde-e-equ-pf4} follows from \eqref{eq:opt-e-tilde-e-equ-pf0.5}.

Similarly, 
\begin{align}
\mD(W^{-}_{R,P}||Q^{\ast}_{R,P}|P) & = \sum_{x \in \cS(P)} P(x) \sum_{y \in \cS(W(\cdot|x))} W^{-}_{R,P}(y|x) \log \frac{W^{-}_{R,P}(y|x) }{Q^{\ast}_{R,P}(y)} \nonumber \\
 & = - \sum_{x \in \cS(P)} P(x) \log Q^{\ast}_{R,P}\{ \cS(W(\cdot|x))\}\sum_{y \in \cS(W(\cdot|x))}W_{R,P}^{-}(y |x) \label{eq:opt-e-tilde-e-equ-pf6} \\
 & = - \sum_{x \in \cS(P)} P(x) \log Q^{\ast}_{R,P}\{ \cS(W(\cdot|x))\}, \label{eq:opt-e-tilde-e-equ-pf7}
\end{align}
where \eqref{eq:opt-e-tilde-e-equ-pf6} follows from the fact that $Q^{\ast}_{R,P} \in \tilde{\cP}_{P,W}(\cY)$ (cf. item (ii) of Proposition~\ref{prop:saddle-point}) and noting $W_{R,P}^{-}(\cdot|x) \equiv W(\cdot|x)$, for all $x \in \cX$. Plugging \eqref{eq:opt-e-tilde-e-equ-pf7} into \eqref{eq:opt-e-tilde-e-equ-pf5} gives the item (i) of the lemma. 

In order to prove the item (ii), observe that $(\rho_{R,P}^{\ast},Q_{R,P}^{\ast})$ is the unique saddle-point of $K_{R,P}(\cdot, \cdot)$. We have
\begin{align}
K_{R,P}(\rho_{R,P}^{\ast},Q_{R,P}^{\ast}) & = \max_{\rho \, \in \, \bbR_{+}}K_{R,P}(\rho,Q_{R,P}^{\ast}) \nonumber \\
 & = \max_{\rho \, \in \, \bbR^{+}}K_{R,P}(\rho,Q_{R,P}^{\ast}), \label{eq:lem-R-new-pf1}
\end{align}
where \eqref{eq:lem-R-new-pf1} follows by noting that $\mE_{\mSP}(R,P) = K_{R,P}(\rho_{R,P}^{\ast},Q_{R,P}^{\ast}) >0$ (cf. \eqref{eq:thrm-hyp-exp-pf6}) and $K_{R,P}(0,Q_{R,P}^{\ast}) = 0$. Observe that $\rho_{R,P}^{\ast} \in \bbR^{+}$ is the unique maximizer of the right side of \eqref{eq:lem-R-new-pf1} and hence 
\begin{equation}
\left. \frac{\partial K_{R,P}(\rho, Q^\ast_{R,P})}{\partial \rho} \right|_{\rho = \rho_{R,P}^{\ast}} = - R  - \Lambda_{Q_{R,P}^{\ast}, P}\left( \frac{\rho_{R,P}^{\ast}}{1+\rho_{R,P}^{\ast}}\right)-\frac{1}{(1+\rho_{R,P}^{\ast})}\Lambda^{\prime}_{Q_{R,P}^{\ast}, P}\left( \frac{\rho_{R,P}^{\ast}}{1+\rho_{R,P}^{\ast}}\right) = 0. \label{eq:lem-R-new-pf2}
\end{equation}

Further,
\begin{align}
\lim_{\lambda \uparrow 1} \Lambda_{Q_{R,P}^{\ast},P}(\lambda) & = \lim_{\lambda \uparrow 1} \sum_{x \in \cS(P)}P(x)\log \sum_{y \in \cS(W(\cdot|x))}W(y|x)^{1-\lambda}Q_{R,P}^{\ast}(y)^{\lambda} \nonumber \\
 & = \sum_{x \in \cS(P)}P(x)\log \lim_{\lambda \uparrow 1}\sum_{y \in \cS(W(\cdot|x))}W(y|x)^{1-\lambda}Q_{R,P}^{\ast}(y)^{\lambda} \nonumber \\
 & = \sum_{x \in \cS(P)}P(x)\log \sum_{y \in \cS(W(\cdot|x))}Q_{R,P}^{\ast}(y) \nonumber \\
 & = - \mD(W_{R,P}^{-}||Q_{R,P}^{\ast}|P), \label{eq:lem-R-new-pf3}
\end{align}
where \eqref{eq:lem-R-new-pf3} follows from \eqref{eq:opt-e-tilde-e-equ-pf7}. 

Moreover, recalling \eqref{eq:opt-w-minus} and \eqref{eq:opt-w-minus-1}, for any $x \in \cS(P)$
\begin{equation}
\lim_{\lambda \uparrow 1}\tilde{W}_{\lambda, Q_{R,P}^{\ast}}(y|x) = W_{R,P}^{-}(y|x), \label{eq:lem-R-new-pf4}
\end{equation}
for all $y \in \cY$. One can check that (e.g. \eqref{eq:opt-der1})
\[
\Lambda^\prime_{Q^\ast_{R,P},P}(\lambda) = \sum_{x \in \cS(P)} P(x) \mE_{\tilde{W}_{\lambda,Q^\ast_{R,P}}(\cdot|x)}\left[ \log \frac{Q^\ast_{R,P}(Y)}{W(Y|x)}\right],
\]
which, coupled with \eqref{eq:lem-R-new-pf4}, implies that
\begin{equation*}
\lim_{\lambda \uparrow 1} \Lambda^{\prime}_{Q_{R,P}^{\ast}, P}(\lambda) = \sum_{x \in \cS(P)}P(x)\sum_{y \in \cS(W(\cdot|x))}W_{R,P}^{-}(y|x)\log\frac{Q_{R,P}^{\ast}(y)}{W(y|x)} \in \bbR,
\end{equation*}
which, in turn, implies that
\begin{equation}
\lim_{\rho \rightarrow \infty}\frac{1}{(1+\rho)}\Lambda^{\prime}_{Q_{R,P}^{\ast},P}\left( \frac{\rho}{1+\rho}\right) = 0. \label{eq:lem-R-new-pf5}
\end{equation}
We have
\begin{align}
0 & > \lim_{\rho \rightarrow \infty}\frac{\partial K_{R,P}(\rho, Q^\ast_{R,P})}{\partial \rho} \label{eq:lem-R-new-pf6} \\
& = \lim_{\rho \rightarrow \infty} -R -\Lambda_{Q_{R,P}^{\ast},P}\left( \frac{\rho}{1+\rho}\right) - \frac{1}{(1+\rho)}\Lambda^{\prime}_{Q_{R,P}^{\ast},P}\left( \frac{\rho}{1+\rho}\right) \nonumber \\
& = \mD(W_{R,P}^{-}||Q_{R,P}^{\ast}|P) - R, \label{eq:lem-R-new-pf7}
\end{align}
where \eqref{eq:lem-R-new-pf6} follows from \eqref{eq:lem-R-new-pf2} and \eqref{eq:opt-cla8-pf8} and \eqref{eq:lem-R-new-pf7} follows from \eqref{eq:lem-R-new-pf3} and \eqref{eq:lem-R-new-pf5}. Hence, we conclude that $R > \mD(W^-_{R,P}||Q^\ast_{R,P}|P)$. 

\section{Proof of Proposition~\ref{prop:SLB}}
\label{app:SLB-pf}

First, the property (i) above ensures that $\Lambda_i$ is $\cC^\infty$ at $\eta$. Moreover, the property (ii) above implies that
\begin{equation}
\Lambda_n^\ast(q) = q \eta - \frac{1}{n}\sum_{i=1}^n\Lambda_i(\eta),
\label{eq:SLB-fenchel-eqv}
\end{equation}
since $\frac{1}{n} \sum_{i=1}^n \Lambda_i(\delta)$ is convex.

Next, from \eqref{eq:SLB-lambda-tilde}, we have
\[
\mE_{\tilde{\lambda}_i}[Z_i] = \frac{1}{M_i(\eta)} \int ze^{\eta z}d\lambda_i(z).
\]
Moreover, since $\Lambda_i$ is $\cC^\infty$ at $\eta$, we also have
\[
\Lambda_i^\prime(\eta) = \frac{M_i^\prime(\eta)}{M_i(\eta)} = \frac{1}{M_i(\eta)} \int ze^{\eta z}d\lambda_i(z). 
\]
And hence, we conclude that
\begin{equation}
\mE_{\tilde{\lambda}_i}[Z_i] = \Lambda_i^\prime(\eta).
\label{eq:SLB-remark2-1}
\end{equation}
Also, basic calculus reveals that
\begin{equation}
\Lambda_i^{\prime \prime}(\eta) = \frac{M_i^{\prime \prime}(\eta)}{M_i(\eta)} - \Lambda^\prime_i(\eta)^2.
\label{eq:SLB-remark2-2}
\end{equation}
Moreover, since $\Lambda_i$ is $\cC^\infty$ at $\eta$, we also have 
\[ 
M_{i}^{\prime \prime}(\eta) = \int z^{2}e^{\eta z} d\lambda_{i}(z), 
\]
which, in turn, implies that (recall \eqref{eq:SLB-lambda-tilde})
\begin{equation}
\mE_{\tilde{\lambda}_{i}}[Z_{i}^{2}] = \frac{M_{i}^{\prime \prime}(\eta)}{M_{i}(\eta)}.
\label{eq:SLB-remark2-3}
\end{equation}
Plugging \eqref{eq:SLB-remark2-1} and \eqref{eq:SLB-remark2-3} into \eqref{eq:SLB-remark2-2} yields
\begin{equation}
\mV_{\tilde{\lambda}_{i}}[Z_{i}] = \Lambda_{i}^{\prime \prime}(\eta). 
\label{eq:SLB-remark2-4}
\end{equation}
Furthermore, recalling \eqref{eq:SLB-lambda-tilde}, it is obvious that $\tilde{\lambda}_{i} \ll \lambda_{i}$. Moreover, since $Z_{i}$ are real-valued and $e^{\eta z - \Lambda_{i}(\eta)} >0$, for all $z \in \bbR$, we have
\[
\frac{d\lambda_{i}}{d\tilde{\lambda}_{i}}(z) = e^{-\eta z + \Lambda_{i}(\eta)},
\]
which, in turn, implies that $\lambda_{i} \ll \tilde{\lambda}_{i}$. Hence, we conclude that $\lambda_{i} \equiv \tilde{\lambda}_{i} $.

Next, we claim that 
\begin{equation}
m_{2,n} >0. 
\label{eq:SLB-remark2-positive-m2}
\end{equation}

To see this, note that for any $i \in \{ 1, \ldots, n\}$,
\begin{align}
\left[ \Lambda_{i}^{\prime \prime}(\eta) = 0 \right]  \Longleftrightarrow & \left[ Z_{i} = \Lambda_{i}^{\prime}(\eta)  \quad \tilde{\lambda_{i}}-\mbox{(a.s.)}\right] \label{eq:SLB-remark2-5} \\
  \Longleftrightarrow & \left[  Z_{i} = \Lambda_{i}^{\prime}(\eta)  \quad \lambda_{i}-\mbox{(a.s.)}\right] \label{eq:SLB-remark2-6} \\
  \Longrightarrow & \left[ \mV[Z_{i}]=0\right], \label{eq:SLB-remark2-7}
\end{align}
where \eqref{eq:SLB-remark2-5} follows from items (i) and (ii) of this remark, \eqref{eq:SLB-remark2-6} follows since $\lambda_{i} \equiv \tilde{\lambda}_{i}$. From the assumption that $\sum_{i=1}^{n}\mV[Z_{i}] >0$ and \eqref{eq:SLB-remark2-7}, we conclude that $\sum_{i=1}^{n}\Lambda_{i}^{\prime \prime}(\eta) >0$, which implies \eqref{eq:SLB-remark2-positive-m2}.  

We continue as follows:
\begin{align}
\mu_{n}([q, \infty)) & = \int_{\{ \hat{S}_{n} \geq q\}}\lambda_{1}(dz_{1})\ldots\lambda_{n}(dz_{n}) \nonumber \\
 & = \int_{\{ \hat{S}_{n} \geq q\}}e^{\sum_{i=1}^{n}[\Lambda_{i}(\eta)-\eta z_{i}]} \tilde{\lambda}_{1} (dz_{1})\ldots\tilde{\lambda}_{n}(dz_{n}) \label{eq:SLB-CofM1} \\
& = e^{\sum_{i=1}^{n}\Lambda_{i}(\eta)}\mE_{\tilde{\mu}_{n}}\left[ \b1_{ \{ \hat{S}_{n} \geq q \} } e^{-n \eta \hat{S}_{n}}\right] \label{eq:SLB-CofM2} \\
& = e^{-n\Lambda^{\ast}_{n}(q)} \mE_{\tilde{\mu}_{n}}\left[ \b1_{ \{ \hat{S}_{n} \geq q \} } e^{-n[\eta \hat{S}_{n} - \eta q]}\right], \label{eq:SLB-CofM3}                                                                 
\end{align}
where \eqref{eq:SLB-CofM1} follows from \eqref{eq:SLB-lambda-tilde}, \eqref{eq:SLB-CofM2} follows by recalling the definition of $\tilde{\mu}_{n}$ and \eqref{eq:SLB-CofM3} follows from \eqref{eq:SLB-fenchel-eqv}.

Note that \eqref{eq:SLB-remark2-1} and \eqref{eq:SLB-remark2-4} imply that
\[
\mE_{\tilde{\lambda}_{i}}[T_{i}]=0, \quad \mV_{\tilde{\lambda}_{i}}[T_{i}]=\Lambda^{\prime \prime}_{i}(\eta).
\]
 and note that
\begin{equation}
\hat{S}_{n} = \sqrt{m_{2,n}}\frac{W_{n}}{n}+q, \label{eq:SLB-CofM4}
\end{equation}
which, in turn, implies that
\begin{equation}
\{ \hat{S}_{n} \geq q\} = \left\{ \sqrt{m_{2,n}} \frac{W_{n}}{n} \geq 0\right\}. \label{eq:SLB-CofM5}
\end{equation}
Plugging \eqref{eq:SLB-CofM4} and \eqref{eq:SLB-CofM5} into \eqref{eq:SLB-CofM3} yields 
\begin{align}
\mu_{n}([q, \infty)) & = e^{-n\Lambda^{\ast}_{n}(q)} \mE_{\tilde{\mu}_{n}}\left[ \b1_{\{W_{n} \geq 0\} } e^{-\eta \sqrt{m_{2,n}}W_{n}}\right] \nonumber \\
 & \geq e^{-n\Lambda^{\ast}_{n}(q)} \mE_{\tilde{\mu}_{n}}\left[ \b1_{ \{ W_{n} \geq 0\} } e^{-\sqrt{m_{2,n}}W_{n}}\right] \label{eq:SLB-CofM6} \\
 & = e^{-n\Lambda^{\ast}_{n}(q)} \int_{0}^{\infty}e^{-x \sqrt{m_{2,n}}}dF_{n}(x) \nonumber \\
 & = e^{-n\Lambda^{\ast}_{n}(q)} \int_{0}^{\infty} \sqrt{m_{2,n}}e^{-x \sqrt{m_{2,n}}}[F_{n}(x) - F_{n}(0)]dx \label{eq:SLB-CofM7} \\
 & = e^{-n\Lambda^{\ast}_{n}(q)} \int_{0}^{\infty} e^{-t}[F_{n}(t/\sqrt{m_{2,n}}) - F_{n}(0) ]dt, \label{eq:SLB-CofM8}
\end{align}
where $F_{n}$ denotes the distribution function of $W_{n}$ when $Z_{i}$ are independent with the marginals $\tilde{\lambda}_{i}, \, $\eqref{eq:SLB-CofM6} follows from the fact that $\eta \leq 1$, \eqref{eq:SLB-CofM7} follows via integration by parts and \eqref{eq:SLB-CofM8} follows by letting $t \eqdef x \sqrt{m_{2,n}}$.

Note that since $\Lambda_{i}$ is $\cC^{\infty}$ at $\eta$, $m_{3,n} < \infty$ and hence (recall \eqref{eq:SLB-remark2-positive-m2}), $K_{n}(\eta) \in \bbR^{+}$. 

Next, Berry-Esseen Theorem (cf. \cite[Theorem III.1]{esseen45}. We use the particular instance of this theorem given by \cite[eq. (III.15), pg. 43]{esseen45}) implies that
\begin{equation}
|F_n(x) - \Phi(x)| \leq c \frac{ m_{3,n}}{m_{2,n}^{3/2}}, \, \forall x \in \bbR, 
\label{eq:SLB-BEB-0}
\end{equation}
where $c$ is an absolute constant and can be chosen as $30/4$. Using \eqref{eq:SLB-BEB-0}, we deduce that
\begin{align}
F_{n}(t/\sqrt{m_{2,n}}) & \geq \Phi(t/\sqrt{m_{2,n}}) -\frac{c m_{3,n}}{m_{2,n}^{3/2}}  \label{eq:SLB-BEB-1}\\
F_{n}(0) & \leq \Phi(0) + \frac{c m_{3,n}}{m_{2,n}^{3/2}}. \label{eq:SLB-BEB-2}
\end{align}

Using \eqref{eq:SLB-BEB-1} and \eqref{eq:SLB-BEB-2} we get
\begin{align}
F_{n}(t/\sqrt{m_{2,n}}) - F_{n}(0) & \geq \Phi(t/\sqrt{m_{2,n}}) - \Phi(0) -\frac{2c m_{3,n}}{m_{2,n}^{3/2}} \nonumber \\
 & \geq \phi(0)\frac{t}{\sqrt{m_{2,n}}} + \frac{\phi^{\prime}(\bar{t})}{2}\left( \frac{t}{\sqrt{m_{2,n}}}\right)^{2} - \frac{2 c m_{3,n}}{m_{2,n}^{3/2}} \label{eq:SLB-BEB3} \\
& \geq \frac{t}{\sqrt{2\pi m_{2,n}}} \left[1 - \frac{2 c m_{3,n}\sqrt{2 \pi}}{t m_{2,n}} \right]  - \frac{1}{2\sqrt{2\pi}}\bar{t}e^{-\bar{t}^2/2}\frac{t^2}{m_{2,n}}, \label{eq:SLB-BEB4} 
\end{align}
where $0 \leq \bar{t} \leq t/\sqrt{m_{2,n}}$, \eqref{eq:SLB-BEB3} follows from Taylor's Theorem and \eqref{eq:SLB-BEB4} follows by noting $\phi^\prime(x) = -\frac{x}{\sqrt{2\pi}}e^{-x^2/2}$.

Observe that $\bbR_+ \ni x \mapsto x e^{-x^2/2} \leq e^{-1/2} <1$, which, in turn, implies that (recall \eqref{eq:SLB-BEB4})
\begin{align}
F_{n}(t/\sqrt{m_{2,n}}) - F_{n}(0) & \geq \frac{t}{\sqrt{2\pi m_{2,n}}} \left[1 - \frac{2 c m_{3,n}\sqrt{2 \pi}}{t m_{2,n}} \right]  - \frac{t^2}{2\sqrt{2 \pi}m_{2,n}}. 
\label{eq:SLB-BEB4.5}
\end{align}

It is easy to check that 
\begin{align}
\int_{K_n(\eta)}^\infty t e^{-t}dt & = e^{-K_n(\eta)}(1+K_n(\eta)), \label{eq:SLB-BEB5} \\
\int_{K_n(\eta)}^\infty t^2 e^{-t}dt & = e^{-K_n(\eta)}[1+(1+K_n(\eta))^2]. \label{eq:SLB-BEB6}
\end{align}

Hence, 
\begin{align}
 \int_{0}^{\infty} e^{-t}\left[F_{n}\left(\frac{t}{\sqrt{m_{2,n}}}\right) - F_{n}(0) \right]dt  & \geq \int_{K_n(\eta)}^\infty e^{-t}\left[F_{n}\left(\frac{t}{\sqrt{m_{2,n}}}\right) - F_{n}(0) \right]dt \nonumber \\
& \geq \frac{e^{-K_n(\eta)}}{\sqrt{2 \pi m_{2,n}}}\left( 1 - \frac{1+(1+K_n(\eta))^2}{2\sqrt{m_{2,n}}}\right), \label{eq:SLB-BEB7}
\end{align}
where \eqref{eq:SLB-BEB7} follows from \eqref{eq:SLB-BEB4.5}, \eqref{eq:SLB-BEB5} and \eqref{eq:SLB-BEB6}. Plugging \eqref{eq:SLB-BEB7} into \eqref{eq:SLB-CofM8} yields
\begin{equation}
\mu_n([q,\infty)) \geq \frac{e^{-n\Lambda_n^\ast(q)} e^{-K_n(\eta)}}{\sqrt{2 \pi m_{2,n}}}\left( 1 - \frac{1+(1+K_n(\eta))^2}{2\sqrt{m_{2,n}}}\right). 
\label{eq:SLB-BEB8}
\end{equation}

Clearly, if \eqref{eq:SLB-final-0} holds, then \eqref{eq:SLB-BEB8} implies \eqref{eq:SLB-final}, which was to be shown.


\section{Proof of Lemma~\ref{thrm:cont-1}}
\label{app:continuity-pf}
Let $(\lambda_{0},P_{0}) \in (0,1]\times \cP_{R}(\cX)$ be arbitrary. Further, consider any $\{ (\lambda_{k},P_{k}) \}_{k \geq 1}$ such that $(\lambda_{k},P_{k}) \in (0,1]\times \cP_{R}(\cX)$, for all $k \in \bbZ^+$ and $\lim_{k \rightarrow \infty}(\lambda_{k},P_{k}) = (\lambda_{0}, P_{0})$.

 Note that for all sufficiently large $k \in \bbZ^+$, $\cS(P_{0}) \subset \cS(P_{k})$. Consider such a $k \in \bbZ^+$. Recalling \eqref{eq:zerovar-0-der1-1} and \eqref{eq:zerovar-0-der1-2}, we have
\begin{equation}
\Lambda_{0,P_{k}}^{\prime}(\lambda_{k}) = \sum_{x \in \cS(P_{0})}P_{k}(x)\mE_{\tilde{W}_{\lambda_{k}, P_{k}}(\cdot|x)}\left[ \log \frac{W_{R,P_{k}}^{-}(Y|x)}{W(Y|x)}\right] + \sum_{x \in \cS(P_{0})^{c}}P_{k}(x)\mE_{\tilde{W}_{\lambda_{k}, P_{k}}(\cdot|x)}\left[ \log \frac{W_{R,P_{k}}^{-}(Y|x)}{W(Y|x)}\right]. \label{eq:thrm-cont-1-pfi1} 
\end{equation}

Using the continuity of the saddle-point proposition, i.e. Proposition~\ref{prop:cont-saddle-point}, \eqref{eq:opt-w-minus}, \eqref{eq:opt-rem2} and the continuity of $\log(\cdot)$, it is easy to see that
\begin{equation*}
\lim_{k \rightarrow \infty} P_{k}(x)\mE_{\tilde{W}_{\lambda_{k}, P_{k}}(\cdot|x)}\left[ \log \frac{W_{R,P_{k}}^{-}(Y|x)}{W(Y|x)}\right] = P_{0}(x)\mE_{\tilde{W}_{\lambda_{0}, P_{0}}(\cdot|x)}\left[ \log \frac{W_{R,P_{0}}^{-}(Y|x)}{W(Y|x)}\right], \, \forall \, x \in \cS(P_{0}) ,
\end{equation*}
which, in turn, implies that 
\begin{equation}
\lim_{k \rightarrow \infty} \sum_{x \in \cS(P_{0})} P_{k}(x)\mE_{\tilde{W}_{\lambda_{k}, P_{k}}(\cdot|x)}\left[ \log \frac{W_{R,P_{k}}^{-}(Y|x)}{W(Y|x)}\right] = \sum_{x \in \cS(P_{0})}P_{0}(x)\mE_{\tilde{W}_{\lambda_{0}, P_{0}}(\cdot|x)}\left[ \log \frac{W_{R,P_{0}}^{-}(Y|x)}{W(Y|x)}\right].
\label{eq:thrm-cont-1-pfi2}
\end{equation}

Next, we claim that 
\begin{equation}
\lim_{k \rightarrow \infty}  P_{k}(x)\mE_{\tilde{W}_{\lambda_{k}, P_{k}}(\cdot|x)}\left[ \log \frac{W_{R,P_{k}}^{-}(Y|x)}{W(Y|x)}\right] = 0, 
\label{eq:thrm-cont-1-pfi3}
\end{equation}
for any $x \in \cS(P_{0})^{c}$. To see this, fix an arbitrary $x \in \cS(P_{0})^{c}$. If $x \in \cS(P_{k})$ for only finite number of $k$, then owing to \eqref{eq:opt-w-minus},  \eqref{eq:thrm-cont-1-pfi3} is trivially true; hence suppose this is not the case. Let $\{ k_{n}\}_{n \geq 1}$ be an arbitrary subsequence such that $x \in \cS(P_{k_{n}})$, for all $n \in \bbZ^{+}$. Owing to the compactness of $\cP(\cY|\cX)$ (swtiching to a subsubsequence if necessary) there exists $W_{0}(\cdot|x) \in \cP(\cY|\cX)$, such that 
\begin{equation}
\lim_{n \rightarrow \infty} W^{-}_{R,P_{k_{n}}}(\cdot|x) = W_{0}(\cdot|x).
\label{eq:thrm-cont-1-pfi4}
\end{equation}
Since $W^{-}_{R,P_{k_{n}}}(\cdot|x) \ll W(\cdot|x)$ for all $n \in \bbZ^{+}$, it is easy to see that (cf. proof of Lemma~\ref{lem:opt-cla9}) $W_{0}(\cdot|x) \ll W(\cdot|x)$. This fact, along with the continuity of $\log(\cdot)$ and \eqref{eq:thrm-cont-1-pfi4}, implies that 
\begin{equation}
\lim_{m \rightarrow \infty}\mE_{\tilde{W}_{\lambda_{k_{n}}, P_{k_{n}}}(\cdot|x)}\left[ \log \frac{W_{R,P_{k_{n}}}^{-}(Y|x)}{W(Y|x)}\right] = \mE_{\tilde{W}_{\lambda_{0}, W_{0}(\cdot|x)}}\left[ \log\frac{W_{0}(Y|x)}{W(Y|x)}\right] < \infty.
\label{eq:thrm-cont-1-pfi5}
\end{equation}
Noting $\lim_{m \rightarrow \infty}P_{k_{n_{m}}}(x)=P_{0}(x) = 0$ and the arbitrariness of the subsequence, \eqref{eq:thrm-cont-1-pfi5} implies \eqref{eq:thrm-cont-1-pfi3}. Plugging \eqref{eq:thrm-cont-1-pfi2} and \eqref{eq:thrm-cont-1-pfi3} into \eqref{eq:thrm-cont-1-pfi1} implies that
\[
\lim_{k \rightarrow \infty}\Lambda_{0,P_{k}}^{\prime}(\lambda_{k}) = \Lambda_{0,P_{0}}^{\prime}(\lambda_{0}),
\]
and hence we conclude $\Lambda_{0,\cdot}^{\prime}(\cdot)$ is continuous on $(0,1] \times \cP_R(\cX)$. 

By following exactly the same steps given above and noting the continuity of $(\cdot)^2$ (resp. $|\cdot|^3$), one can conclude the continuity of $\Lambda_{0,\cdot}^{\prime \prime}(\cdot)$ (resp. $m_{0,3}(\cdot, \cdot)$) on $(0,1] \times \cP_R(\cX)$.

Finally, the proof of the item (iv) follows from the similar arguments given in the proof of the item (i).


\section{Proof of Lemma~\ref{lem:lem-easier-test-2}}
\label{app:regularity-lemma-pf}
Let $s^\ast(R,P,r) \in \bbR_+$ be as defined in \eqref{eq:thrm-sstar-def}. Since it is the unique maximizer of $\tilde{\me}_{\mSP}(R,P,r)$, it should satisfy 
\begin{equation}
r = \left.\frac{\partial e_o(s, P)}{\partial s}\right|_{s^{\ast}_{R,P,r}}. 
\label{eq:zerovar-3.b.2}
\end{equation}
It is easy to verify that
\begin{equation}
\frac{\partial e_o(s,P)}{\partial s}  = -\Lambda_{0,P}\left(\frac{s}{1+s}\right) - \frac{1}{1+s}\Lambda_{0,P}^\prime\left(\frac{s}{1+s}\right), \label{eq:zerovar-3.b.3} 
\end{equation}
Owing to \eqref{eq:zerovar-3.b.2} and \eqref{eq:zerovar-3.b.3}, we have
\begin{equation}
r = -\Lambda_{0,P}\left(\frac{s^\ast(R,P,r)}{1+s^\ast(R,P,r)}\right) - \frac{1}{(1+s^\ast(R,P,r))}\Lambda_{0,P}^\prime\left( \frac{s^\ast(R,P,r)}{1+s^\ast(R,P,r)}\right).
\label{eq:lem-easier-test-2-pf1}
\end{equation}
By noting (recall \eqref{eq:opt-e0-def})
\[
e_o(s^\ast(R,P,r), P) = -(1+s^\ast(R,P,r))\Lambda_{0,P}\left( \frac{s^\ast(R,P,r)}{1+s^\ast(R,P,r)}\right),
\]
Lemma~\ref{lem:opt-lem5}, Corollary~\ref{cor:subdiff}, \eqref{eq:thrm-sstar-def} and \eqref{eq:lem-easier-test-2-pf1} imply that
\begin{equation}
\tilde{\me}_{\mSP}(R,P,r) = \frac{s^\ast(R,P,r)}{1+s^\ast(R,P,r)}\Lambda_{0,P}^\prime\left( \frac{s^\ast(R,P,r)}{1+s^\ast(R,P,r)}\right) - \Lambda_{0,P}\left( \frac{s^\ast(R,P,r)}{1+s^\ast(R,P,r)}\right).
\label{eq:lem-easier-test-2-pf2}
\end{equation}
Due to \eqref{eq:lem-easier-test-2-pf1} and \eqref{eq:lem-easier-test-2-pf2}, we get
\begin{equation}
\Lambda_{0,P}^\prime\left( \frac{s^\ast(R,P,r)}{1+s^\ast(R,P,r)}\right) = \tilde{\me}_{\mSP}(R,P,r) - r.
\label{eq:lem-easier-test-2-pf3}
\end{equation} 

Using \eqref{eq:fenchel-0}, \eqref{eq:lem-easier-test-2-pf2} and \eqref{eq:lem-easier-test-2-pf3}, it is easy to see that (recall \eqref{eq:SLB-fenchel-eqv})
\[
\Lambda^\ast_{0,P}(\tilde{\me}_{\mSP}(R,P,r) - r) = \tilde{\me}_{\mSP}(R,P,r),
\]
which proves the item (i). 

Item (ii) follows immediately follows from \eqref{eq:zerovar-1-der1}, \eqref{eq:zerovar-1-der2}, \eqref{eq:fenchel-0}, \eqref{eq:fenchel-1} and the item (i).

In order to see the item (iii), first note that $\tilde{\me}_{\mSP}(R,P,\cdot)$ is a non-increasing function. Further, it is clear that $\tilde{\me}_{\mSP}(R,P,0) = \mD(W_{R,P}^{-}||W|P)$ and $\tilde{\me}_{\mSP}(R,P,\mD(W||W_{R,P}^{-}|P)) = 0$. These observations, along with \eqref{eq:zerovar-der1-0}, \eqref{eq:zerovar-der1-1} and the positive variance lemma, i.e. Lemma~\ref{thrm:zerovar-prop4}, suffice to conclude the existence and uniqueness of $\eta{R,P,r} \in (0,1)$ with the stated property. Finally, recalling \eqref{eq:lem-easier-test-2-pf3}, one can see that $\eta(R,P,r) = \frac{s^{\ast}(R,P,r)}{1+s^{\ast}(R,P,r)}$, which completes the proof of the lemma.


\end{document}